\documentclass[11pt]{article}

\pdfoutput=1

\usepackage[utf8x]{inputenc}

\newcommand*{\colorboxed}{}
\def\colorboxed#1#{%
  \colorboxedAux{#1}%
}
\newcommand*{\colorboxedAux}[3]{%
  \begingroup
    \colorlet{cb@saved}{.}%
    \color#1{#2}%
    \boxed{%
      \color{cb@saved}%
      #3%
    }%
  \endgroup
}

\usepackage[dvipsnames]{xcolor}
\usepackage{amssymb,amsmath,amscd}
\usepackage{epsfig}
\usepackage{epstopdf}
\usepackage{latexsym}
\usepackage{graphicx}
\usepackage{booktabs}
\usepackage{bbm}
\usepackage[margin=15pt,small]{caption}
\usepackage{subcaption}
\usepackage{enumitem}
\usepackage{float}
\usepackage[toc]{appendix}
\usepackage[numbers,compress]{natbib}
\usepackage{tikz}
\usepackage[labelfont=bf]{caption}
\usepackage[export]{adjustbox}

\usepackage{braket,cellspace}
\usepackage{mathrsfs}
\usepackage{braket}
\usepackage{chngpage}
\usepackage{tabu}

\allowdisplaybreaks

\usetikzlibrary{positioning}
\definecolor{darkspringgreen}{rgb}{0.09, 0.45, 0.27}

\usepackage{datetime}
\usepackage[
      colorlinks=true,
      linkcolor=darkspringgreen,  
      urlcolor=darkspringgreen,    
      filecolor=darkspringgreen,     
      citecolor=darkspringgreen,
      linktocpage=true,
      pdfstartview=FitV,
      bookmarksopen=true     
      ]{hyperref}

\setlist{nolistsep}
\ifpdf
\fi

\let\oldbibliography\thebibliography
\renewcommand{\thebibliography}[1]{\oldbibliography{#1}
\setlength{\itemsep}{0pt}} 

\numberwithin{equation}{section} 

\begin{document}  

\begin{titlepage}

\begin{center} 

\vspace*{12mm}

{\LARGE \bf 
Quantum BMS transformations in conformally
\\[7pt]
flat space-times and holography}

\bigskip
\bigskip
\bigskip
\bigskip

{\bf Laura Donnay,${}^{(1)}$ Gaston Giribet${}^{(2)}$ and Felipe Rosso${}^{(3,4)}$\\ }
\bigskip
${}^{(1)}$ Institute for Theoretical Physics\\ 
Vienna University of Technology \\ 
A-1040 Vienna, Austria
\vskip 2mm
${}^{(2)}$ Physics Department, University of Buenos Aires  \\ and IFIBA-CONICET Ciudad Universitaria \\
Pabell\'on 1, Buenos Aires, 1428, Argentina
\vskip 2mm
${}^{(3)}$ Department of Physics and Astronomy 
\\
University of Southern California \\
Los Angeles, CA 90089, USA  \\
\vskip 2mm
${}^{(4)}$ Kavli Institute for Theoretical Physics \\
University of California, Santa Barbara, CA 93106, USA \\
\bigskip
\tt{
laura.donnay@tuwien.ac.at,
giribet@gmail.com,
felipero@usc.edu}  \\

\end{center}

\bigskip

\begin{abstract}
\noindent We define and study asymptotic Killing and conformal Killing vectors in $d$-dimensional Minkowski, (A)dS, $\mathbb{R}\times S^{d-1}$ and ${\rm AdS}_2\times S^{d-2}$. We construct the associated quantum charges for an arbitrary CFT and show they satisfy a closed algebra that includes the BMS as a sub-algebra (i.e. supertranslations and superrotations) plus a novel transformation  we call `superdilations'. We study representations of this algebra in the Hilbert space of the CFT, as well as the action of the finite transformations obtained by exponentiating the charges. In the context of the AdS/CFT correspondence, we propose a bulk holographic description in semi-classical gravity that reproduces the results obtained from CFT computations. We discuss the implications of our results regarding quantum hairs of asymptotically flat (near-)extremal black holes.
\end{abstract}

\vfill

\end{titlepage}


\newpage

\setcounter{tocdepth}{2}
\tableofcontents

\pagebreak

\section{Introduction}

Back in the sixties, Bondi, van der Burg, Metzner and Sachs (BMS) studied the symmetry algebra of asymptotically flat space-times at future and past null infinity $\mathcal{I}^\pm$ \cite{Sachs:1962wk,Sachs:1962zza,Bondi:1962px}, and to their surprise, found that instead of the finite dimensional Poincar\'e algebra, space-time translations were enhanced to an infinite dimensional sub-algebra they called supertranslations. It was later realized that by relaxing certain technical conditions, the Lorentz transformations could also be enhanced into what is commonly referred as superrotations \cite{Barnich:2010eb, Barnich:2011mi, Barnich:2016lyg,Compere:2018ylh}. In this paper, we call BMS algebra to the enhanced version containing both supertranslations and superrotations.\footnote{In the four dimensional case there are two different infinite dimensional extensions of the Lorentz algebra at null infinity, one involving two copies of the Virasoro algebra and the other ${\rm Diff}(S^2)$. In this work we shall mostly consider the extension involving ${\rm Diff}(S^{2})$. While a more appropriate term for these elements would be `super-Lorentz', we shall follow most of the literature and call them superrotations.} Infinite-dimensional symmetry algebras should be taken seriously, as they have shown to play an important role in other setups, such as in the AdS$_3$/CFT$_2$~\cite{Brown:1986nw} and Kerr/CFT \cite{Guica:2008mu} correspondence.

In recent years, BMS asymptotic symmetries have been investigated from several different promising perspectives. Studies of the gravitational scattering matrix in Minkowski \cite{Strominger:2013jfa} have lead to interesting relations between BMS symmetry, soft theorems \cite{He:2014laa}, and the so-called gravitational memory effects \cite{Strominger:2014pwa} (see \cite{Strominger:2017zoo} for a review and further references). The structure of BMS symmetry has also appeared at the horizon of classical black holes solutions \cite{Donnay:2015abr, Donnay:2016ejv}, and it has been suggested it supplies the necessary additional structure to provide a possible resolution of the black hole information paradox \cite{Hawking:2016msc,Hawking:2016sgy}.

The purpose of this paper is to study in detail quantum aspects of BMS symmetry in conformally flat space-times, aiming towards possible applications to holography and black hole physics. To do so, we consider a simple system obtained by placing an arbitrary conformal field theory (CFT) on a fixed $d$-dimensional space-time. Using conformal symmetry, we are able to study BMS transformations that are defined not only on the asymptotic null regions of Minkowski $\mathcal{I}^\pm$, but on certain Killing horizons $H_\pm$ on a variety of conformally flat space-times, including (A)dS, $\mathbb{R}\times S^{d-1}$ and ${\rm AdS}_2\times S^{d-2}$.\footnote{The BMS transformations we construct on the Killing horizons $H_\pm$ are closely related to previous work on horizon symmetries \cite{Donnay:2015abr, Donnay:2016ejv}.} All the results of this paper hold for arbitrary CFTs on any of these space-times and involve the following quantum aspects of BMS symmetry: the algebra satisfied by the charge operators, its representations on the Hilbert space of the CFT as well as the group action obtained by exponentiating the charges. For holographic CFTs that are well described by semi-classical Einstein gravity, we propose and provide evidence in favor of a holographic description of our CFT computations. Towards the end of this work we apply our construction to study some aspects of asymptotically flat (near-)extremal black holes.

Our analysis does not only involve the ordinary supertranslation and superrotation BMS transformation, but also a novel transformation that we call `superdilation'. We show how this asymptotic transformation naturally arises when considering conformal theories and include it in all our analysis throughout this paper.\footnote{See \cite{Haco:2017ekf,Adami:2020ugu,Nguyen:2020hot} for previous studies on the relation between asymptotic and conformal symmetries.} While we point out some issues associated to this novel superdilation transformation, it is still an interesting asymptotic symmetry that might be valuable in the appropriate setting.

\subsection{Summary of results}

We start in \textbf{section \ref{sec:1}} by describing a set of conformal transformations that map Minkowski to (A)dS$_d$, $\mathbb{R}\times S^{d-1}$ and ${\rm AdS}_2\times S^{d-2}$ (see figure \ref{fig:1} for a sketch of some of their Penrose diagrams). The future null boundary of Minkowski $\mathcal{I}^+$ is mapped to a Killing horizon $H_+$ in the curved space-time, e.g. for ${\rm AdS}_2\times S^{d-2}$ the surface $H_+$ corresponds to the future boundary of the Poincar\'e patch of ${\rm AdS}_2$ (left diagram in figure \ref{fig:1}). All the results of this paper apply to any of these space-times and their associated surfaces $\mathcal{I}^+$ or $H_+$.

We then construct asymptotic Killing and conformal Killing vectors on the surface $\mathcal{I}^+$ for Minkowski, and $H_+$ in the curved space-times. These vectors are a natural generalizations of translations, Lorentz transformations, special conformal transformations and dilations 
\begin{equation}\label{eq:232}
\big\lbrace
\xi_T(f),
\xi_R(Y)
\big\rbrace
\cup
\big\lbrace
\xi_S(h),
\xi_D(g)
\big\rbrace\ .
\end{equation}
The first two vectors depend on a function $f$ and a vector $Y^A$, and generate the ordinary BMS transformations, i.e. supertranslations and superrotations. The remaining vectors generalize special conformal transformations and dilations respectively and depend on two functions $h$ and $g$. These are given in (\ref{eq:159}) for $d=3$ and (\ref{eq:178}) for arbitrary dimensions, with the metric fall of conditions preserved by them near $\mathcal{I}^+$ ($H_+$) shown in tables \ref{table:6} and \ref{table:5}. Evaluating the vectors (\ref{eq:232}) on the surface $\mathcal{I}^+$ ($H_+$) we compute their algebra and find that a subset of these vectors given by
\begin{equation}\label{eq:172}
{\rm Asymptotic\,\,symmetries}
=
\big\lbrace
\xi_T(f),
\xi_R(Y)
\big\rbrace
\cup
\big\lbrace
\xi_D(g)
\big\rbrace\ ,
\end{equation}
satisfy the closed algebra in (\ref{eq:165}) and (\ref{eq:189}). This is an extension of the ordinary BMS algebra that includes the superdilation vector $\xi_D(g)$. It is the asymptotic transformations generated by the three vectors in (\ref{eq:172}) the one we study in the rest of the paper.

At this point we should issue a word of warning regarding the superdilation vector $\xi_D(g)$, since the metric fall-off conditions preserved by this vector are more singular than those obtained for the ordinary BMS vectors $\xi_T(f)$ and $\xi_R(Y)$. This means that we should be careful when studying superdilation transformations, as complications can (and will) arise when computing physical quantities associated to $\xi_D(g)$. More precisely, in the holographic analysis of section \ref{sec:4} we find that certain conserved charges associated to $\xi_D(g)$ diverge.

Despite these issues we have decided to study superdilations throughout this paper, since it is a novel and interesting transformation that might be useful in the appropriate setting. Nevertheless, we should stress that \textit{all} the results in this paper involving the BMS transformations generated by $\xi_T(f)$ and $\xi_R(Y)$ are not affected by any issue that might arise regarding $\xi_D(g)$. Readers that do not like superdilations can simply ignore our analysis involving $\xi_D(g)$.  

The quantum analysis of the transformation generated by the vectors in (\ref{eq:172}) starts in \textbf{section~\ref{sec:2}}, where we consider an arbitrary CFT in any of the conformally flat space-times introduced in section~\ref{sec:1}. We write the conserved charges in terms of the stress tensor operator $T_{\mu \nu}$, that in the Minkowski case are given by
\begin{equation}\label{eq:173}
\mathcal{T}(f)=\int_{\mathcal{I}^+}
dS^\mu T_{\mu \nu}\xi_T^\nu(f)\ ,
\qquad 
\mathcal{R}(Y)=
\int_{\mathcal{I}^+}
dS^\mu T_{\mu \nu}\xi_R^\nu(Y)\ ,
\qquad
\mathcal{D}(g)=
\int_{\mathcal{I}^+}
dS^\mu T_{\mu \nu}\xi_D^\nu(g)\ .
\end{equation}
We show how these operators can be mapped by a CRT transformation\footnote{${\rm CRT}$ is a discrete transformation analogous to ${\rm CPT}$, but instead of a complete spatial reflection $\vec{x}\rightarrow -\vec{x}$, it involves only a single component $x_1\rightarrow -x_1$. While in even space-time dimensions both ${\rm CRT}$ and ${\rm CPT}$ are symmetries of any QFT, ${\rm CPT}$ is not when the space-time dimensions is odd (see subsection 5.1 of \cite{Witten:2018lha}).} to the charges defined in the past region $\mathcal{I}^-$. Using the conformal transformation they can also be mapped to the charges at $H_+$ defined for the CFT in the curved space-times. We finish this section using the results in \cite{Cordova:2018ygx} (see also \cite{Kologlu:2019bco}) to prove the operators in (\ref{eq:173}) satisfy the same algebra as the associated vectors (\ref{eq:172}).

In \textbf{section \ref{sec:3}} we study the action of the quantum charges (\ref{eq:173}) on the Hilbert space of the CFT. Applying the transformations on the vacuum allows us to study the following states
\begin{equation}\label{eq:129}
\ket{f}\equiv e^{-i\mathcal{T}(f)}\ket{0}\ ,
\qquad \qquad
\ket{Y}\equiv e^{-i\mathcal{R}(Y)}\ket{0}\ ,
\qquad \qquad
\ket{g}\equiv e^{-i\mathcal{D}(g)}\ket{0}\ ,
\end{equation}
that are defined on $\mathcal{I}^\pm$ in Minkowski (or $H_\pm$ in the curved space-times). Remarkably, using the algebra satisfied by the charges together with some other natural ingredients, we are able to compute several features of these states in full generality, summarized in table~\ref{table:3}. Perhaps the stronger of these results is that the supertranslated vacuum $\ket{f}$ is equivalent to the vacuum $\ket{f}=\ket{0}$ for \textit{any} function $f$. This follows from the achronal Averaged Null Energy Condition (ANEC), proven for arbitrary QFTs in \cite{Faulkner:2016mzt,Hartman:2016lgu,Rosso:2020cub}. In the remaining of section \ref{sec:3} we study the algebra satisfied by the charges (\ref{eq:173}) and construct representations in the Hilbert space of the CFT, for three and four space-time dimensions. In both cases we are able to make concrete statements that provide further insight into the action of these asymptotic transformations on the Hilbert space.

In \textbf{section \ref{sec:4}} we propose a holographic description of the states (\ref{eq:129}), in the context of the AdS$_{d+1}$/CFT$_{d}$ correspondence where the bulk is well described by semi-classical gravity. As a first step we extend the boundary vectors (\ref{eq:172}) into the bulk 
\begin{equation}
{\rm Bulk \,\,vectors}=
\big\lbrace
\chi_T(f),
\chi_R(Y)
\big\rbrace
\cup
\big\lbrace
\chi_D(g)
\big\rbrace\ .
\end{equation}
While in principle there is an infinite number of ways of doing so, we fix them by imposing the following conditions:
\begin{itemize}
\item[1.] As we approach the boundary we must recover the boundary vectors $\chi_p\rightarrow \xi_p$ where $p=T,R,D$.
\item[2.] For some particular values of the functions $\left\lbrace f_0,Y_0^A,g_0 \right\rbrace$ entering in the definition of $\xi_p$ in (\ref{eq:172}), the vectors generate ordinary conformal isometries. When fixing the functions in this way for $\chi_p$, we require the bulk vectors to generate exact isometries of the ${\rm AdS}_{d+1}$ bulk space-time.
\item[3.] The algebra satisfied by the bulk vectors $\chi_p$ must be exactly the same as the one obtained for the boundary vectors $\xi_p$.
\end{itemize}
These conditions allow us to completely fix the bulk vectors according to (\ref{eq:151}) and (\ref{eq:230}), for a three and arbitrary dimensional boundary respectively.

Our proposal is that the boundary states (\ref{eq:129}) are described by a bulk geometry obtained  by acting on the pure ${\rm AdS}_{d+1}$ metric with the (finite) transformation generated by $\chi_p$. We denote the resulting metric as $g_{\mu \nu}(\chi_p)\equiv e^{\chi_p}(g_{\mu \nu}^{\rm AdS})$. The ordinary AdS/CFT dictionary then gives the usual mapping between boundary expectation values and gravitational bulk Noether charges. In table \ref{table:1} we summarize this holographic proposal, that we put to test in section \ref{sec:4} by computing the resulting bulk metrics and Noether charges, the final results shown in tables \ref{table:2} and \ref{table:8}. Comparing with the boundary CFT computations of section \ref{sec:3} (given in table \ref{table:3}) we find perfect agreement for all the quantities involving the ordinary BMS transformations $\chi_T(f)$ and $\chi_R(Y)$. This is strong evidence in favor of our holographic description of the states $\ket{f}$ and $\ket{Y}$ in (\ref{eq:129}).

\begin{table}[t]
\setlength{\tabcolsep}{10 pt} 
\centering
\begin{tabular}{  Sc | Sc  }
\specialrule{.13em}{0em}{0em}
Boundary CFT$_d$  &
Semi-classical gravity dual
\\
\specialrule{.05em}{0em}{0em}
$\displaystyle \ket{0}$  &
$\displaystyle g_{\mu \nu}^{\rm AdS}$ 
\\
$\displaystyle \xi_p$  &
$\displaystyle \chi_p$ 
\\
$\displaystyle 
\ket{\xi_p}\equiv e^{-i\widehat{Q}[\xi_p]}\ket{0}$  &
$\displaystyle 
g_{\mu \nu}(\chi_p)\equiv
e^{\chi_p}(g_{\mu \nu}^{\rm AdS})	$ 
\\
$\displaystyle
\braket{\xi_p|
\widehat{Q}[\xi_q]
|\xi_p}$  &
$\displaystyle 
Q_{g_{\mu \nu}(\chi_p)}[\chi_q]$ 
\\
\specialrule{.13em}{0em}{0em}
\end{tabular}
\caption{
Summary of our proposal for the holographic description of the boundary states (\ref{eq:129}), that correspond to $\ket{\xi_p}$ on the first column with $p=T,R,D$ respectively. The boundary charge $\widehat{Q}$ in the first column are written in (\ref{eq:173}), where we add a hat to remind ourselves it is an operator. The metric $g_{\mu \nu}(\chi_p)$ in the second column is obtained by acting on the pure ${\rm AdS}_{d+1}$ metric with the (finite) transformation generated by $\chi_p$. $Q_{g_{\mu \nu}(\chi_p)}[\chi_q]$ corresponds to the Noether charge associated to the vector $\chi_q$ computed in the metric $g_{\mu \nu}(\chi_p)$.
}\label{table:1}
\end{table}

For superdilations we get a very different result, as the bulk and boundary computations disagree on several instances. In particular, some of the bulk Noether charges associated to the bulk vector $\chi_D(g)$ diverge. A divergent charge associated to an asymptotic transformation is usually a sign that the metric fall-off condition preserved by the vector is too permissive, that is precisely what we previously noticed in the analysis of section \ref{sec:1} for the boundary vector $\xi_D(g)$ (see tables \ref{table:5} and \ref{table:3}). As a result, we do not interpret the disagreement between bulk and boundary computations as a failure of the holographic prescription, but as evidence that superdilation is not a well behaved asymptotic symmetry. It would be interesting to understand how this issue arises directly from the boundary CFT perspective. 

We end in \textbf{section \ref{sec:5}}, where we discuss the implications of our work regarding quantum hairs of asymptotically flat (near-)extremal black holes. Building on our computations and focusing on a CFT in ${\rm AdS}_2\times S^{d-2}$, we argue it is possible to construct an infinite family of zero energy quantum states on both the future and past horizons, and asymptotic regions. The states on these surfaces are not independent but related in a precise way by conformal and CRT symmetry. Several appendices include important technical results used throughout the paper.

\section{Asymptotic (conformal) Killing vectors}
\label{sec:1}

In this section we construct and study asymptotic Killing and conformal Killing vectors in $d$-dimensional Minkowski and a number of conformally flat space-times (see table \ref{table:4}). The algebra satisfied by these vectors includes the BMS as a sub-algebra (supertranslations and superrotations) together with a novel transformation that we call `superdilation'.

\subsection{Conformally flat space-times}
\label{subsec:1}

Let us start by considering the $d$-dimensional Minkowski metric written as
\begin{equation}\label{eq:174}
ds^2=-du^2+\frac{2dud\rho+d\Omega_{d-2}^2}{\rho^2}\ ,
\end{equation}
where $\rho=1/r\in \mathbb{R}_{\geq 0}$ and $u=t-r$, with $t$ and $r$ the ordinary time and radial coordinates in Minkowski. In these coordinates the null surface $\rho=0$ corresponds to future null infinity $\mathcal{I}^+$. An analogous coordinate system allows us to describe past null infinity. The metric in the unit sphere, $S^{d-2}$, can be parametrized in stereographic coordinates $\vec{y}\in \mathbb{R}^{d-2}$ as
\begin{equation}\label{eq:176}
d\Omega_{d-2}^2=
\frac{4\,d\vec{y} \cdot d\vec{y}}{(1+|\vec{y}\,|^2)^2}
\ ,
\end{equation}
with $|\vec{y}\,|=0,\infty$ corresponding to the Poles of the sphere.

There are several interesting conformal transformations we can apply to the Minkowski metric, obtained by rewriting (\ref{eq:174}) as
\begin{equation}
ds^2=w^2(x^\mu)
\left[
\frac{-\rho^2du^2+2dud\rho+d\Omega_{d-2}^2}
{\rho^2w^2(x^\mu)}
\right]\ ,
\end{equation}
and performing a Weyl rescaling that removes the conformal factor $w^2(x^\mu)$, so that the resulting space-time is given by
\begin{equation}\label{eq:175}
d\bar{s}^2=
\frac{-\rho^2du^2+2dud\rho+d\Omega_{d-2}^2}
{\rho^2w^2(x^\mu)}\ .
\end{equation}
Taking the conformal factor as indicated in table \ref{table:4}, we obtain a variety of conformally flat space-times. Since in each case the connection is not entirely obvious, let us comment on each case separately.

\begin{table}[t]
\setlength{\tabcolsep}{10 pt} 
\centering
\begin{tabular}{ Sl || Sc | Sc | Sc | Sc }
\specialrule{.13em}{0em}{0em}
  &
${\rm AdS}_2\times S^{d-2}$ &
${\rm dS}_d$ &
$\mathbb{R}\times S^{d-1}$ &
${\rm AdS}_d$ 
\\
\specialrule{.05em}{0em}{0em}
$w^2(x^\mu)$  &
$\displaystyle
\frac{1}{\rho^2}$ &
$\displaystyle
\frac{(1+\rho u)^2}{\rho^2}$ &
$\displaystyle
\frac{(1+\rho u)(1+u^2)+(\rho/2)^2(1+u^2)^2}
{\rho^2}$ & 
$\displaystyle
\frac{\sin^2(\psi)}{\rho^2}$
\\
\specialrule{.13em}{0em}{0em}
\end{tabular}
\caption{Different choices for the conformal factor $w^2(x^\mu)$ in (\ref{eq:175}) that result in a variety of interesting space-times.}\label{table:4}
\end{table}

\begin{figure}
\centering
\includegraphics[scale=0.27]{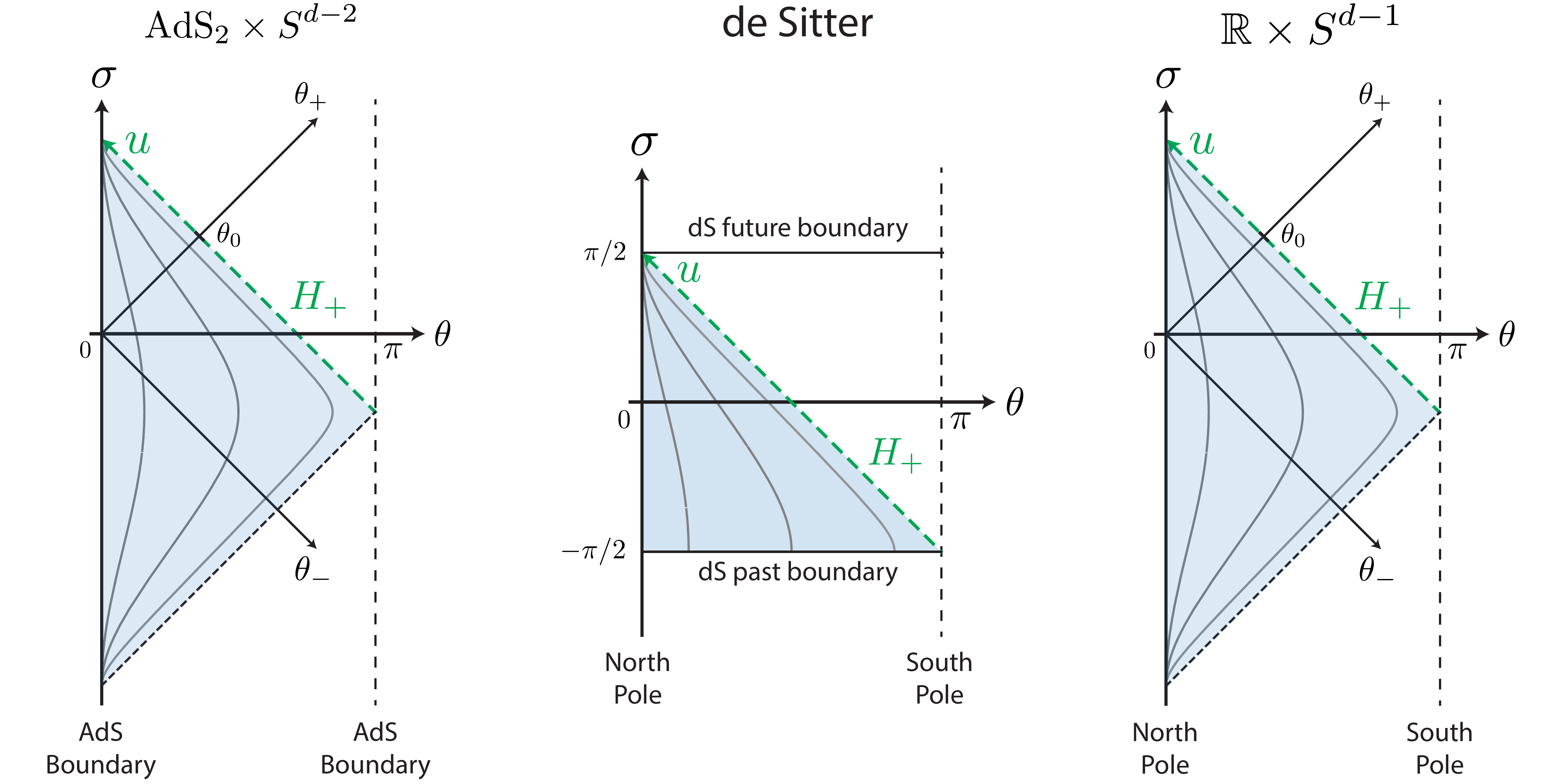}
\caption{Penrose diagrams associated to ${\rm AdS}_2\times S^{d-2}$, de Sitter and the Lorentzian cylinder ${\mathbb{R}\times S^{d-1}}$. The coordinates $(u,\rho)$ only cover the shaded blue region in each case, where several constant $\rho$ trajectories are sketched in gray. Future null infinity at $\rho=0$ in Minkowski is mapped to the future horizon $H_+$ indicated with a dashed green line.}\label{fig:1}
\end{figure}

When the conformal factor is given by $w^2(\rho)=1/\rho^2$ we obtain ${\rm AdS}_2\times S^{d-2}$, where the ${\rm AdS}_2$ factor is in Poincar\'e coordinates. We can see this more clearly by going to global coordinates $(\sigma,\theta)\in\mathbb{R}\times (0,\pi)$, defined according to
\begin{equation}\label{eq:116}
2/\rho=
\tan\left[\frac{\theta_++(\pi-\theta_0)}
{2}\right]+
\tan\left[\frac{\theta_--(\pi-\theta_0)}
{2}\right]\ ,
\qquad 
u=-\tan\left[
\frac{\theta_--(\pi-\theta_0)}{2}\right]\ ,
\end{equation}
where $\theta_\pm=\theta\pm \sigma$. In these coordinates the rescaled metric $d\bar{s}^2$ in (\ref{eq:175}) becomes
\begin{equation}\label{eq:78}
d\bar{s}^2_{{\rm AdS}_2\times S^{d-2}}=
-\rho^2du^2+2dud\rho+d\Omega_{d-2}^2
=
  \frac{-d\sigma^2+d\theta^2}
  {\sin^2(\theta)}+d\Omega_{d-2}^2\ ,
\end{equation}
that we recognize as ${\rm AdS}_2\times S^{d-2}$ in global coordinates, with the two ${\rm AdS}_2$ boundaries located at $\theta=0,\pi$, see figure \ref{fig:1}. The original coordinates $(\rho,u)$ do not cover the whole ${\rm AdS}_2$ space-time but only its Poincar\'e patch
\begin{equation}\label{eq:126}
{\rm Poincar\acute{e}\,\,patch}:
\qquad
-\pi 
\le \theta_\pm \pm (\pi-\theta_0)\le \pi\ ,
\end{equation}
which corresponds to the shaded blue region in the left diagram in figure \ref{fig:1}. Depending on the value of the parameter $\theta_0\in (0,\pi)$ appearing in the change of coordinates (\ref{eq:116}), the coordinates cover a different region of global ${\rm AdS}_2$. The transformation maps the Minkowski asymptotic null infinity $\mathcal{I}^+$ at $\rho=0$ to the future Poincar\'e horizon of ${\rm AdS}_2\times S^{d-2}$ at $\theta_+=\theta_0$.

Next, we can analyze the de Sitter case, which corresponds to taking the adequate conformal factor indicated in table \ref{table:4}. Same as in the previous case, it is instructive to rewrite the metric in the coordinates $(\sigma,\theta)$ in (\ref{eq:116}) but in this case with $\theta_0=\pi/2$, so that the rescaled metric (\ref{eq:175}) becomes
\begin{equation}
d\bar{s}^2_{{\rm dS}_d}=
\frac{-\rho^2du^2+2dud\rho+d\Omega_{d-2}^2}
{(1+\rho u)^2}
=
\frac{-d\sigma^2+d\theta^2+\sin^2(\theta)d\Omega_{d-2}^2}{\cos^2(\sigma)}\ .
\end{equation} 
We recognize this as global de Sitter, with the space-like boundaries at $|\sigma|=\pi/2$. The coordinates $(\rho,u)$ do not cover the whole space-time but only the flat slicing of de Sitter, see figure \ref{fig:1}. The future null infinity of Minkowski $\mathcal{I}^+$ is mapped to the cosmological horizon $H_+$ at $\rho=0$.

To obtain the Lorentzian cylinder $\mathbb{R}\times S^{d-1}$ we perform the Weyl rescaling in table \ref{table:4} and change to the coordinates in (\ref{eq:116}) with an arbitrary value of $\theta_0$, so that we find
\begin{equation}\label{eq:192}
d\bar{s}^2_{\mathbb{R}\times S^{d-1}}=
\frac{-\rho^2du^2+2dud\rho+d\Omega_{d-2}^2}
{(1+\rho u)(1+u^2)+(\rho/2)^2(1+u^2)^2}=
-d\sigma^2+d\theta^2+\sin^2(\theta)d\Omega_{d-2}^2\ .
\end{equation}
The region covered by the coordinates $(u,\rho)$ is indicated in the right diagram of figure \ref{fig:1}, where we see that $\rho=0$ now corresponds to the horizon $H_+$ at $\theta_+=\theta_0$.

Finally, to make the connection between Minkowski and ${\rm AdS}_d$ clear, we must explain the meaning of the conformal factor in the last column of table \ref{table:4}. The angle $\psi\in[-\pi,\pi]$ is obtained by writing the metric in the unit sphere $S^{d-2}$ in (\ref{eq:176}) as
\begin{equation}\label{eq:187}
d\Omega_{d-2}^2=d\psi^2+\cos^2(\psi)d\Omega_{d-3}^2\ .
\end{equation}
The ordinary spherical angle is obtained by shifting $\psi \rightarrow \psi+\pi/2$. The rescaled metric (\ref{eq:175}) in these coordinates becomes
\begin{equation}\label{eq:188}
d\bar{s}^2_{{\rm AdS}_d}=
\frac{-\rho^2du^2+2dud\rho+d\psi^2+\cos^2(\psi)d\Omega_{d-3}^2}{\sin^2(\psi)}\ ,
\end{equation}
where now the range of $\psi$ is restricted to $\psi \in (0,\pi/2]$, with the ${\rm AdS}_d$ boundary being at $\psi = 0$. These coordinates do not cover the full space-time but only the Poincar\'e patch, with the Poincar\'e horizon being at $\rho=0$. See section \ref{sec:4} for a construction of these coordinates from the embedding description of AdS.

\subsection{Three dimensions}

We now consider asymptotic Killing and conformal Killing vectors defined in three dimensional Minkowski at future null infinity, $\rho=0$ in (\ref{eq:174}). Using the conformal relations explained in the previous subsection means the asymptotic transformations are also defined for the horizons $H_+$ in the conformally flat space-times.

Let us start by writing the exact Killing and conformal Killing vectors of Minkowski in the coordinates $(u,\rho,\phi)$, where the periodic angle $|\phi|\le \pi$ is defined as $y(\phi)=\tan(\phi/2)$, so that ${d\Omega_1^2=d\phi^2}$. Conformal transformations of Minkowski in these coordinates are generated by the following vectors
\begin{equation}\label{eq:159}
\begin{aligned}
\xi_T(f)&=
f(\phi)\partial_u
-\rho f'(\phi)\partial_\phi
-\rho^2 f''(\phi)\partial_\rho\ , \\
\xi_R(Y)&=
uY'(\phi)\partial_u
+\left[
Y(\phi)-\rho uY''(\phi)
\right]\partial_\phi
+\rho\left[
Y'(\phi)-\rho u Y'''(\phi)
\right]
\partial_\rho\ , \\
\xi_S(h)&=
u^2h(\phi)\partial_u
-u(2+\rho u)h'(\phi)\partial_\phi
-\left[
2(1+\rho u)h(\phi)+(2+\rho u)^2h''(\phi)
\right]
\partial_\rho\ ,\\
\xi_D(g)&=ug(\phi)\partial_u
-\rho u g'(\phi)\partial_\phi
-\rho \left[ g(\phi)+\rho u g''(\phi) \right]\partial_\rho\ ,
\end{aligned}
\end{equation}
that correspond to translations, Lorentz transformations, special conformal transformations, and dilation respectively, where the four functions of $\phi$ are fixed according to
\begin{equation}\label{eq:160}
\begin{aligned}
f_0(\phi)&=a_0+a_1\cos(\phi)+a_2\sin(\phi)\ ,\\
Y_0(\phi)&=b_0+b_1\cos(\phi)+b_2\sin(\phi)\ ,\\
h_0(\phi)&=c_0+c_1\cos(\phi)+c_2\sin(\phi)\ ,\\
g_0(\phi)&=d_0\ .
\end{aligned}
\end{equation}
This gives the ten independent transformations of the conformal group in three dimensions, ${\rm SO}(3,2)$. Let us denote the ordinary and the ``conformal" Lie derivatives of the metric as
\begin{equation}\label{eq:177}
\begin{aligned}
\mathcal{L}_\xi(g_{\mu \nu})&=
\nabla_\mu \xi_\nu+\nabla_\nu \xi_\mu\ ,\\
\widehat{\mathcal{L}}_\xi(g_{\mu \nu})&=
\nabla_\mu \xi_\nu+\nabla_\nu \xi_\mu-
\frac{2}{d}(\nabla \cdot \xi)g_{\mu \nu}\ ,
\end{aligned}
\end{equation}
where $d$ is the space-time dimension ($d=3$ in this case). The vectors $\xi_T(f)$ and $\xi_R(Y)$ ($\xi_S(h)$ and $\xi_D(g)$) have vanishing (conformal) Lie derivative when the functions are fixed according to (\ref{eq:160}).

Looking at the form (\ref{eq:159}), it is natural to consider the more general class of transformation generated by arbitrary functions, not necessarily those given in (\ref{eq:160}), and compute the associated (conformal) Lie derivatives. In this case, there are two components of the metric that do not vanish but instead satisfy the fall-off conditions in $\rho$ indicated in the first four columns in table \ref{table:6}. In other words, table \ref{table:6} gives the asymptotic boundary conditions at $\rho =0$ that are preserved by the asymptotic conformal Killing vectors (\ref{eq:159}) defined by arbitrary functions $f(\phi )$, $Y(\phi )$, $h(\phi )$, and $g(\phi )$.

\begin{table}[t]
\setlength{\tabcolsep}{8.5 pt} 
\centering
\begin{tabular}{ Sl || Sc | Sc | Sc | Sc || Sc | Sc }
\specialrule{.13em}{0em}{0em}
  &
$\mathcal{L}_\xi(g_{uu})$ &
$\mathcal{L}_\xi(g_{u\phi})$ &
$\widehat{\mathcal{L}}_\xi(g_{uu})$ &
$\widehat{\mathcal{L}}_\xi(g_{u\phi})$ &
$\widehat{\mathcal{L}}_\xi(\bar{g}_{uu})$ &
$\widehat{\mathcal{L}}_\xi(\bar{g}_{u\phi})$ 
\\
\specialrule{.05em}{0em}{0em}
$\xi_T(f)$  &
$\displaystyle 0$ &
$\displaystyle \mathcal{O}(1)$ &
$\displaystyle 0$ & 
$\displaystyle \mathcal{O}(1)$ &
$\displaystyle 0$ &
$\displaystyle \mathcal{O}(\rho^2)$ 
\\
$\xi_R(Y)$  &
$\displaystyle \mathcal{O}(1)$ &
$\displaystyle \mathcal{O}(1)$ &
$\displaystyle \mathcal{O}(1)$ & 
$\displaystyle \mathcal{O}(1)$ &
$\displaystyle \mathcal{O}(\rho^2)$ & 
$\displaystyle \mathcal{O}(\rho^2)$
\\
$\xi_S(h)$  &
$\displaystyle -$ &
$\displaystyle -$ &
$\displaystyle 0$ & 
$\displaystyle \mathcal{O}(1/\rho^2)$ &
$\displaystyle 0$ & 
$\displaystyle \mathcal{O}(1)$
\\
$\xi_D(g)$  &
$\displaystyle -$ &
$\displaystyle -$ &
$\displaystyle \mathcal{O}(1)$ & 
$\displaystyle \mathcal{O}(1/\rho)$ &
$\displaystyle \mathcal{O}(\rho^2)$ & 
$\displaystyle \mathcal{O}(\rho)$
\\
\specialrule{.13em}{0em}{0em}
\end{tabular}
\caption{Non vanishing ordinary and conformal Lie derivatives of the metric $g_{\mu \nu}$ and $\bar{g}_{\mu \nu}=g_{\mu \nu}/w^2(x^\mu)$ when considering the vectors in (\ref{eq:159}) for \textit{arbitrary} functions $f(\phi)$, $Y(\phi)$, $h(\phi)$ and $g(\phi)$. Since the vectors $\xi_T(f)$ and $\xi_R(Y)$ have vanishing divergence we can replace the ordinary Lie derivatives in the first two columns by the conformal derivatives.}\label{table:6}
\end{table}

Note that the way in which we have written the vectors in (\ref{eq:159}) in order to obtain the conformal transformations is non-unique. For instance, using that $Y'(\phi)=-Y'''(\phi)$ is satisfied by (\ref{eq:160}) or the fact that $g'(\phi)=0$, we can find other ways of writing extensions of such vectors. The reason we have chosen this precise way among others is that when we promote (\ref{eq:160}) to arbitrary functions, the vectors satisfy the simple fall-off conditions given in table \ref{table:6}. In particular, note the (conformal) Lie derivative of the metric components $g_{\rho \rho}$, $g_{\rho \phi}$ and $g_{\phi \phi}$ vanish exactly, corresponding to the components that are fixed exactly when writing an arbitrary asymptotically flat metric in the Bondi gauge.

Let us now consider the action of the vectors in (\ref{eq:159}) on the rescaled metric $\bar{g}_{\mu \nu}=g_{\mu \nu}/w^2$ in (\ref{eq:175}) after the Weyl transformation. While conformal Killing vectors are preserved under conformal transformations, this is not the case for exact Killing vectors. This becomes clear by noting the divergence of a vector $\xi^\mu$ behaves in the following way under a Weyl transformation
\begin{equation}
(\bar{\nabla}\cdot \xi)=
(\nabla \cdot \xi)
+d(\xi \cdot \partial)\ln(w)\ ,
\end{equation}
where $\bar{\nabla}$ is the covariant derivative with respect to the rescaled metric (\ref{eq:175}). Since both vectors $\xi_T(f)$ and $\xi_R(Y)$ have vanishing divergence computed with respect to the Minkowski metric, we can trivially replace the ordinary Lie derivative by the conformal version (\ref{eq:177}). Therefore, with respect to the rescaled metric it makes sense to compute conformal Lie derivatives of all the vectors. Doing so, we find the Minkowski fall of conditions (shown in the first four columns of table~\ref{table:6}) simply get rescaled by the conformal factor as $1/w^2(\rho)$. Given that all of the conformal factors we are considering in table~\ref{table:4} have the same scaling behavior for small $\rho$, i.e. $w^2(\rho)\sim 1/\rho^2$, all the conformally mapped space-time satisfy the fall-off conditions given in the last two columns in table~\ref{table:6}.

It is worth mentioning that, depending in the context, we can also consider the ordinary Lie derivative of the vectors in (\ref{eq:159}) for the curved space-times. For instance, the near horizon symmetries studied in \cite{Donnay:2016ejv} can be recovered from this perspective by considering the ${{\rm AdS}_2\times S^{d-2}}$ case. The asymptotic Killing vectors evaluated at $\rho=0$ constructed in \cite{Donnay:2016ejv} that preserve some particular boundary conditions take the following form\footnote{See equations (67) and (85) in \cite{Donnay:2016ejv} for the boundary conditions and (68) and (86) for the vectors.}
\begin{equation}\label{eq:122}
\zeta_1\big|_{\rho=0}=T(\phi)\partial_u\ ,
\qquad \qquad
\zeta_2\big|_{\rho=0}=X(\phi)u\partial_u
\ ,
\qquad \qquad
\zeta_3\big|_{\rho=0}=Y(\phi)\partial_\phi\ .
\end{equation}
Comparing with the vectors defined in this work at $\rho=0$ (see (\ref{eq:171}) below), we see they can be obtained from (\ref{eq:122}) by considering simple linear combinations. 

Let us now consider the algebra satisfied by the vectors (\ref{eq:159}) for arbitrary functions. Since the vectors are in general pretty complicated, it is useful to first evaluate them at $\rho=0$, where they have the following simpler structure
\begin{equation}\label{eq:171}
\begin{aligned}
\xi_T(f)\big|_{\rho=0}&=
f(\phi)\partial_u\ , \\
\xi_R(Y)\big|_{\rho=0}&=
Y'(\phi)u\partial_u
+Y(\phi)\partial_\phi\ , \\
\xi_S(h)\big|_{\rho=0}&=
h(\phi)u^2\partial_u
-2h'(\phi)u\partial_\phi
-2\left[
h(\phi)+2h''(\phi)
\right]
\partial_\rho\ ,\\
\xi_D(g)\big|_{\rho=0}&=g(\phi)u\partial_u\ .
\end{aligned}
\end{equation}
Note that the vector $\xi_S(h)$ is the only one with non-vanishing component in the $\rho$ direction, meaning the associated transformation makes the $\rho=0$ surface fluctuate transversely. This is somehow expected as special conformal transformations leave the origin of the space-time fixed while shift the asymptotic region. This feature has the consequence that the associated algebra does not close.\footnote{If we try restrict to functions $h(\phi)$ such that the vector $\xi_S(h)\big|_{\rho=0}$ has no component in the $\rho$ direction, then we find the function $h(\phi)$ is not periodic in $\phi$. It is possible that a closed algebra can be obtained by considering a modified version of the Lie brackets that takes into account the variation of the metric \cite{Barnich:2010eb,Barnich:2011mi}. This might also be useful to show the algebra (\ref{eq:165}) is satisfied by the vectors away from $\rho=0$.} The algebra closes if we consider the remaining vectors at $\rho=0$, so that we find
\begin{equation}\label{eq:165}
\begin{aligned}
\big[
\xi_T(f_1),\xi_T(f_2)
\big]&=0\ , \\
\big[
\xi_T(f),\xi_R(Y)
\big]&=\xi_T(\widehat{f}\,)\ ,
\qquad \qquad
\widehat{f}(\phi)=f(\phi)Y'(\phi)-f'(\phi)Y(\phi)\ ,\\
\big[
\xi_R(Y_1),\xi_R(Y_2)
\big]&=
\xi_R(\widehat{Y}\,)\ ,
\qquad \quad \,\,\,
\widehat{Y}(\phi)=Y_1(\phi)Y_2'(\phi)-Y_1'(\phi)Y_2(\phi)\ , \\
\big[
\xi_D(g_1),\xi_D(g_2)
\big]&=0\ , \\
\big[
\xi_T(f),\xi_D(g)
\big]&=\xi_T(\widehat{f}\,)\ ,
\qquad \qquad
\widehat{f}(\phi)=g(\phi)f(\phi)\ , \\
\big[
\xi_R(Y),\xi_D(g)
\big]&=\xi_D(\widehat{g}\,)\ ,
\qquad \qquad
\widehat{g}(\phi)=g'(\phi)Y(\phi)\ .
\end{aligned}
\end{equation}
The first three relations in (\ref{eq:165}) give the BMS algebra, with $\xi_T(f)$ and $\xi_R(Y)$ generating supertranslations and superrotations respectively. This algebra is naturally extended by incorporating the vector $\xi_D(g)$, that generates the novel transformation we call `superdilation'. Note that these vectors and their associated algebra are not only defined on Minkowski but for all the conformally flat space-times.

Before generalizing to higher dimensions let us highlight a feature of the vectors $\xi_S(h)$ and $\xi_D(g)$ in (\ref{eq:159}). As we can see from the fourth column in table \ref{table:6}, the conformal Killing equation for $g_{u\phi}$ behaves like $\mathcal{O}(1/\rho^2)$ and $\mathcal{O}(1/\rho)$ respectively. This is an asymptotic behavior that is much more singular than the conditions satisfied by the ordinary BMS vectors $\xi_T(f)$ and $\xi_R(Y)$. Therefore we should be careful when studying the transformations generated by $\xi_S(h)$ and $\xi_D(g)$ as complications can arise when extracting physical quantities associated to these vectors. This issue arises for superdilations in the holographic analysis of section \ref{sec:4}.

\subsection{Arbitrary dimensions}

Let us now generalize the previous discussion to arbitrary space-time dimensions $d$, where the Minkowski metric is given in (\ref{eq:174}) with the unit sphere $S^{d-2}$ describe in stereographic coordinates $\vec{y}\in \mathbb{R}^{d-2}$ (\ref{eq:176}). The vectors generating conformal transformations that generalize (\ref{eq:159}) are given by
\begin{equation}\label{eq:178}
\begin{aligned}
\xi_T(f)&=f \partial_u
-\rho (D^Af) \partial_A
-\frac{\rho^2 }{d-2}(D^2 f)
\partial_\rho\ ,\\
\xi_R(Y)&=
\frac{(D \cdot Y)}{d-2}
u\partial_u+
\left[
Y^A-\frac{\rho u }{d-2}
D^A(D \cdot Y)
\right]\partial_A+
\frac{\rho}{d-2}
\left[
(D \cdot Y)-
\frac{\rho u}{d-2}
D^2(D \cdot Y)
\right]
\partial_\rho\ ,\\
\xi_S(h)&=
h\,u^2\partial_u-
u(2+\rho u)(D^A h)\partial_A
-\left[
2(1+\rho u)h+
\frac{(2+\rho u)^2}{d-2}
(D^2 h)
\right]\partial_\rho\ ,\\
\xi_D(g)&=
g \,u\partial_u
-\rho u (D^A g)\partial_A
-\rho
\left[
g
+
\frac{\rho u}{d-2}
(D^2 g)
\right]\partial_\rho\ ,
\end{aligned}
\end{equation}
where $D_A$ is the covariant derivative on the unit sphere $S^{d-2}$. The vectors generating ordinary conformal vectors are obtained by taking the functions according to
\begin{equation}\label{eq:169}
\begin{aligned}
f_0(\vec{y}\,)&=a_0+
\sum_{B=1}^{d-2} a_B
\left(
\frac{y^B}{|\vec{y}\,|^2+1}
\right)+
a_{d-1}
\left(
\frac{|\vec{y}\,|^2-1}
{|\vec{y}\,|^2+1}
\right)\ ,\\
Y^A(\vec{y}\,)&=
b_0y^A+\sum_{B=1}^{d-2}
\Big\lbrace
w^A_{\,\,\,\,B}y^B+
p_B\left[
2y^By^A-\delta^{AB}(|\vec{y}\,|^2+1)
\right]+
\tilde{p}_B\left[
2y^By^A-\delta^{AB}(|\vec{y}\,|^2-1)
\right]
\Big\rbrace
\ ,\\
h_0(\vec{y}\,)&=c_0+
\sum_{B=1}^{d-2} c_B
\left(
\frac{y^B}{|\vec{y}\,|^2+1}
\right)+
c_{d-1}
\left(
\frac{|\vec{y}\,|^2-1}
{|\vec{y}\,|^2+1}
\right)\ ,\\
g_0(\vec{y}\,)&=d_0\ ,
\end{aligned}
\end{equation}
where $w^A_{\,\,\,B}=-w^B_{\,\,\,A}$. The functions $f(\vec{y}\,)$ and $h(\vec{y}\,)$ give $d$ independent transformations corresponding to space-time translations and special conformal transformations, while $g(\vec{y}\,)=d_0$ is the dilation. Lorentz transformations generated by $Y^A(\vec{y}\,)$ are determined by the parameters $\left\lbrace b_0,w^A_{\,\,\,B},p_B,\tilde{p}_B \right\rbrace$, that give the appropriate number of independent transformations corresponding to ${\rm SO}(d-1,1)$, namely
\begin{equation}
\begin{aligned}
\text{dim}\, [Y^A]&=1+
\frac{(d-2)(d-3)}{2}+
(d-2)+(d-2)=\frac{d(d-1)}{2}\ .
\end{aligned}
\end{equation}

Let us now consider vectors of the form (\ref{eq:178}) but defined with arbitrary functions, not necessarily given by (\ref{eq:169}). When doing this for the conformal vectors $\xi_S(h)$ and $\xi_D(g)$, we find the conformal Lie derivatives have non-vanishing components that explicitly violate some of the Bondi gauge conditions. It is therefore convenient not to consider completely arbitrary function $h(\vec{y}\,)$ and $g(\vec{y}\,)$ but to restrict them to those with the following dependence
\begin{equation}\label{eq:179}
g(\vec{y}\,)=g(y^A/y^1)\ ,
\qquad \qquad
h(\vec{y}\,)=h(y^A/y^1)\ .
\end{equation}
These functions do not depend in an arbitrary way on all the $(d-2)$ components of $y^A$, but only on the $(d-1)$ coordinates obtained as $y^A/y^1$ (there is no special role played by $y^1$, as we can change this by any other component). As an example, for $d=5$ we have $g(\vec{y}\,)=g(y^2/y^1,y^3/y^1)$. With this restriction, the conformal Lie derivative of the metric satisfy much nicer relations that do not violate the Bondi gauge conditions; these are given in the first six columns in table \ref{table:5}.\footnote{The Bondi gauge conditions are $g_{\rho \rho}=g_{\rho A}=\partial_{\rho}[\rho^2{\rm det}(g_{AB})]=0$. The first two are satisfied by the fall-off conditions given in table  \ref{table:5}, while we have not checked the third condition. However if we consider the Newman-Unti gauge condition \cite{Newman:1962cia,Barnich:2011ty}, the third condition is replaced by $g_{u\rho}=2/\rho^2$, that is preserved by the vectors in (\ref{eq:178}).} If we consider the transformations in the curved space-times obtained through the Weyl rescaling in~(\ref{eq:175}), the conformal Lie derivatives are replaced by the fall-off conditions in the last three columns in table~\ref{table:5} (we have used that all the conformal factors in table \ref{table:4} scale as $w^2(\rho)\sim 1/\rho^2$).

\begin{table}[t]
\resizebox{\columnwidth}{!}{
\setlength{\tabcolsep}{4.5 pt} 
\centering
\begin{tabular}{ Sl || Sc | Sc | Sc | Sc | Sc | Sc || Sc | Sc | Sc }
\specialrule{.13em}{0em}{0em}
  &
$\mathcal{L}_\xi(g_{uu})$ &
$\mathcal{L}_\xi(g_{uA})$ &
$\mathcal{L}_\xi(g_{AB})$ &
$\widehat{\mathcal{L}}_\xi(g_{uu})$ &
$\widehat{\mathcal{L}}_\xi(g_{uA})$ &
$\widehat{\mathcal{L}}_\xi(g_{AB})$ &
$\widehat{\mathcal{L}}_\xi(\bar{g}_{uu})$ &
$\widehat{\mathcal{L}}_\xi(\bar{g}_{uA})$ &
$\widehat{\mathcal{L}}_\xi(\bar{g}_{AB})$ 
\\
\specialrule{.05em}{0em}{0em}
$\xi_T(f)$  &
$\displaystyle 0$ &
$\displaystyle \mathcal{O}(1)$ &
$\displaystyle \mathcal{O}(1/\rho)$ & 
$\displaystyle 0$ &
$\displaystyle \mathcal{O}(1)$ &
$\displaystyle \mathcal{O}(1/\rho)$ &
$\displaystyle 0$ &
$\displaystyle \mathcal{O}(\rho^2)$ &
$\displaystyle \mathcal{O}(\rho)$
\\
$\xi_R(Y)$  &
$\displaystyle \mathcal{O}(1)$ &
$\displaystyle \mathcal{O}(1)$ &
$\displaystyle \mathcal{O}(1/\rho^2)$ & 
$\displaystyle \mathcal{O}(1)$ &
$\displaystyle \mathcal{O}(1)$ &
$\displaystyle \mathcal{O}(1/\rho^2)$ &
$\displaystyle \mathcal{O}(\rho^2)$ &
$\displaystyle \mathcal{O}(\rho^2)$ &
$\displaystyle \mathcal{O}(1)$ 
\\
$\xi_S(h)$  &
$\displaystyle -$ &
$\displaystyle -$ &
$\displaystyle -$ & 
$\displaystyle 0$ &
$\displaystyle \mathcal{O}(1/\rho^2)$ & 
$\displaystyle \mathcal{O}(1/\rho^2)$ &
$\displaystyle 0$ &
$\displaystyle \mathcal{O}(1)$ & 
$\displaystyle \mathcal{O}(1)$
\\
$\xi_D(g)$  &
$\displaystyle -$ &
$\displaystyle -$ &
$\displaystyle -$ & 
$\displaystyle \mathcal{O}(1)$ &
$\displaystyle \mathcal{O}(1/\rho)$ & 
$\displaystyle \mathcal{O}(1/\rho)$ &
$\displaystyle \mathcal{O}(\rho^2)$ &
$\displaystyle \mathcal{O}(\rho)$ & 
$\displaystyle \mathcal{O}(\rho)$
\\
\specialrule{.13em}{0em}{0em}
\end{tabular}
}
\caption{Non vanishing ordinary and conformal Lie derivatives of the metric $g_{\mu \nu}$ and $\bar{g}_{\mu \nu}=g_{\mu \nu}/w^2(x^\mu)$ when considering the vectors in (\ref{eq:178}) for \textit{arbitrary} functions $f(\vec{y}\,)$ and $Y^A(\vec{y}\,)$, while $h(\vec{y}\,)$ and $g(\vec{y}\,)$ are restricted to (\ref{eq:179}).}\label{table:5}
\end{table}

Same as in the three dimensional case, we obtain the associated algebra by first evaluating the vectors at $\rho=0$; namely
\begin{equation}\label{eq:180}
\begin{aligned}
\xi_T(f)\big|_{\rho=0}&=f \partial_u\ ,\\
\xi_R(Y)\big|_{\rho=0}&=
\frac{(D \cdot Y)}{d-2}
u\partial_u+
Y^A\partial_A\ ,\\
\xi_S(h)\big|_{\rho=0}&=
h\,u^2\partial_u-
2u(D^A h)\partial_A
-2\left[
h+
\frac{2}{d-2}(D^2 h)
\right]\partial_\rho\ ,\\
\xi_D(g)\big|_{\rho=0}&=g\,u\partial_u\ .
\end{aligned}
\end{equation}
Since the vector $\xi_S(h)$ contains a non-trivial component in the $\rho$ direction, the full algebra of these vectors does not close. However, if we consider the algebra of the remaining vectors, it does close and is given by
\begin{equation}\label{eq:189}
\begin{aligned}
\big[
\xi_T(f_1),\xi_T(f_2)
\big]&=0\ ,\\
\big[
\xi_T(f),\xi_R(Y)
\big]&=\xi_T(\widehat{f}\,)\ ,
\qquad \qquad
\widehat{f}=\frac{(D \cdot Y)}{d-2}f-
Y^AD_A f\ , \\
\big[
\xi_R(Y_1),\xi_R(Y_2)
\big]&=\xi_R(\widehat{Y}\,)\ ,
\qquad \qquad
\widehat{Y}^A=Y_1^BD_BY_2^A-Y_2^BD_B Y_1^A\ , \\
\big[
\xi_D(g_1),\xi_D(g_2)
\big]&=0\ ,\\
\big[
\xi_T(f),\xi_D(g)
\big]&=
\xi_T(\widehat{f}\,)\ ,
\qquad \qquad
\widehat{f}=f\,g\ ,\\
\big[
\xi_R(Y),\xi_D(g)
\big]&=\xi_D(\widehat{g}\,)\ ,
\qquad \qquad
\widehat{g}=Y^AD_A g\ .
\end{aligned}
\end{equation}
The first three relations form a subalgebra that corresponds to the ordinary BMS algebra, obtained from supertranslations and superrotations generated by $\xi_T(f)$ and $\xi_R(Y)$ respectively. The full algebra also includes the superdilation vector $\xi_D(g)$. Same as in the three dimensional case, these vectors and their algebra are defined in Minkowski as well as in any of the conformally flat space-times discussed in subsection \ref{subsec:1}.

\section{CFT charges and algebra}
\label{sec:2}

In this section, we consider an arbitrary CFT and construct the quantum charges associated to the asymptotic (conformal) Killing vectors in (\ref{eq:178}). We start in subsection \ref{subsec:2.1} by showing how the charges for the CFT in the various conformally flat space-times are not independent but related between themselves by conformal symmetry. In subsection \ref{subsec:2.2} we use a discrete symmetry of the CFT to map the charges in the future regions (either $\mathcal{I}^+$ or $H_+$) to the regions in the past ($\mathcal{I}^-$ or $H_-$). We finish in subsection \ref{subsec:2.3} where we use the results of \cite{Cordova:2018ygx} to show all these charges satisfy the same algebra as the associated vectors in (\ref{eq:189}). 

\subsection{Conformal transformation of the charges}
\label{subsec:2.1}

Let us consider an arbitrary CFT in $d$-dimensional Minkowski space-time. The charge associated to a vector $\xi$ can be written in terms of the stress tensor operator, $T_{\mu \nu}$, as follows
\begin{equation}\label{eq:182}
Q[\xi]\equiv 
\int_{\Sigma}dS^\mu \, T_{\mu \nu}\, \xi^\nu\ ,
\end{equation}
where $\Sigma$ is a Cauchy surface in Minkowski with surface element $dS^\mu=dS\,n^\mu$, with $n^\mu$ the future directed unit normal. When constructing a conserved charge $Q$ with $\xi^\nu$ being an exact conformal Killing vector, the Cauchy surface $\Sigma$ we choose to write the operator is unimportant, as different choices for $\Sigma$ result in the same operator. For this reason, the charge is often said to be a topological operator. In this case, however, we are interested in constructing the charges for \textit{asymptotic} conformal Killing vectors, which do not necessarily share this property precisely because they do not generate exact symmetries of the theory. As a result, the Cauchy surface $\Sigma$ we use to write the operator becomes important and turns out to be part of the prescription. 

While the full expression of the vectors in (\ref{eq:178}) is quite complicated, we consider the simpler case $\rho=0$, so that the charges obtained from (\ref{eq:180}) take the form
\begin{equation}\label{eq:181}
\begin{aligned}
Q[\xi_T]&\equiv \mathcal{T}(f)=
\lim_{\rho \rightarrow 0}
\frac{1}{\rho^{d-2}}
\int_{S^{d-2}}
d\Omega(\vec{y}\,)
f(\vec{y}\,)
\mathcal{E}(\vec{y}\,)\ ,\\
Q[\xi_R]&\equiv \mathcal{R}(Y)=
\lim_{\rho \rightarrow 0}
\frac{1}{\rho^{d-2}}
\int_{S^{d-2}}
d\Omega(\vec{y}\,)
\left[
\frac{(D\cdot Y)}{d-2}
\mathcal{K}(\vec{y}\,)+
Y^A\mathcal{N}_A(\vec{y}\,)
\right]\ , \\
Q[\xi_D]&\equiv \mathcal{D}(g)=
\lim_{\rho \rightarrow 0}
\frac{1}{\rho^{d-2}}
\int_{S^{d-2}}
d\Omega(\vec{y}\,)
g(\vec{y}\,)
\mathcal{K}(\vec{y}\,)\ ,\\
\end{aligned}
\end{equation}
where $d\Omega(\vec{y}\,)$ is the volume element of $S^{d-2}$ and where we have defined the following light-ray operators
\begin{equation}\label{eq:184}
\mathcal{E}(\vec{y}\,)\equiv \int_{-\infty}^{+\infty}
du\,T_{uu}(u,\rho=0,\vec{y}\,)\ ,
\qquad \qquad
\mathcal{K}(\vec{y}\,)\equiv
\int_{-\infty}^{+\infty}
du\,u\,T_{uu}(u,\rho=0,\vec{y}\,)\ ,
\end{equation}
and
\begin{equation}\label{eq:191}
\mathcal{N}_A(\vec{y}\,)\equiv
\int_{-\infty}^{+\infty}
du\,T_{uA}(u,\rho=0,\vec{y}\,)\ .
\end{equation}

For the CFT defined in the curved conformally flat space-times we can apply exactly the same procedure to write the charges. However, instead of writing the charges in these space-times from scratch, it is convenient to apply the conformal transformations and map the Minkowski charges (\ref{eq:181}) to the other space-times. This has the advantage that the functions $f(\vec{y}\,)$, $Y^A(\vec{y}\,)$ and $g(\vec{y}\,)$ in Minkowski space-time happen to determine the corresponding charges in the other space-times.

To apply the mapping we use that the stress tensor transforms under a conformal transformation in the following way
\begin{equation}\label{eq:195}
UT_{\mu \nu}U^\dagger=
\frac{\partial \bar{x}^\alpha}
{\partial x^\mu}
\frac{\partial \bar{x}^\beta}
{\partial x^\nu}
\left(
\frac{\bar{T}_{\alpha \beta}-
\langle 
\bar{T}_{\alpha \beta}
\rangle_0}{w(\bar{x})^{d-2}}
\right)\ ,
\end{equation}
where $U$ is the unitary operator implementing the conformal transformation $U:\mathcal{H}\rightarrow \bar{\mathcal{H}}$, and we add a bar over quantities after the mapping. Using this on the general expression for the charge $Q[\xi]$ in (\ref{eq:182}), we find
\begin{equation}\label{eq:183}
\bar{Q}[\xi]\equiv U Q[\xi]U^\dagger=
\int_{H_+}d\bar{S}^\alpha\left(
\bar{T}_{\alpha \beta}-\langle \bar{T}_{\alpha \beta} \rangle_0
\right)
\xi^\beta\ ,
\end{equation}
where both vectors $n^\alpha$ and $\xi^\beta$ are now written in the new coordinates $\bar{x}^\alpha$ and we have defined ${d\bar{S}^\alpha \equiv dS n^\alpha/w^{d-2}}$. For each of the different conformal factors given in table \ref{table:4} the surface element can be written as
\begin{equation}\label{eq:207}
d\bar{S}^\alpha=\delta^\alpha_u
\,d\Omega(\vec{y}\,)\,du \times
\begin{cases}
\qquad \quad 1 \qquad \,\,\,\ ,
\qquad {\rm AdS}_2\times S^{d-2}
\,\,{\rm and}\,\,{\rm dS}_d\ ,\\
\quad (1+u^2)^{\frac{2-d}{2}}\ ,
\qquad \mathbb{R}\times S^{d-1}\ , \\
\quad \,\,\,\sin(\psi)^{2-d}\,\,\ ,
\qquad {\rm AdS}_d\ .
\end{cases}
\end{equation}

The term $\langle \bar{T}_{\alpha \beta} \rangle_0$ in (\ref{eq:183}) corresponds to the vacuum expectation value and appears due to the anomalous transformation of the stress tensor (for $d=2$, it is fixed by the Schwartzian derivative). To write the mapped charges in (\ref{eq:183}) we must compute the components $\langle \bar{T}_{uu} \rangle_0$ and $\langle \bar{T}_{uA} \rangle_0$ at $\rho=0$, that are highly constrained by symmetry. Consider the vacuum state of a QFT defined on a geometry $\mathcal{M}$ obtained as the product of two maximally symmetric manifolds $\mathcal{M}=\mathcal{M}_1\times \mathcal{M}_2$. Using the isometries in each factor and the fact that the vacuum state $\ket{0}$ remains invariant, we can reduce $\langle \bar{T}_{\alpha \beta} \rangle_0$ to\footnote{See the appendix B of \cite{Rosso:2020cub} for a detailed discussion. The constants $(a_1,a_2)$ can be related by the vanishing of the trace of the stress tensor, since the space-times we are considering are conformally flat.}
\begin{equation}\label{eq:206}
\langle \bar{T}_{\alpha \beta} \rangle_0=
a_1 \bar{g}^{(1)}_{ij}+
a_2 \bar{g}^{(2)}_{ab}\ ,
\end{equation}
where $(a_1,a_2)$ are constants and $(\bar{g}^{(1)}_{ij},\bar{g}^{(2)}_{ab})$ are the metrics in each maximally symmetric manifold.\footnote{Since the coordinates $(u,\rho,\vec{y}\,)$ we are using to describe the curved manifold do not cover the whole space-time (see blue regions in figure \ref{fig:1}), one might ask whether the vacuum state appearing in the expectation value in (\ref{eq:183}) is the same as the global vacuum in (\ref{eq:206}). For instance, this is the case when comparing the usual Minkowski and Rindler vacuum, which are distinct states. The global vacuum is most conveniently defined by analytically continuing the Lorentzian time $t$ to $t_E=it$ and considering the path integral over half of the Euclidean manifold $t_E<0$, namely $\ket{0}=
\int_{\Sigma_t,t_E<0} \mathcal{D}\phi\,\exp {(-S_E[\phi])}$, where $\phi$ are the fields of the QFT and $S_E[\phi]$ the Euclidean action. From this description of the vacuum we see that as long as there is a Cauchy surface $\Sigma$ that fits in the region covered by the coordinates, the vacuum state will be equivalent to the global vacuum. From the diagrams in figure \ref{fig:1} we see we can always choose a Cauchy surface of the whole manifold that fits entirely in the blue region. For the Rindler region in Minkowski this is not the case, and therefore the vacuum states defined in each case are different. While this is a formal argument, it can be shown explicitly for a free scalar by comparing the vacuum two-point functions in each quantization scheme. See \cite{Spradlin:1999bn} for ${\rm AdS}_2$ and section 3 of \cite{Akhmedov:2013vka} for dS$_d$.\label{foot:2}} We can use this to compute the vacuum contribution of the stress tensor in (\ref{eq:183}). Since (\ref{eq:206}) applies for the metric written in terms of the global coordinates $(\sigma,\theta)$ in (\ref{eq:116}), we must first compute the appropriate components in the global coordinates and then translate to $(u,\rho)$, so that the final result gives
\begin{equation}\label{eq:208}
\langle \bar{T}_{uu} \rangle_0=
\begin{cases}
\begin{aligned}
0\ ,&
\qquad {\rm for} \qquad
{\rm AdS}_2\times S^{d-2},
{\rm dS}_d,
{\rm AdS}_d\ ,\\
\neq 0\ ,&
\qquad {\rm for} \qquad
\mathbb{R}\times S^{d-1}
\end{aligned}
\end{cases}\ ,
\qquad \qquad
\langle \bar{T}_{uA} \rangle_0=0\ .
\end{equation}
The component $\langle \bar{T}_{uA} \rangle_0$ vanishes since all the metrics have zero non-diagonal components $\bar{g}_{\sigma A}$ and $\bar{g}_{\theta A}$. The contributions to $\langle \bar{T}_{uu} \rangle_0$ vanish when they are obtained by projecting along a \textit{null} tangent vector. This is not the case for the CFT on the Lorentzian cylinder since the null geodesic moving along $H_+$ at $(\theta_+,\vec{y}\,)=(\theta_0,\vec{y}_0)$ in the global coordinates (\ref{eq:116}) has a tangent vector with non-trivial components along $\mathbb{R}$ and $S^{d-1}$ which does not vanish because the constants $(a_1,a_2)$ in (\ref{eq:206}) are different. This component can be computed explicitly using the results in \cite{Herzog:2013ed} (see eq. (2.24) in \cite{Rosso:2019txh}).

We can now use (\ref{eq:207}) and (\ref{eq:208}) to write the mapped charge in (\ref{eq:183}) explicitly. For the ${\rm AdS}_2\times S^{d-2}$ and ${\rm dS}_d$ cases the final result is the same as in (\ref{eq:181}) but without the prefactor $1/\rho^{d-2}$. Note that the coordinate $u$ is an affine parameter for the null geodesics in Minkowski, defined as $\rho=0$ with constant $\vec{y}$ . This follows from checking that the geodesic equation is satisfied in the Minkowski metric (\ref{eq:174}); namely
\begin{equation}\label{eq:186}
\frac{d^2x^\mu }{du^2}+
\Gamma^\mu_{\rho \sigma}
\frac{dx^\rho}{du}
\frac{dx^\sigma}{du}=0\ .
\end{equation}
If $u$ was not affine, then the right-hand side of (\ref{eq:186}) would be proportional to the tangent vector along the curve. Having an affine parameter $u$ is important since it allows us to identify the light-ray operator $\mathcal{E}(\vec{y}\,)$ in (\ref{eq:184}) as the ANEC, meaning $\mathcal{E}(\vec{y}\,)\ge 0$.

When we apply a conformal transformation, the geodesic equation (\ref{eq:186}) is not invariant since the connection transforms with an anomalous term due to the Weyl rescaling. The coordinate $u$ remains affine only if the following condition is satisfied
(see section 2.1  of \cite{Rosso:2019txh} for details) 
\begin{equation}\label{eq:185}
u\,\,\,{\rm is\,\,\,affine\,\,\,in}\,\,\,\bar{g}_{\alpha \beta}
\qquad \Longleftrightarrow \qquad
\left.
\frac{dw(x)}{du}\right|_{\rho=0}=0\ .
\end{equation}
From the first line in (\ref{eq:207}) we see this is satisfied in all space-times except for the Lorentzian cylinder. In this case, we can define a new parameter $\lambda=\lambda(u)$ that is affine in $\mathbb{R}\times S^{d-1}$ (i.e. it satisfies the geodesic equation as written in (\ref{eq:186})) according to
\begin{equation}
u(\lambda)=\tan(\lambda)\ ,
\qquad \qquad |\lambda|\le \pi/2\ .
\end{equation}
The charges in $\mathbb{R}\times S^{d-1}$ are then given by (\ref{eq:181}) without the factor $1/\rho^{d-2}$ and with the following light-ray operators
\begin{equation}\label{eq:194}
\begin{aligned}
\mathcal{\bar{E}}(\vec{y}\,)&=\int_{-\pi/2}^{\pi/2}
d\lambda
\cos^d(\lambda)
\left(
\bar{T}_{\lambda \lambda}-
\langle \bar{T}_{\lambda \lambda} \rangle_0
\right)\ ,\\
\mathcal{\bar{K}}(\vec{y}\,)&=
\int_{-\pi/2}^{\pi/2}
d\lambda
\sin(\lambda)\cos^{d-1}(\lambda)
\left(
\bar{T}_{\lambda \lambda}-
\langle \bar{T}_{\lambda \lambda} \rangle_0
\right)\ ,\\
\mathcal{\bar{N}}_A(\vec{y}\,)&=
\int_{-\pi/2}^{\pi/2}
d\lambda
\cos^{d-2}(\lambda)
\bar{T}_{\lambda A }\ ,
\end{aligned}
\end{equation}
where we have also considered the non-vanishing vacuum contribution in (\ref{eq:208}). The operator $\bar{\mathcal{E}}(\vec{y}\,)$ in the Lorentzian cylinder and its connection to the ANEC has been recently studied in \cite{Rosso:2020cub,Rosso:2019txh,Iizuka:2019ezn}.

The mapping to ${\rm AdS}_d$ is slightly more involved since in that case we must not only consider a Weyl rescaling of the metric but also a change of coordinates in the sphere $S^{d-2}$, according to (\ref{eq:187}). This requires us to change the angular coordinates of the vectors $\xi$ generating the transformations (\ref{eq:180}). While this is certainly a straightforward computation, we shall only write the three dimensional case explicitly, where the vectors $\xi$ (\ref{eq:171}) are already written in terms of the angle $\phi$. The Weyl rescaling in this case is given by $w^2(\rho,\phi)=\sin^2(\phi)/\rho^2$, so that the ${\rm AdS}_3$ metric (\ref{eq:187}) is
\begin{equation}
d\bar{s}^2_{{\rm AdS}_3}=
\frac{-\rho^2du^2+2dud\rho+d\phi^2}{\sin^2(\phi)}\ ,
\end{equation}
where the range of $\phi$ in this case is given by $\phi\in(0,\pi)$, with $\phi\rightarrow 0,\pi$ corresponding to different ways of approaching the same ${\rm AdS}_3$ boundary (see figure 7 in \cite{Rosso:2020cub}). The charges can be then written from (\ref{eq:183}), (\ref{eq:207}) and (\ref{eq:208}) as
\begin{equation}
\begin{aligned}
\bar{\mathcal{T}}(f)&=\int_0^\pi
\frac{d\phi}{\sin(\phi)}f(\phi)
\int_{-\infty}^{+\infty}du\,
\bar{T}_{uu}\ ,\\
\bar{\mathcal{R}}(Y)&=\int_0^\pi
\frac{d\phi}{\sin(\phi)}
\int_{-\infty}^{+\infty}du
\left[
Y'(\phi)u
\bar{T}_{uu}+
Y(\phi)
\bar{T}_{u\phi}
\right]\ ,
\\
\bar{\mathcal{D}}(g)&=\int_0^\pi
\frac{d\phi}{\sin(\phi)}g(\phi)
\int_{-\infty}^{+\infty}du\,u\,
\bar{T}_{uu}\ .
\end{aligned}
\end{equation}
The takeaway from this subsection is that given the charges (\ref{eq:181}) defined in $\mathcal{I}^+$ and determined by the functions $\left\lbrace f(\vec{y}\,),Y^A(\vec{y}\,),g(\vec{y}\,) \right\rbrace$, the charges in $H_+$ are not written in terms of new functions, but the same ones as in the Minkowski.

\subsection{Discrete transformation between future and past regions}
\label{subsec:2.2}

Our analysis so far applies to the regions in the future $\mathcal{I}^+$ for Minkowski and the horizon $H_+$ for the curved space-times (see figure~\ref{fig:1}). A completely analogous analysis can be performed for the corresponding asymptotic surfaces located in the past regions. In this subsection we show the charges in the future and past regions are not independent but related in a very precise and interesting way through a discrete symmetry of the CFT.

Given a QFT in the space-time $g_{\mu \nu}$ the vacuum is invariant under the action generated by the charges associated to the isometries of the space-time that are smoothly connected to the identity. Generically, discrete isometries such as a time reversal in Minkowski, are not symmetries of the ground state. However, there are certain combinations of discrete symmetries that leave the vacuum invariant. In the case of Minkowski, a transformation that leaves the vacuum invariant is given by the following transformation\footnote{While in even space-time dimensions the ${\rm CRT}$ symmetry is equivalent to the more standard ${\rm CPT}$ transformation, for odd dimensions the ${\rm CPT}$ is not a symmetry of the QFT, see subsection 5.1 in \cite{Witten:2018lha}}
\begin{equation}\label{eq:220}
{\rm CRT}: \qquad
(t,x_1,\vec{x}\,)
\longrightarrow 
(-t,-x_1,\vec{x}\,)\ ,
\end{equation}
where $(t,x_1,\vec{x}\,)$ are ordinary Cartesian coordinates, $\vec{x}=(x_2,...,x_{d-1})$. We can use this symmetry to relate the charges in the future and past regions. 

\begin{figure}
\centering
\includegraphics[scale=0.28]{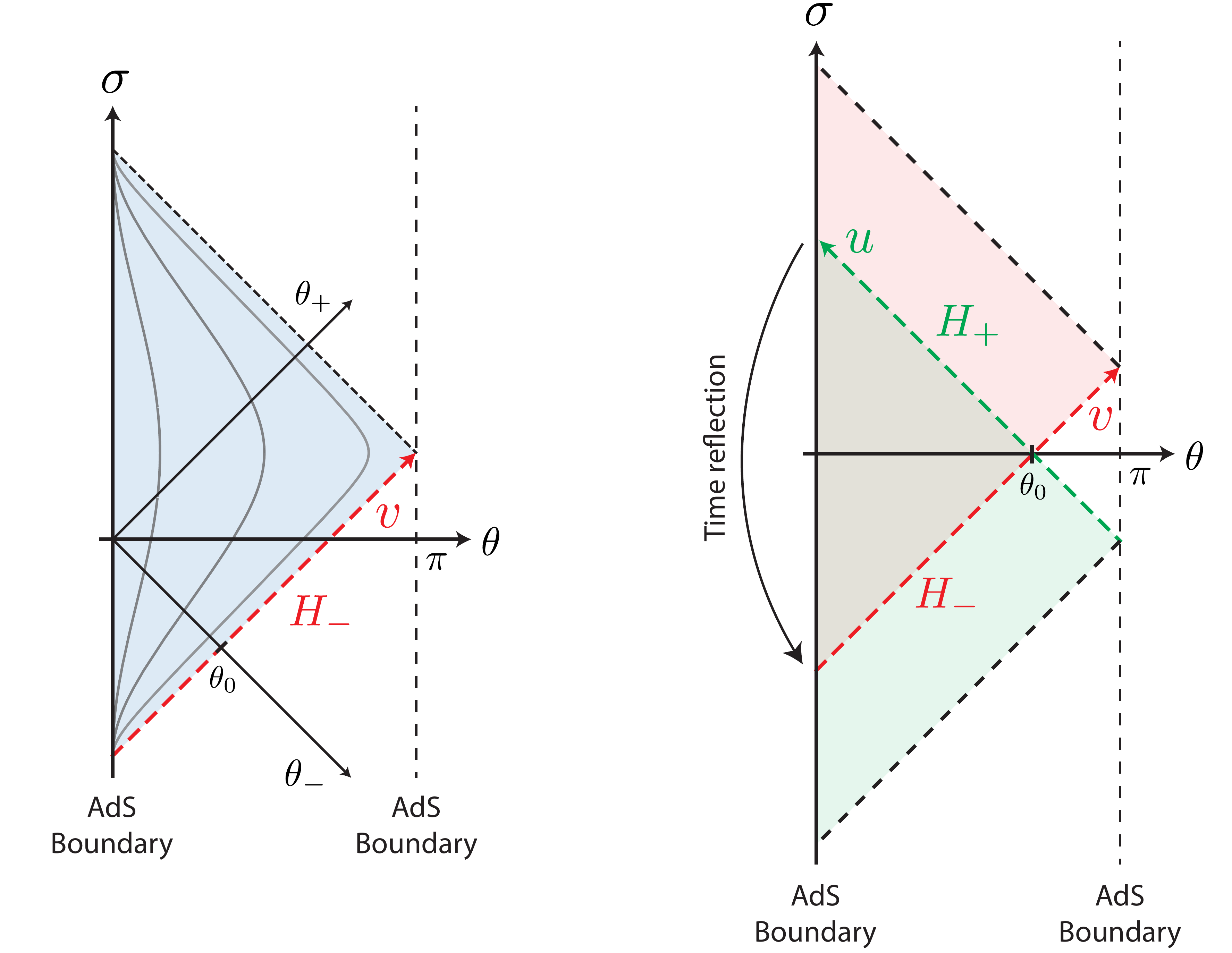}
\caption{On the left we have the region of ${\rm AdS}_2$ covered by the coordinates $(v,\varrho)$ defined in (\ref{eq:125}). The past horizon (marked in red) is obtained as $\varrho=0$ and $v\in \mathbb{R}$. On the right we plot the region of ${\rm AdS}_2$ covered by the coordinates $(u,\rho)$ and $(v,\varrho)$ in green and red respectively. The ${\rm CRT}$ transformation in (\ref{eq:128}) gives a map between these regions, most importantly mapping between the horizons $H_+\leftrightarrow H_-$.}\label{fig:3}
\end{figure}

Instead of considering the Minkowski case, let us focus on the more interesting setup of a CFT on ${\rm AdS}_2\times S^{d-2}$, where the future Poincar\'e horizon $H_+$ we have been considering so far is located at ${\theta_+=\theta_0\in[0,\pi]}$ in terms of the global coordinates in~(\ref{eq:78}). We define the past horizon $H_-$ as the surface ${\theta_-=\theta_0}$, which can be conveniently described in terms of a new set of coordinates~$(v,\varrho)$ defined similarly to~(\ref{eq:116}) as
\begin{equation}\label{eq:125}
2/\varrho=
\tan\left[
\frac{\theta_+-(\pi-\theta_0)}{2}
\right]
+
\tan\left[
\frac{\theta_-+(\pi-\theta_0)}
{2}
\right]\ ,
\qquad 
v=\tan\left[
\frac{\theta_+-(\pi-\theta_0)}{2}
\right]\ .
\end{equation}
These coordinates only cover a Poincar\'e patch of ${\rm AdS}_2$ given by
\begin{equation}
{\rm Poincar\acute{e}\,\,patch}:
\qquad
-\pi \le \theta_\pm \mp (\pi-\theta_0)\le \pi\ ,
\end{equation}
and plotted in the left diagram of figure \ref{fig:3}. Note that for general $\theta_0\in [0,\pi]$ this is a different Poincar\'e patch that the one covered by the coordinates $(u,\rho)$ in (\ref{eq:126}), shown in green on the right diagram of the same figure. It is only for $\theta_0=\pi$ that both set of coordinates cover the same region. The ${\rm AdS}_2\times S^{d-2}$ metric (\ref{eq:78}) in these coordinates becomes
\begin{equation}
d\bar{s}_{{\rm AdS}_2\times S^{d-2}}^2=
-\varrho^2dv^2
-2dvd\varrho
+\frac{4\,d\vec{z} \cdot d\vec{z}}{(1+|\vec{z}\,|)^2}=
  \frac{-d\sigma^2+d\theta^2}
  {\sin^2(\theta)}+d\Omega_{d-2}^2\ ,
\end{equation}
where we have taken stereographic coordinates $\vec{z}\in \mathbb{R}^{d-2}$ to describe the sphere $S^{d-2}$. The past horizon $H_-$ at $\theta_-=\theta_0$ is located at $\varrho=0$.

It is easy to check the asymptotic (conformal) Killing vectors on the past horizon $H_-$ are completely analogous to those given in (\ref{eq:178}) after replacing $(u,\rho,\vec{y}\,)\rightarrow (-v,\varrho,\vec{z}\,)$. The charges constructed from these vectors evaluated at $\varrho=0$ are similar to (\ref{eq:181}) and given by\footnote{The charges in (\ref{eq:219}) do not have the prefactor $1/\rho^{d-2}$ in (\ref{eq:181}) since the ${\rm AdS}_2\times S^{d-2}$ surface element is given by (\ref{eq:207}). After making the replacement $(u,\rho,\vec{y}\,)\rightarrow (-v,\varrho,\vec{z}\,)$ on the vectors (\ref{eq:180}) at $\varrho=0$ we see the supertranslation vector picks up a minus sign, that appears in the supertranslation charge in (\ref{eq:219}).}
\begin{equation}\label{eq:219}
\begin{aligned}
\bar{Q}[\xi_{T_-}]&\equiv
\bar{\mathcal{T}}_-(f_-)=-\int_{S^{d-2}}
d\Omega(\vec{z}\,)
f_-(\vec{z}\,)
\bar{\mathcal{E}}(\vec{z}\,)\ ,\\
\bar{Q}[\xi_{R_-}]&\equiv
\bar{\mathcal{R}}_-(Y_-)=
\int_{S^{d-2}}d\Omega(\vec{z}\,)
\left[
\frac{(D\cdot Y_-)}{d-2}
\bar{\mathcal{K}}(\vec{z}\,)
+Y^A_-\bar{\mathcal{N}}_A(\vec{z}\,)
\right]\ , \\
\bar{Q}[\xi_{D_-}]&\equiv
\bar{\mathcal{D}}_-(g_-)=
\int_{S^{d-2}}d\Omega(\vec{z}\,)
g_-(\vec{z}\,)
\bar{\mathcal{K}}(\vec{z}\,) \ ,
\end{aligned}
\end{equation}
where we have added a minus subscript on the charges and functions to distinguish these from those associated to the future regions, which should now contain a plus subscript. The light-ray operators in (\ref{eq:219}) are analogous to those defined previously, namely
\begin{equation}
\bar{\mathcal{E}}(\vec{z}\,)\equiv 
\int_{-\infty}^{+\infty}dv\,\bar{T}_{vv}(v,\varrho=0,\vec{z}\,)\ ,
\qquad \qquad
\bar{\mathcal{K}}(\vec{z}\,)\equiv 
\int_{-\infty}^{+\infty}dv\,v\,\bar{T}_{vv}(v,\varrho=0,\vec{z}\,)\ ,
\end{equation}
and 
\begin{equation}
\bar{\mathcal{N}}_A(\vec{z}\,)\equiv 
\int_{-\infty}^{+\infty}dv\,\bar{T}_{vA}(v,\varrho=0,\vec{z}\,)\ .
\end{equation}

Naively, it might seen that the charges in each of the horizons are unrelated, meaning the functions $f_\pm$, $Y^A_\pm$ and $g_{\pm}$ are completely independent. However, this is not the case since an analogous ${\rm CRT}$ symmetry in ${\rm AdS}_2\times S^{d-2}$ relates these functions in a precise way. In appendix \ref{zapp:2} we show that any QFT in ${\rm AdS}_2\times S^{d-2}$ is invariant under the following discrete transformation
\begin{equation}\label{eq:128}
{\rm CRT}:
\qquad
(\sigma,\theta,\vec{y}\,)
\quad \longrightarrow \quad
\left(
-\sigma,\theta,\vec{z}=\frac{\vec{y}}{|\vec{y}\,|^2}
\right)\ ,
\end{equation}
realizing $({\rm CRT})\ket{\bar{0}}=\ket{\bar{0}}$ in this case. Apart from the time reflection we have an inversion in the stereographic coordinates on the sphere $S^{d-2}$.\footnote{This discrete transformation is completely analogous to the one in Minkowski (\ref{eq:220}). Taking spherical coordinates in Minkowski $(t,r,\vec{y}\,)$, with $\vec{y}$ parametrizing the $S^{d-2}$, one can check the transformation (\ref{eq:220}) is equivalent to (\ref{eq:128}) replacing $\sigma\rightarrow t$, see appendix \ref{zapp:2} for details.} The usefulness of this transformations comes from the fact that it maps between the future horizon $H_+$ at $\theta_+=\theta_0$ to $H_-$ at $\theta_-=\theta_0$ (see right diagram in figure \ref{fig:3}).

Applying the adjoint action of the ${\rm CRT}$ operator on the stress tensor we find\footnote{While the charge conjugation operator ${\rm C}$ implements a hermitian conjugate it makes no difference in this case since the stress tensor is Hermitian.}
$$({\rm CRT})\bar{T}_{\mu \nu}(x)({\rm CRT})^{-1}=
  \frac{\partial \tilde{x}^\alpha}
  {\partial x^\mu}
  \frac{\partial \tilde{x}^\beta}
  {\partial x^\nu}
  \bar{T}_{\alpha \beta}(\tilde{x})\ ,$$
where $\tilde{x}^\alpha$ are the transformed coordinate. Using this we can map the charge on $H_+$ to $H_-$ in the following way
\begin{equation}\label{eq:112}
({\rm CRT})\bar{Q}_+({\rm CRT})^{-1}=
({\rm CRT})
\left[
\int_{H_+}dS\,n^\mu \bar{T}_{\mu \nu}\xi^\nu
\right]
({\rm CRT})^{-1}=
\int_{H_-}dS\,n^\alpha \bar{T}_{\alpha \beta}\xi^\beta\ ,
\end{equation}
where the vectors $n^\alpha$ and $\xi^\beta$ are the ones defining $Q_+$ but written in the transformed coordinates in (\ref{eq:128}). Writing the final expression in terms of the $(v,\varrho,\vec{z}\,)$ coordinates we find
\begin{equation}\label{eq:221}
\begin{aligned}
({\rm CRT})\bar{\mathcal{T}}_+(f_+)({\rm CRT})^{-1}
=\bar{\mathcal{T}}_-(f_-)\ ,
\qquad &{\rm where} \qquad
f_-(\vec{z}\,)=-f_+(\vec{z}/|\vec{z}\,|^2)
\ , \\
({\rm CRT})\bar{\mathcal{R}}_+(Y_+)({\rm CRT})^{-1}
=\bar{\mathcal{R}}_-(Y_-)\ ,
\qquad &{\rm where} \qquad
Y^A_-(\vec{z}\,)=
-
\left(
\frac{\partial z^A}{\partial y^B}
\right)
Y^B_+(\vec{z}/|\vec{z}\,|^2)
\ , \\
({\rm CRT})\bar{\mathcal{D}}_+(g_+)({\rm CRT})^{-1}
=\bar{\mathcal{D}}_-(g_-)\ ,
\qquad &{\rm where} \qquad
g_-(\vec{z}\,)=-g_+(\vec{z}/|\vec{z}\,|^2)
\ , \\
\end{aligned}
\end{equation}
where the charges in the past region are given in (\ref{eq:219}). The minus sign appearing in the relation between the functions comes from the fact that the normal vector $n^\mu=\delta^\mu_u$ in the coordinates on $H_-$ is $n^\alpha=-\delta^\alpha_v$. 

This shows the charges in each horizon are not independent but related in an interesting way. Through the conformal map explained in the previous subsection, this analysis together with the relations in (\ref{eq:221}) are not exclusive to the CFT in ${\rm AdS}_2\times S^{d-2}$ but to any of the conformally flat space-times. Using the invariance of the vacuum under ${\rm CRT}$ transformation also allows us to map states, for instance
\begin{equation}\label{eq:242}
({\rm CRT})\ket{Y_+}
\equiv
({\rm CRT})e^{-i\bar{\mathcal{R}}(Y_+)}\ket{\bar{0}}
=
({\rm CRT})e^{-i\bar{\mathcal{R}}(Y_+)}({\rm CRT})\ket{\bar{0}}=
e^{-i\mathcal{\bar{R}}(Y_-)}\ket{\bar{0}}=\ket{Y_-}\ .
\end{equation}
In section \ref{sec:3}, we study features of states such as $\ket{Y_+}$ in detail.

\subsection{Charge algebra}
\label{subsec:2.3}

We finish this section by considering the algebra of the charges associated to the asymptotic (conformal) Killing vectors. Since the charges in the different setups are related through the adjoint action of $U$ in (\ref{eq:183}) or $({\rm CRT})$ in (\ref{eq:112}), which preserves the structure of any algebra, we know from the beginning that the algebra satisfied by the charges in Minkowski or any of the conformally flat space-times must be exactly the same. Since the asymptotic vectors at $\rho=0$ in (\ref{eq:180}) satisfy the algebra (\ref{eq:189}), the expectation is that the charges realize the same algebra, which can be written as
\begin{equation}\label{eq:190}
\begin{aligned}
\big[
\mathcal{T}(f_1),
\mathcal{T}(f_2)
\big]&=0\ ,\\
\big[
\mathcal{T}(f),\mathcal{R}(Y)
\big]&=i\mathcal{T}(\widehat{f}\,)\ ,
\qquad \qquad
\widehat{f}=\frac{(D\cdot Y)}{d-2}f-Y^AD_Af\ ,\\ 
\big[
\mathcal{R}(Y_1),
\mathcal{R}(Y_2)
\big]&=i\mathcal{R}(\widehat{Y}\,)\ ,
\qquad \qquad
\widehat{Y}=Y_1^BD_BY_2^A-Y_2^BD_BY_1^A\ ,\\
\big[
\mathcal{D}(g_1),
\mathcal{D}(g_2)
\big]&=0\ , \\
\big[
\mathcal{T}(f),
\mathcal{D}(g)
\big]&=i\mathcal{T}(\widehat{f}\,)\ ,
\qquad \qquad
\widehat{f}=f\,g\ ,\\
\big[
\mathcal{R}(Y),
\mathcal{D}(g)
\big]&=i\mathcal{D}(\widehat{g}\,)\ ,
\qquad \qquad
\widehat{g}=Y^AD_Ag\ ,
\end{aligned}
\end{equation}
where the factor $i$ arises since the charges are Hermitian operators. Showing this operator algebra is satisfied is highly non-trivial, since it involves computing complicated commutators involving the stress tensor. We can still show (\ref{eq:190}) is satisfied by building on the results of \cite{Cordova:2018ygx}, where the algebra of light-ray operators on the Minkowski null plane where computed for arbitrary CFTs. In appendix \ref{zapp:1} we apply a conformal transformation (worked out in \cite{Rosso:2019txh}) to the algebra of light-ray operators of \cite{Cordova:2018ygx} from the Minkowski null plane to ${\rm AdS}_2\times S^{d-2}$ and use it to show (\ref{eq:190}) is indeed satisfied.

\section{Quantum transformations on the Hilbert space}
\label{sec:3}

In previous sections we have considered asymptotic (conformal) Killing vectors in several conformally flat space-times, constructed the quantum charges and shown they satisfy the extended BMS algebra in (\ref{eq:190}). In this section, we study the action of these transformations on the Hilbert space, implemented through the charges $\mathcal{T}(f)$, $\mathcal{R}(Y)$ and $\mathcal{D}(g)$, corresponding to supertranslations, superrotations, and superdilations respectively. For concreteness, here we focus on the CFT in Minkowski with the charges defined at future null infinity $\mathcal{I}^+$, even though the results apply in much more generality. In subsection \ref{subsec:3.1}, we consider the group action obtained by exponentiating the charges on the vacuum state $\ket{0}$ of the CFT. In subsection \ref{subsec:3.2}, we study vacuum and non-vacuum representations of the algebra (\ref{eq:190}) on the Hilbert space of the CFT, where we focus on the three and four dimensional cases.

\subsection{Group action}
\label{subsec:3.1}

Let us start by recalling some basic notions of symmetries and conserved charges in CFTs. For any vector $\xi^\mu$, there is an associated conserved charge $Q[\xi]$ that can be written in terms of the stress tensor as in (\ref{eq:182}). The finite transformation is then implemented on the states $\ket{\psi}\in \mathcal{H}$ by exponentiating the charges in the standard way $\ket{\psi_s}=e^{-isQ}\ket{\psi}$, with $s\in \mathbb{R}$. For CFTs, if the vector $\xi^\mu$ satisfies the Killing or conformal Killing equation, the vacuum state is invariant under the transformation, which implies $\ket{0}$ is an eigenstate of $Q_0$. In this case, the charge annihilates the vacuum $Q_0\ket{0}=0$ since the vacuum expectation value of the stress tensor vanishes $\braket{0|T_{\mu \nu}|0}=0$. The charges that annihilate the vacuum are given in (\ref{eq:181})
\begin{equation}\label{eq:214}
Q_0=\Big\lbrace
\mathcal{T}(f_0),
\mathcal{R}(Y_0),
\mathcal{D}(g_0)
\Big\rbrace
\ ,
\end{equation}
where the functions $f_0(\vec{y}\,)$, $Y_0^A(\vec{y}\,)$ and $g_0(\vec{y}\,)$ are fixed by (\ref{eq:169}).

In this work we are considering more general charges that are associated to \textit{asymptotic} (conformal) Killing vectors, obtained by taking arbitrary functions not necessarily given by (\ref{eq:169}). This means the charges are not required to annihilate the vacuum but instead induce some non-trivial transformation on the states. Applying the finite transformations on the vacuum allows us to define the following states
\begin{equation}\label{eq:197}
\ket{f}\equiv e^{-i\mathcal{T}(f)}\ket{0}\ ,
\qquad \qquad
\ket{Y}\equiv e^{-i\mathcal{R}(Y)}\ket{0}\ ,
\qquad \qquad
\ket{g}\equiv e^{-i\mathcal{D}(g)}\ket{0}\ .
\end{equation}
Since the charges are evaluated at $\rho=0$, these states are as well defined on this surface.

While in principle the action induced by the three charges generate a state that is different from the vacuum itself, a distinct role is played by the supertranslated vacuum $\ket{f}$. Note the supertranslation charge $\mathcal{T}(f)$ in (\ref{eq:181}) is obtained by integrating the light-ray operator $\mathcal{E}(\vec{y}\,)$ over the transverse sphere $S^{d-2}$. The achronal ANEC, proven for general QFTs in Minkowski \cite{Faulkner:2016mzt,Hartman:2016lgu} and for (A)dS and ${\rm AdS}_2\times S^{d-2}$ in \cite{Rosso:2020cub}, implies this particular light-ray operator is positive $\mathcal{E}(\vec{y}\,)\ge 0$.\footnote{The corresponding operator $\bar{\mathcal{E}}(\vec{y}\,)$ in the Lorentzian cylinder (\ref{eq:194}) is not the achronal ANEC but has been proven to also be a positive operator for arbitrary CFTs \cite{Rosso:2019txh} and holographic CFTs \cite{Iizuka:2019ezn}.} Using this together with the fact that $\braket{0|\mathcal{E}(\vec{y}\,)|0}=0$ and the Cauchy-Schwarz inequality, we find
\begin{equation}\label{eq:198}
|\braket{\psi|\mathcal{E}(\vec{y}\,)|0}|^2
\le 
\braket{\psi|\mathcal{E}(\vec{y}\,)|\psi}
\braket{0|\mathcal{E}(\vec{y}\,)|0}
  =0
  \qquad \Longrightarrow \qquad
  \mathcal{E}(\vec{y}\,)\ket{0}= 0\ .
\end{equation}
Since the supertranslation charge $\mathcal{T}(f)$ is obtained by integrating $\mathcal{E}(\vec{y}\,)$ over the sphere $S^{d-2}$ (\ref{eq:181}), we have $\mathcal{T}(f)\ket{0}=0$ for \textit{any} function $f(\vec{y}\,)$, meaning finite supertranslations act trivially on the vacuum state
\begin{equation}
\ket{f}=e^{-i\mathcal{T}(f)}\ket{0}=
  \sum_{n=0}^{\infty}
  \frac{(-i)^n}{n!}
  \mathcal{T}^{\, n}(f)\ket{0}=\ket{0}\ .
\end{equation}
This is a very general and strong result that holds for all QFTs defined in the conformally flat space-times in any dimension. Although $f(\vec{y}\,)\partial_u$ is not an exact Killing vector, the vacuum state is still invariant under the associated transformation. This is certainly not the case for the superrotation and superdilation charges, $\mathcal{R}(Y)$ and $\mathcal{D}(g)$, which are built from the light-ray operators $\mathcal{K}(\vec{y}\,)$ and $\mathcal{N}_A(\vec{y}\,)$ (\ref{eq:181}) that are not positive definite. Notice, however, that these light-ray operators still have vanishing expectation value since they are built from integrating the stress tensor $T_{\mu \nu}$. These results are summarized in the first row of table \ref{table:3}.

\begin{table}[t]
\resizebox{\columnwidth}{!}{
\setlength{\tabcolsep}{2.5pt} 
\centering
\begin{tabular}{ Sl || Sc | Sc | Sc || Sc | Sc | Sc  || Sc | Sc | Sc }
\specialrule{.13em}{0em}{0em}
$\ket{\psi}$  &
$\mathcal{T}(f_0)\ket{\psi}$ &
$\mathcal{R}(Y_0)\ket{\psi}$ &
$\mathcal{D}(g_0)\ket{\psi}$ &
$\mathcal{T}(f)\ket{\psi}$ &
$\mathcal{R}(Y)\ket{\psi}$ &
$\mathcal{D}(g)\ket{\psi}$ &
$\langle \mathcal{T}(f) \rangle_\psi$ & 
$\langle \mathcal{R}(Y) \rangle_\psi$ & 
$\langle \mathcal{D}(g) \rangle_\psi$ 
\\
\specialrule{.05em}{0em}{0em}
$\ket{0}$  &
\color{ForestGreen}{$0$} &
\color{ForestGreen}{$0$} &
\color{ForestGreen}{$0$} &
\color{ForestGreen}{$0$} &
\color{Red}{$\neq 0$} &
\color{Red}{$\neq 0$} &
\color{ForestGreen}{$0$} & 
\color{ForestGreen}{$0$} & 
\color{ForestGreen}{$0$} 
\\
$\ket{Y}$  &
\color{ForestGreen}{$0$} &
\color{Red}{$\neq 0$} &
\color{ForestGreen}{$0$} &
\color{ForestGreen}{$0$} &
\color{Red}{$\neq 0$} &
\color{Red}{$\neq 0$} &
\color{ForestGreen}{$0$} & 
\color{ForestGreen}{$0$} & 
\color{ForestGreen}{$0$} 
\\
$\ket{g}$  &
\color{ForestGreen}{$0$} &
\color{Red}{$\neq 0$} &
\color{ForestGreen}{$0$} &
\color{ForestGreen}{$0$} &
\color{Red}{$\neq 0$} &
\color{Red}{$\neq 0$} &
\color{ForestGreen}{$0$} & 
\color{ForestGreen}{$0$} & 
\color{ForestGreen}{$0$} 
\\
\specialrule{.13em}{0em}{0em}
\end{tabular}
}
\caption{Action of the charges (\ref{eq:181}) on the vacuum state $\ket{0}$ and the states $\ket{Y}=e^{-i\mathcal{R}(Y)}\ket{0}$ and $\ket{g}=e^{-i\mathcal{D}(g)}\ket{0}$. The first three columns correspond to the charges (\ref{eq:214}) that generate ordinary conformal transformations. By ``$\neq 0$" we mean the state is not an eigenstate of the charge, but it has a non-trivial action.}\label{table:3}
\end{table}

We can then study features of the states $\ket{Y}$ and $\ket{g}$ in (\ref{eq:197}). To do so, we use the following two algebraic identities that hold for arbitrary operators $V$ and $W$
\begin{equation}\label{eq:53}
\begin{aligned}
We^{-iV}\ket{0}&=
\left[
e^{-iV}W+
\sum_{n=1}^\infty
\sum_{m=1}^n
\binom{n}{m}
  \frac{(-i)^n(-1)^{m}}{n!}
  V^{n-m}
  \mathscr{L}_{V}^{m}(W)
\right]
\ket{0}\ ,\\
\braket{0|
e^{iV}We^{-iV}
|0}&=
\sum_{n=0}^{\infty}
\frac{i^n}{n!}
\braket{0|
\mathscr{L}_V^n(W)
|0}\ ,
\end{aligned}
\end{equation}
where $\mathscr{L}_V(W)=[V,W]$, $\mathscr{L}^2_V(W)=[V,[V,W]]$ and so on. While the second identity is standard, we prove the first one in appendix \ref{zapp:4}. Using these relations, together with the algebra satisfied by the charges (\ref{eq:190}) and the transformation properties of the vacuum $\ket{0}$ given in the first row in table \ref{table:3}, we obtain the second and third rows, that determine the transformation properties of $\ket{Y}$ and $\ket{g}$. 

Let us now comment on the most salient features of this table. In the first three columns we have the action of the charges associated to ordinary conformal transformations on the transformed states $\ket{Y}$ and $\ket{g}$. One of the charges in $\mathcal{T}(f_0)$ is obtained from $f_0(\vec{y}\,)=1$, which corresponds to rigid translations in the $u$ time-like coordinate of the Minkowski metric (\ref{eq:174}), meaning that $H_u\equiv \mathcal{T}(f_0=1)$ is the associated Hamiltonian. Hence, the first column in table \ref{table:3} implies
\begin{equation}
H_u\ket{0}=H_u\ket{Y}=H_u\ket{g}=0\ .
\end{equation}
This shows the states $\ket{Y}$ and $\ket{g}$ are an infinite number of eigenstates of the Hamiltonian $H_u$ with minimum eigenvalue and thus can be regarded as soft modes. While the states are also invariant under rigid dilations generated by $\mathcal{D}(g_0)$, Lorentz transformations induce non-trivial transformations. This means there is not a true infinite degeneracy of the vacuum state $\ket{0}$, since we can distinguish it from the states $\ket{Y}$ and $\ket{g}$ by applying ordinary Lorentz transformations.

From the fourth to the sixth columns in table \ref{table:3}, we see that all the states are invariant under arbitrary supertranslations, same as the vacuum state $\ket{0}$. This follows from the ANEC $\mathcal{E}(\vec{y}\,)\ket{0}=0$ and the first identity in (\ref{eq:53}), where the second term is computed using the algebra (\ref{eq:190}). On the other hand, we see that all the states have a non-trivial transformation under superrotation and superdilation transformations. However, using the second identity in (\ref{eq:53}) together with the algebra, we can show the vacuum expectation of the charges in all these states vanish, as indicated in the last three columns in table \ref{table:3}.

All the results given in table \ref{table:3} apply to arbitrary CFTs in Minkowski space-time as well as on any of the conformally related space-times defined by performing the Weyl rescaling in (\ref{eq:175}). While one can study the action of the finite transformations on non-vacuum states, the setup becomes more complicated and it is difficult to make concrete statements.

\subsection{Algebra representations}
\label{subsec:3.2}

We can get a more detailed characterization of the action of the charges on the Hilbert space by studying the algebra representations of (\ref{eq:190}). Since the algebra greatly depends on the space-time dimension, we focus on the three and four dimensional cases.

\subsubsection{Three dimensions}

For the three dimensional case the vectors generating the transformations at $\rho=0$ (\ref{eq:171}) only depend on the angle $\phi$. To study its algebra, it is convenient to expand the functions $f(\phi)$, $Y(\phi)$ and $g(\phi)$ in terms of their Fourier series
\begin{equation}
f(\phi)=\sum_{n\in \mathbb{Z}}f_ne^{in\phi}\ ,
\qquad \qquad
Y(\phi)=\sum_{n\in \mathbb{Z}}
Y_ne^{in\phi}\ ,
\qquad \qquad
g(\phi)=\sum_{n\in \mathbb{Z}}
g_ne^{in\phi}\ ,
\end{equation}
where real functions demand the coefficient expansion satisfy $\bar{c}_n=c_{-n}$. Using this in the definition of the charges in (\ref{eq:181}) we find
\begin{equation}\label{eq:164}
\begin{aligned}
\mathcal{T}(f)&=
\sum_{n\in \mathbb{Z}}
f_n\left[
\lim_{\rho \rightarrow 0}
\int_{S^1}\frac{d\phi}{\rho}\,e^{in\phi}\mathcal{E}(\phi)
\right]\equiv 
\sum_{n \in \mathbb{Z}}f_nT_n\ ,\\
\mathcal{R}(Y)&=
\sum_{n\in \mathbb{Z}}Y_n\left[
\lim_{\rho \rightarrow 0}
\int_{S^1}\frac{d\phi}{\rho}\,
e^{in\phi}\left(
in\mathcal{K}(\phi)+\mathcal{N}_A(\phi)
\right)
\right]\equiv
\sum_{n\in \mathbb{Z}}Y_nR_n\ ,\\
\mathcal{D}(g)&=
\sum_{n\in \mathbb{Z}}
g_n\left[
\lim_{\rho\rightarrow 0}
\int_{S^1}\frac{d\phi}{\rho} \,
e^{in\phi}\mathcal{K}(\phi)
\right]\equiv
\sum_{n\in \mathbb{Z}}g_nD_n\ .
\end{aligned}
\end{equation}
Since the charges are Hermitian we have that all the mode operators verify $P^\dagger_n=P_{-n}$. Using this expansion on the algebra in (\ref{eq:190}) we find
\begin{equation}\label{eq:163}
\begin{aligned}
\left[
T_n,T_m
\right]&=\left[
D_n,D_m
\right]=0\ ,\\
\left[
T_{n},R_m
\right]&=(n-m)T_{m+n}\ ,
\qquad \qquad
\left[
R_n,R_m
\right]=(n-m)R_{n+m}\ ,\\
\left[
T_n,D_m
\right]&=iT_{n+m}\ ,
\qquad \qquad \qquad \quad \,
\left[
R_n,D_m
\right]=-mD_{n+m}\ .
\end{aligned}
\end{equation}
From these relations we identify several interesting sub-algebras. The subsets $\left\lbrace T_n \right\rbrace$, $\left\lbrace D_n \right\rbrace$ and $\left\lbrace R_n \right\rbrace$ give two abelian and a Witt sub-algebra respectively, while $\left\lbrace T_n,R_m \right\rbrace$ corresponds to the standard ${\rm BMS}_3$ algebra, obtained as a semi-direct sum of the abelian and Witt sub-algebras \cite{Oblak:2016eij}. The subset $\left\lbrace D_n \right\rbrace,\, \left\lbrace R_m \right\rbrace$ is an algebra that has appeared in other contexts, like in \cite{Donnay:2016ejv}. The subset $\left\lbrace D_n \right\rbrace,\, \left\lbrace T_m \right\rbrace$ can be regarded as an infinite-dimensional extension of the Borel subalgebra of $sl(2)$. In the following we study vacuum and non-vacuum representations of this algebra (see \cite{Campoleoni:2016vsh} for related work on the ${\rm BMS}_3$ algebra).

\paragraph{Vacuum representation:} The starting point for the vacuum representation is its invariance under the action generated by the following modes 
\begin{equation}\label{eq:199}
\begin{aligned}
T_n\ket{0}&=0\ , \qquad n\in \mathbb{Z}\ ,\\
R_{n}\ket{0}&=0\ , \qquad n=0,\pm 1\ , \\
D_0\ket{0}&=0\ .
\end{aligned}
\end{equation}
The first relation is implied by the ANEC operator $\mathcal{E}(\phi)$, that gives $\mathcal{E}(\phi)\ket{0}=0$ as in (\ref{eq:198}). The remaining conditions are a consequence of the invariance of the vacuum under ordinary Lorentz and dilation transformations, that correspond to taking $Y_0(\phi)$ and $g_0(\phi)$ in (\ref{eq:160}).

The vector space of the representation is spanned by acting successively with the operators $R_n$ and $D_n$ on the vacuum state $\ket{0}$, which gives
\begin{equation}\label{eq:201}
\ket{\lbrace n_i \rbrace;\lbrace m_j \rbrace}\equiv 
\prod_{i=1}^\ell D_{n_i}
\prod_{j=1}^k R_{m_j}\ket{0}\ ,
\qquad m_{j}\ge m_{j+1}\ .
\end{equation}
States with different ordering can be put in this form using the algebra (\ref{eq:163}). From the conditions in (\ref{eq:199}) and the algebra we can use induction to prove these states satisfy the following properties
\begin{equation}\label{eq:200}
\begin{aligned}
T_n\ket{\lbrace n_i \rbrace;\lbrace m_j \rbrace}&=D_0\ket{\lbrace n_i \rbrace;\lbrace m_j \rbrace}=0
\ ,  \\
R_0\ket{\lbrace n_i \rbrace;\lbrace m_j \rbrace}&=M_\phi
\ket{\lbrace n_i \rbrace;\lbrace m_j \rbrace}\ ,
\qquad \qquad
M_\phi\equiv -\sum_{i=1}^\ell n_i-
\sum_{j=1}^k m_j\ .
\end{aligned}
\end{equation}
We see that all supertranslations $T_n$ and rigid dilations $D_0$ annihilate all the states in the vacuum representation, in agreement with the previous analysis in general dimensions, summarized in table~\ref{table:3}. Since the hermitian operator $R_0$ is the angular momentum generator in the $\phi$ direction, the states $\ket{\lbrace n_i \rbrace;\lbrace m_j \rbrace}$ are angular momentum eigenstates with integer eigenvalue $M_\phi$ in (\ref{eq:200}), that gets contributions both from the modes $D_n$ and $R_m$. For a fixed eigenvalue $M_\phi$ there is a huge degeneracy, since there is a (very large) infinite number of ways to obtain $M_\phi$ by summing integer numbers.

It is instructive to compare with heighest weight representations of the Virasoro algebra. In that setup, the operator playing a similar role to the angular momentum $R_0$ is the energy $L_0$. Since a well defined CFT requires the energy to be bounded from below, we get that half of the operators in the Virasoro algebra $L_n$ annihilate the vacuum. For the representation in (\ref{eq:200}) the situation is very different, as there is no reason for the angular momentum to have a bounded spectrum, given that a state can have arbitrary angular momentum in either direction. This is reflected in the fact that the eigenvalue $M_\phi$ in (\ref{eq:200}) gets non-trivial contributions for \textit{integer} values of $n_i$ and $m_j$.

While the angular momentum eigenstates $\ket{\lbrace n_i \rbrace;\lbrace m_j \rbrace}$ provide a clear picture for the vector space of the vacuum representation, it is not the whole story since we cannot compute their inner product. The algebra together with the conditions in (\ref{eq:199}) are not enough to do so. However, we can proceed similarly as in \cite{Campoleoni:2016vsh} and construct states with well defined norm by instead considering the states
\begin{equation}\label{eq:140}
\mathcal{H}=
\Big\lbrace \,\,
\ket{g,Y}\equiv 
\braket{0|0}^{-1/2}
e^{-i(
\mathcal{D}(g)+\mathcal{R}(Y))}
\ket{0}
\quad :
\quad
g(\phi),Y(\phi)\in {\rm Diff}(S^1)
\,\,
\Big\rbrace\ ,
\end{equation}
where $\mathcal{D}(g)$ and $\mathcal{R}(Y)$ are the full charges given in (\ref{eq:164}). The crucial feature of $\mathcal{H}$ is that all of these states have unit norm by construction.\footnote{While the states $\ket{g,Y}$ have well defined norm we would like to have states that are orthonormal, i.e. $\braket{g',Y'|g,Y}=\delta(Y',Y)\delta(g,g')$, where $\delta(\cdot,\cdot)$ is a Dirac delta defined with respect to an appropriate measure in the space of functions. For a similar setup in \cite{Campoleoni:2016vsh} it was argued in favor of the existence of such measure and the irreducibility of the representation.} Although we are exponentiating the charges as we do when considering the group representation, we should think of (\ref{eq:140}) as a particular change of basis from the angular momentum eigenstates in (\ref{eq:201}).

Before moving on to consider non-vacuum representations, let us write a simple element of the Hilbert space in (\ref{eq:140}) explicitly, obtained by taking $(g,Y)=(0,2\varepsilon \cos(k\phi))$ for some integer $k$, so that we get
\begin{equation}
\ket{0,2\varepsilon\cos(k\phi)}=
e^{-i\varepsilon(R_k+R_{-k})}\ket{0}=
\ket{0}-i\varepsilon
(R_k+R_{-k})\ket{0}+\mathcal{O}(\varepsilon^2)\ .
\end{equation}
From this expression we see that the hermiticity condition on the charges forces that any particular mode $R_k$ must be accompanied by a mode with the same magnitude but inverse direction. Every time angular momentum is added in one direction of the circle, another excitation must be considered in the opposite direction. From this perspective it is reasonable to obtain a vanishing expectation value of the angular momentum $\braket{g,Y|R_0|g,Y}=0$ when taking the average, as previously obtained in table \ref{table:3}.

\paragraph{Non-vacuum representations:} We now consider non-vacuum representations of the algebra (\ref{eq:163}), where we start from an excited state $\ket{\psi}$ that is an eigenstate of a commuting subset of the operators that generate the ordinary conformal transformations
\begin{equation}
Q_0=\left\lbrace
T_0,T_{\pm 1},
R_0,R_{\pm 1},
D_0
\right\rbrace\ .
\end{equation}
There are three commuting subsets of operators we can chose from
\begin{equation}
S_1=\left\lbrace
T_0,T_{\pm 1}
\right\rbrace\ ,
\qquad
S_2=\left\lbrace
T_0,R_0
\right\rbrace\ ,
\qquad
S_3=\left\lbrace
R_0,D_0
\right\rbrace\ .
\end{equation}
We find $S_2$ to be the most convenient, so that the $\ket{\psi}$ corresponds to an energy and angular momentum eigenstate\footnote{We have also studied representations built from $S_1$, that can be extended $\left\lbrace T_{n}
\right\rbrace $ for all $n\in \mathbb{Z}$, but did not find this approach useful. The subset $S_3$ is also interesting, in which $R_0\ket{\psi}=\mu_\phi\ket{\psi}$ and $D_0\ket{\psi}=\Delta\ket{\psi}$. In Lorentzian signature the eigenvalue $\Delta$ of $D_0$ \textit{does not} correspond to the discrete and positive scaling dimensions of the operators in the CFT. We can see this by considering the state $T_0\ket{\psi}$ and using the algebra (\ref{eq:163}) to show it is also an eigenstate of $D_0$ but with complex eigenvalue $\Delta-i$. This is consistent with $D_0$ being hermitian only if the state $\ket{\psi}$ is not normalizable. Exactly this same issue arises when considering representations of the ordinary conformal algebra, starting from the state $\ket{\mathcal{O}}\equiv \mathcal{O}(0)\ket{0}$ where $\mathcal{O}(x^\mu)$ is a primary hermitian operator. Using the Ward identites and the conformal algebra one finds that while the state $\ket{\mathcal{O}}$ has the same issues as explained for $\ket{\psi}$, it is not normalizable since $\braket{\mathcal{O}|\mathcal{O}}=\braket{0|\mathcal{O}(0)^2|0}={\rm undefined}$. We thank David Simmons-Duffin for pointing us towards an explanation of this issue by Petr Kravchuk.\label{foot:1}}
\begin{equation}\label{eq:213}
T_0\ket{\psi}=E_u\ket{\psi}\ ,
\qquad \qquad
R_0\ket{\psi}=\mu_\phi\ket{\psi}\ ,
\end{equation}
where $E_u>0$ and $\mu_\phi \in \mathbb{Z}$. A difference with respect to the vacuum representation is that the state $\ket{\psi}$ is not invariant under arbitrary supertranslations $T_n$, as $\braket{\psi|\mathcal{E}(\phi)|\psi}\neq 0$ so that the argument in (\ref{eq:198}) does not apply. 

States in this representation are obtained by acting on $\ket{\psi}$ with any element of the algebra
\begin{equation}\label{eq:203}
\ket{\lbrace p_r \rbrace;\lbrace n_i \rbrace;\lbrace m_j \rbrace;\psi}\equiv 
\prod_{r=1}^s T_{p_r}
\prod_{i=1}^\ell D_{n_i}
\prod_{j=1}^k R_{m_j}\ket{\psi}\ ,
\qquad m_{j}\ge m_{j+1}\ .
\end{equation}
Acting on these states with the Hamiltonian $T_0$ and the angular momentum $R_0$ we can use the algebra (\ref{eq:163}) to show 
\begin{equation}
\begin{aligned}
R_0\ket{\lbrace p_r \rbrace;\lbrace n_i \rbrace;\lbrace m_j \rbrace;\psi}=
M_\phi\ket{\lbrace p_r \rbrace;\lbrace n_i \rbrace;\lbrace m_j \rbrace;\psi}\ ,
\quad \quad
M_\phi\equiv \mu_\phi 
-\sum_{r=1}^sp_r-
\sum_{i=1}^\ell n_i-
\sum_{j=1}^k m_j
\end{aligned}
\end{equation}
together with a cumbersome expression for $T_0\ket{\lbrace p_r \rbrace;\lbrace n_i \rbrace;\lbrace m_j \rbrace;\psi}$ that does not give an energy eigenstate. On the other hand, all the states in (\ref{eq:203}) are angular momentum eigenstates, where the action of all the elements in the algebra contribute to the eigenvalue $M_\phi$. This gives a generalization of the second relation in (\ref{eq:200}) for the vacuum representation. Similarly, while we cannot compute the norm of the states in (\ref{eq:203}), we can consider the normalized states constructed as
\begin{equation}
\mathcal{H}=\left\lbrace\,\,
\ket{f,g,Y}\equiv
\braket{\psi|\psi}^{-1/2}
e^{-i(\mathcal{T}(f)+\mathcal{D}(g)+\mathcal{R}(Y))}\ket{\psi}
\quad : \quad
f(\phi),g(\phi),Y(\phi)\in 
{\rm Diff}(S^1)\,\,
\right\rbrace \ ,
\end{equation}
where we are assuming the state $\ket{\psi}$ the representation is built from is normalizable (see footnote \ref{foot:1} for an example in which this is not the case).

\subsubsection{Four dimensions}

Let us now study representations of the algebra (\ref{eq:190}) in the four dimensional case, where it is convenient to use the complex coordinate $z=y^1+iy^2$ to parametrize the unit two-sphere, so that the metric (\ref{eq:176}) becomes
\begin{equation}
d\Omega^2_2=\frac{4dzd\bar{z}}{(1+z\bar{z})^2}\ .
\end{equation}
The arbitrary functions $f(z,\bar{z})$ and $g(z,\bar{z})$ on $S^2$ that define the asymptotic (conformal) Killing vectors in (\ref{eq:180}) can be expanded in a Laurent expansion as
\begin{equation}\label{eq:204}
f(z,\bar{z})=
  \frac{1}{1+|z|^2}
  \sum_{n,m\in \mathbb{Z}}
  f_{(n,m)}
  z^n\bar{z}^{m}\ ,
\qquad \qquad
g(z,\bar{z})=\sum_{n,m\in \mathbb{Z}}
g_{(n,m)}z^n\bar{z}^m\ .
\end{equation}
We have added the factor $(1+|z^2|)^{-1}$ to the expansion of $f(z,\bar{z})$ (as in \cite{Barnich:2011mi}) so that we can recover the functions in (\ref{eq:169}) that result in rigid space-time translations by considering a finite number of modes. Demanding the functions are real constraints the coefficients to satisfy $\bar{f}_{(n,m)}=f_{(m,n)}$ and $\bar{g}_{(n,m)}=g_{(m,n)}$.

For the vector $Y^A(z,\bar{z})$ we should in principle apply a similar procedure, in which we expand the two components in a Laurent series in both $z$ and $\bar{z}$. However, as a first step we constraint ourselves to a holomorphic ansatz given by 
\begin{equation}\label{eq:75}
Y^z(z,\bar{z})=Y(z)\ ,
\qquad \qquad
Y^{\bar{z}}(z,\bar{z})=\bar{Y}(\bar{z})\ .
\end{equation}
This is the usual approach taken in the BMS literature when instead of considering `super-Lorentz' transformations (obtained by ${\rm Diff}(S^2)$) one considers superrotations by restricting to (\ref{eq:75}). The six rigid Lorentz transformations given in (\ref{eq:169}) in terms of the stereographic coordinates $\vec{y}$, are obtained in complex coordinates by taking $Y(z)=a+bz+cz^2$ with $a$, $b$ and $c$ complex parameters. The holomorphic function $Y(z)$ can be expanded in a Laurent series as
\begin{equation}\label{eq:108}
  Y(z)=i
  \sum_{m\in \mathbb{Z}}
  Y_{m}z^{m+1}\ .
\end{equation}

We can now use the expansions in (\ref{eq:204}) and (\ref{eq:108}) in the Minkowski charges (\ref{eq:181}) at null infinity to define the following mode operators
\begin{equation}\label{eq:211}
\begin{aligned}
\mathcal{T}(f)&=
\sum_{n,m\in \mathbb{Z}}
f_{(n,m)}
\left[
\lim_{\rho\rightarrow 0}
\int_{S^2}
\frac{d\Omega_2}{\rho^{2}}
  \frac{z^n\bar{z}^{m}}
  {1+|z|^2}
  \mathcal{E}(z,\bar{z})
\right]\equiv
\sum_{n,m\in \mathbb{Z}}
f_{(n,m)}
T_{(n,m)}\ ,
\\
\mathcal{R}(Y)&=
\sum_{m\in \mathbb{Z}}
Y_m
\left[i
\lim_{\rho \rightarrow 0}
\int_{S^{2}}
\frac{d\Omega_2}{\rho^2}
\left\lbrace
\left[
\frac{m+1}{2}
-\frac{|z|^2}{1+|z|^2}
\right]
z^{m}
\mathcal{K}(z,\bar{z})+
z^{m+1}\mathcal{N}_z(z,\bar{z})
\right\rbrace
\right]
+{\rm h.c.}=\\
&\equiv \sum_{m\in \mathbb{Z}}  \left[  Y_m
R_m+\bar{Y}_mR^\dagger_m \right]\ ,
\\
\mathcal{D}(g)&=
\sum_{n,m\in \mathbb{Z}}
g_{(n,m)}
\left[
\lim_{\rho\rightarrow 0}
\int_{S^2}
\frac{d\Omega_2}{\rho^2}
z^n\bar{z}^m
\mathcal{K}(z,\bar{z})
\right]\equiv
\sum_{n,m\in \mathbb{Z}}
g_{(n,m)}
D_{(n,m)}\ ,
\end{aligned}
\end{equation}
where the additional term in the superrotation modes $R_m$ comes from the connection on $S^2$ when computing the divergence of $Y^A(z,\bar{z})$. Since the full charges are hermitian operators, we get the following conditions on the modes
\begin{equation}
T_{(n,m)}^\dagger=T_{(m,n)}\ ,
\qquad \qquad
D^\dagger_{(n,m)}=
D_{(m,n)}\ ,
\end{equation}
while $R_m$ and $R_m^\dagger$ are independent. Using the algebra (\ref{eq:190}) we can work out the following algebra satisfied by the modes
\begin{equation}\label{eq:109}
\begin{alignedat}{2}
\left[T_{(n,m)},T_{(k,r)}\right]&=
\left[D_{(n,m)},D_{(k,r)}\right]=0\ ,
\qquad \qquad \qquad \qquad
\big[R_n,R_m^\dagger\big]&&=0
\ , \\
\left[T_{(n,m)},R_k\right]&=
\Big(n-\frac{k+1}{2}\Big)
T_{(n+k,m)}\ ,
\qquad \qquad \qquad \quad \,\,
\big[R_n,R_m\big]&&=
(n-m)R_{n+m}
\ ,\\
\left[T_{(n,m)},D_{(k,r)}\right]&=
iT_{(n+k,m+r)}\ ,
\qquad \qquad \qquad \qquad \qquad \,\,\,\, 
\left[R_n,D_{(k,r)}\right]&&=
-kD_{(n+k,r)}\ .
\end{alignedat}
\end{equation}
Similar expressions involving the independent operator $R_n^\dagger$ are obtained by taking the Hermitian conjugate. The sub-algebra generated by $\big\lbrace T_{(n,m)},R_n,R^\dagger_n \big\rbrace$ is the BMS$_4$ algebra built from supertranslations and superrotations.\footnote{Matching with the ${\rm BMS}_4$ algebra as written in \cite{Barnich:2011mi} involves redefining some generators by adding some additional minus signs.} Although the algebra is more complicated than the three dimensional case in (\ref{eq:163}), there are several qualitative similarities that result in similar representations.

\paragraph{Vacuum representation:} The starting point of the vacuum representation is its invariance under ordinary conformal transformations, that gives
\begin{equation}\label{eq:212}
\begin{aligned}
T_{(n,m)}\ket{0}&=0\ , \qquad \qquad
(n,m)\in \mathbb{Z}\times \mathbb{Z}\ , \\
R_{n}\ket{0}=
R^\dagger_{n}\ket{0}&=0\ , \qquad \qquad n=0,\pm 1\ ,\\
D_{(0,0)}\ket{0}&=0\ .
\end{aligned}
\end{equation}
All the supertranslation modes annihilate the vacuum due to the argument involving the ANEC in (\ref{eq:198}). The Hamiltonian operator that generates rigid $u$ translations is written in terms of the supertranslation modes as
\begin{equation}\label{eq:215}
H_u=
\lim_{\rho \rightarrow 0}
  \frac{1}{\rho^2}
  \int_{S^2}d\Omega_2\,
  \mathcal{E}(z,\bar{z})
=T_{(0,0)}+T_{(1,1)}
  \ ,
\end{equation}
while the three rigid rotations are generated by
\begin{equation}\label{eq:216}
\begin{alignedat}{2}
Y_1(z)&=iz\ ,
\qquad \qquad \qquad \qquad \qquad
J_1&&=R_0+R_0^\dagger\ , \\
Y_2(z)&=\frac{z^2+1}{2}\ ,
\qquad \qquad \qquad \qquad
J_2&&=\frac{-i}{2}\left[
(R_1-R_1^\dagger)+(R_{-1}-R_{-1}^\dagger)
\right]\ ,\\
Y_3(z)&=\frac{i(z^2-1)}{2}\ ,
\qquad \qquad \qquad \,\,\,\,
J_3&&=
\frac{1}{2}
\left[
(R_1+R_1^\dagger)-(R_{-1}+R_{-1}^\dagger)
\right]\ .
\end{alignedat}
\end{equation}
From the algebra (\ref{eq:109}) one can easily show these operators satisfy the appropriate ${\rm SO}(3)$ algebra $[J_i,J_j]=i\epsilon_{ijk}J_k$. The vectors space in the vacuum representation can be spanned by acting with all the elements of the algebra on the vacuum state $\ket{0}$, which results in the following states
\begin{equation}
\ket{
\left\lbrace n_i,m_j \right\rbrace;
\left\lbrace p_r \right\rbrace;
\left\lbrace \bar{p}_{\bar{r}}\right\rbrace}\equiv
\prod_{i=1}^s
\prod_{j=1}^\ell
D_{(n_i,m_j)}
\prod_{r=1}^k
R_{p_r}
\prod_{\bar{r}=1}^{\bar{k}}
R_{\bar{p}_{\bar{r}}}^\dagger
\ket{0}\ .
\end{equation}
Using the algebra (\ref{eq:109}) together with the relations in (\ref{eq:212}) we can show these states satisfy the following properties
\begin{equation}\label{eq:218}
\begin{aligned}
T_{(n,m)}
\ket{
\left\lbrace n_i,m_j \right\rbrace;
\left\lbrace p_r \right\rbrace;
\left\lbrace \bar{p}_{\bar{r}}\right\rbrace}&=
D_{(0,0)}
\ket{
\left\lbrace n_i,m_j \right\rbrace;
\left\lbrace p_r \right\rbrace;
\left\lbrace \bar{p}_{\bar{r}}\right\rbrace}=0\\
J_1
\ket{
\left\lbrace n_i,m_j \right\rbrace;
\left\lbrace p_r \right\rbrace;
\left\lbrace \bar{p}_{\bar{r}}\right\rbrace}&=
M_1\ket{
\left\lbrace n_i,m_j \right\rbrace;
\left\lbrace p_r \right\rbrace;
\left\lbrace \bar{p}_{\bar{r}}\right\rbrace}\ ,
\end{aligned}
\end{equation}
where the angular momentum eigenvalue $M_1$ in the direction of $J_1$ is given by
\begin{equation}
M_1\equiv 
-\sum_{i=1}^sn_i
-\sum_{r=1}^kp_r
+\sum_{j=1}^\ell m_j
+\sum_{\bar{r}=1}^{\bar{k}}
\bar{p}_{\bar{r}}\ .
\end{equation}
These relations are very similar to (\ref{eq:200}) for the vacuum representations in three dimensions: all the states are angular momentum eigenstates and annihilated by supertranslations and rigid dilations. There is a difference coming from the fact that in the four dimensional case some of the operators contribute with a minus sign to the angular momentum, while others with a plus. We can define states in the representation with a well defined norm in an analogous way as done in (\ref{eq:140}).

\paragraph{Non-vacuum representations:} Non-vacuum representations of the algebra (\ref{eq:109}) are obtained by starting from a state $\ket{\psi}$, which is an energy and rotation eigenstate 
\begin{equation}\label{eq:217}
H_u\ket{\psi}=E_u\ket{\psi}\ ,
\qquad \qquad
J_1\ket{\psi}=\mu_1\ket{\psi}\ ,
\qquad \qquad
J^2\ket{\psi}=j(j+1)\ket{\psi}\ ,
\end{equation}
where $E_u>0$, $\mu_1\in \mathbb{Z}$ and $j\in \mathbb{N}_0$. The Hamiltonian $H_u$ is given by (\ref{eq:215}) while $J^2=J_1^2+J_2^2+J_3^2$ is the Casimir of ${\rm SO}(3)$ built from the modes in (\ref{eq:216}). A straightforward computation using the algebra (\ref{eq:109}) shows that these operators commute, as expected. The vector space of the representation is spanned by the states obtained by acting on $\ket{\psi}$ with all the elements all the algebra, which gives
\begin{equation}\label{eq:223}
\prod_{a=1}^t
\prod_{b=1}^f
T_{(q_a,w_b)}
\prod_{i=1}^s
\prod_{j=1}^\ell
D_{(n_i,m_j)}
\prod_{r=1}^k
R_{p_r}
\prod_{\bar{r}=1}^{\bar{k}}
R_{\bar{p}_{\bar{r}}}^\dagger\ket{\psi}\ .
\end{equation}
Acting on these states with the angular momentum $J_1$, we find that it is an eigenstate as in (\ref{eq:218}) with eigenvalue given by
\begin{equation}
M_1\equiv
\mu_1
-\sum_{a=1}^tq_a
-\sum_{i=1}^sn_i
-\sum_{r=1}^kp_r
+\sum_{b=1}^f w_b
+\sum_{j=1}^\ell m_j
+\sum_{\bar{r}=1}^{\bar{k}}
\bar{p}_{\bar{r}}
\ ,
\end{equation}
which generalizes the second relation in (\ref{eq:218}) for the vacuum representation. Similarly as in the three dimensional case, the states in (\ref{eq:223}) are not energy eigenstates. All in all, we see the representations of the four dimensional algebra have the same qualitative features as the three dimensional case.

\section{Holography}
\label{sec:4}

In previous sections, we studied several quantum aspects of the asymptotic transformations introduced in section \ref{sec:1}. We showed the associated charges satisfy the expected algebra, and studied its representation in the Hilbert space of the CFT, as well as the action of finite transformations on the state space obtained by exponentiating the charges. In this section, we present a holographic description in the context of the AdS$_{d+1}$/CFT$_d$ correspondence. More precisely, we give a bulk geometric realization of the states $\ket{f}$, $\ket{Y}$ and $\ket{g}$ defined in (\ref{eq:197}) and studied in section \ref{sec:3}.

Our starting point is the bulk description of the vacuum state $\ket{0}$ of the CFT, which, as usual, is identified with pure ${\rm AdS}_{d+1}$. It is first convenient to write this bulk metric using an appropriate set of coordinates such that the boundary is given by (\ref{eq:175}), described by the boundary coordinates $(u,\rho,\vec{y}\,)$. We can do this by resorting to the embedding description of ${\rm AdS}_{d+1}$, starting from the surface
\begin{equation}\label{eq:157}
-(X^0)^2-(X^1)^2+\sum_{i=2}^{d+1}(X^i)^2=-1\ ,
\end{equation} 
on the $(d+2)$-dimensional space
\begin{equation}\label{eq:146}
ds^2=
-(dX^0)^2-(dX^1)^2+\sum_{i=2}^{d+1}(dX^i)^2\ .
\end{equation}
The appropriate parametrization of this surface that results in the wanted boundary metric is given by
\begin{equation}\label{eq:224}
X^0=\frac{1+\rho u}{\sin(\psi)}\ ,
\qquad \qquad
\begin{aligned}
X^1&=\frac{\rho-u(2+\rho u)}{2\sin(\psi)}\ ,\\
X^2&=\frac{\rho+u(2+\rho u)}{2\sin(\psi)}\ ,
\end{aligned}
\qquad \qquad
\begin{aligned}
X^3&=\cot(\psi)
\frac{|\vec{y}\,|^2-1}{|\vec{y}\,|^2+1}\ ,\\
X^{3+A}&=\cot(\psi)
  \frac{2y^A}{|\vec{y}\,|^2+1}\ ,
\end{aligned}
\end{equation}
where $y^A$ with $A=1,\dots,d-2$. The constraint (\ref{eq:157}) is automatically satisfied by the parametrization, while the induced metric (\ref{eq:146}) gives the ${\rm AdS}_{d+1}$ space-time
\begin{equation}\label{eq:158}
ds^2_{{\rm AdS}_{d+1}}=
\frac{-\rho^2du^2+2dud\rho+d\psi^2+\cos^2(\psi)
 d\Omega_{d-2}^2}{\sin^2(\psi)}\ ,
\end{equation}
where $d\Omega_{d-2}^2$ is determined by $\vec{y}$ as in (\ref{eq:176}). The ${\rm AdS}_{d+1}$ boundary of the metric is obtained by taking the limit $\psi\rightarrow 0$ of the coordinate $\psi \in (0,\pi/2]$ together with an appropriate rescaling, so that the boundary metric becomes
\begin{equation}
\lim_{\psi \rightarrow 0}
\left(
\frac{\sin(\psi)}{\rho\, w(x^\mu)}
\right)^2ds^2_{{\rm AdS}_{d+1}}=
\frac{-\rho^2du^2+2dud\rho+
 d\Omega_{d-2}^2}{\rho^2w(x^\mu)^2}\ .
\end{equation}
Depending on the value of the function $w(x^\mu)$ (that corresponds to choosing a conformal frame) we obtain a different metric at the boundary, matching  with (\ref{eq:175}).\footnote{Note that to recover an ${\rm AdS}_d$ boundary from the bulk ${\rm AdS}_{d+1}$, the conformal factor is given by $w(x^\mu)^2=\sin^2(\hat{\psi})/\rho^2$ where $\hat{\psi}$ is different from the bulk coordinate in (\ref{eq:158}).} 

\begin{figure}
\centering
\qquad \qquad \quad
\includegraphics[scale=0.45]{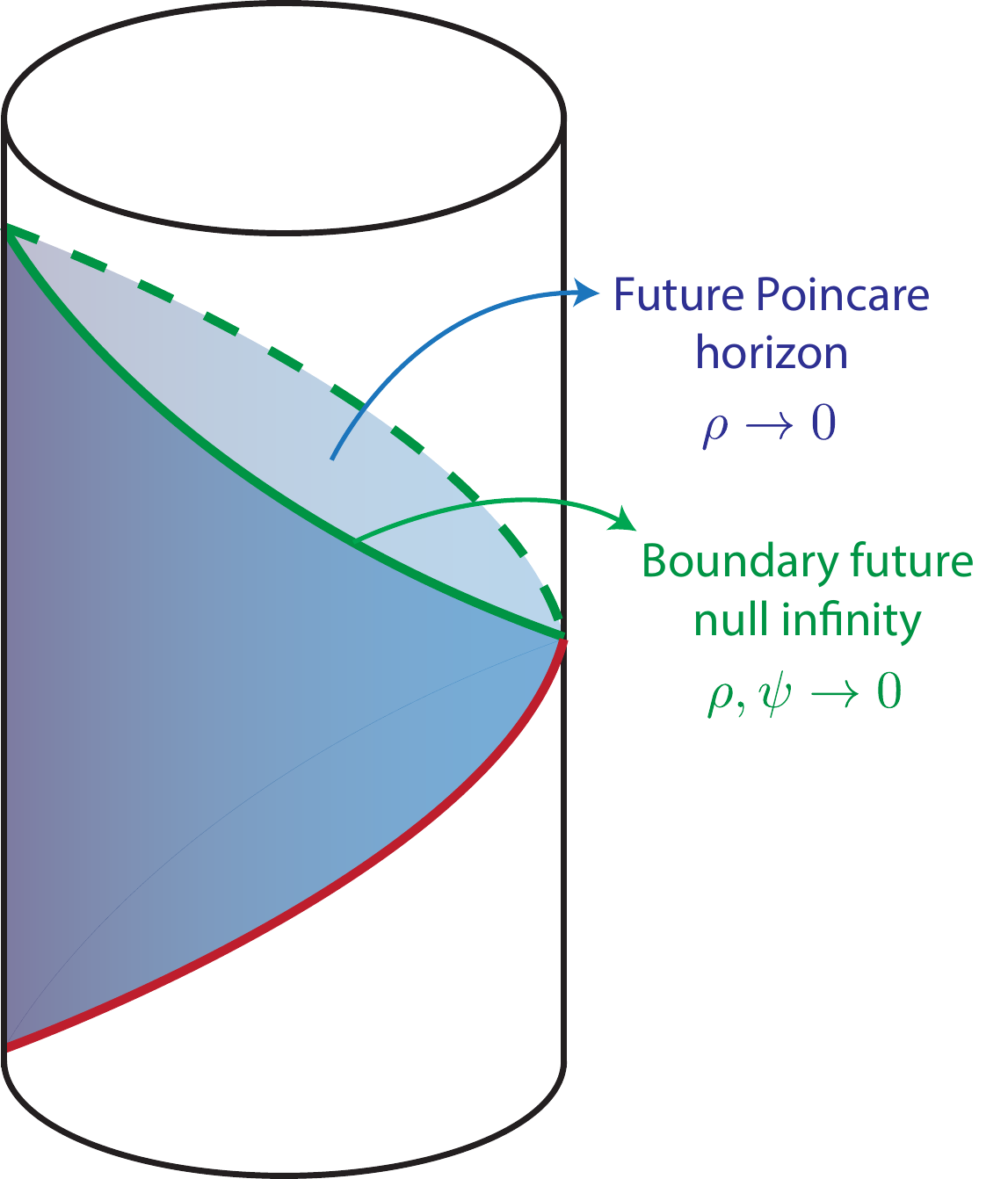}
\caption{Diagram representing the region of global ${\rm AdS}_3$ (solid cylinder) covered by the Poincar\'e patch (solid blue region) in the coordinates in (\ref{eq:158}). While the future boundary of the bulk Poincar\'e patch is given by $\rho \rightarrow 0$, the future null infinity of the AdS boundary (the boundary of the boundary) is located at $\rho,\psi\rightarrow 0$.}\label{fig:5}
\end{figure}

The unusual ${\rm AdS}_{d+1}$ coordinates (\ref{eq:158}) give the bulk description of the ground state $\ket{0}$. The bulk horizon at $\rho=0$ corresponds to the (future) Poincar\'e horizon of ${\rm AdS}_{d+1}$, meaning the coordinates $(u,\psi,\rho,\vec{y}\,)$ only cover the Poincar\'e patch of anti-de Sitter. This is clear from the discussion in subsection 3.2 of \cite{Rosso:2020cub}, where essentially the same coordinates where constructed for a different purpose. See figure~\ref{fig:5} for a diagram of the Poincar\'e patch embedded in the cylinder representing global ${\rm AdS}_3$. 

Now that we have an appropriate bulk description of the CFT ground state, we would like to extended the asymptotic (conformal) Killing vectors (\ref{eq:178}) for $\psi \neq 0 $, i.e. away from the boundary and into the bulk. For the case in which the boundary vectors generate exact conformal transformations (which amounts to taking the functions as in (\ref{eq:169})) we already know how to do this, as the isometries of the bulk metric ${\rm AdS}_{d+1}$ give the boundary conformal transformations. The bulk Killing vectors are most easily described in terms of the embedding coordinates (\ref{eq:224}) as
\begin{equation}\label{eq:149}
\chi_{ab}=X_a\frac{\partial}{\partial X^b}-
X_b\frac{\partial}{\partial X^a}\ ,
\qquad \qquad
a,b=0,1,\dots,(d+1)\ ,
\end{equation}
where the index in $X^a$ is lowered using the embedding metric (\ref{eq:146}). Similarly as in section \ref{sec:1}, let us consider the cases $d=3$ and $d>3$ separately.

\subsection{Three dimensional boundary}

For a three dimensional boundary ($d=3$), it is convenient to write $y(\phi)=\tan(\phi/2)$ so that the ${\rm AdS}_4$ bulk metric (\ref{eq:158}) takes the form
\begin{equation}\label{eq:224n}
ds^2_{{\rm AdS}_{4}}=
\frac{-\rho^2du^2+2dud\rho+d\psi^2+\cos^2(\psi)
 d\phi^2}{\sin^2(\psi)}\ .
\end{equation}
The ten Killing vectors of this bulk geometry can be written in a compact way using (\ref{eq:149}); namely
\begin{equation}\label{eq:151}
\begin{aligned}
\chi_T(f)&=
\cos(\psi)
\left[f(\phi)\partial_u-
\rho^2f''(\phi)\partial_\rho
\right]
-\rho
\frac{f'(\phi)}{\cos(\psi)}\partial_\phi
-\rho\sin(\psi)f''(\phi)\partial_\psi\ ,
\\
\chi_R(Y)&=
\cos^2(\psi)
\left[
Y'(\phi)
u\partial_u
+\rho
\left( Y'(\phi)
-\rho Y'''(\phi) u 
\right)\partial_\rho
\right]+
\left[Y(\phi)-\rho Y''(\phi) u\right]
\partial_\phi
+\\
&\hspace{17mm}
+\frac{1}{2}\sin(2\psi)
\left[Y'(\phi)-\rho Y'''(\phi) u\right]
\partial_\psi\ , \\
\chi_S(h)&=
\cos(\psi)h(\phi)u^2\partial_u
-h'(\phi)
\frac{u(2+\rho u)}
{\cos(\psi)}\partial_\phi
-\cos(\psi)
\left[
h(\phi)2(1+\rho u)
+h''(\phi)(2+\rho u)^2
\right]\partial_\rho+\\
& \hspace{17mm}- h''(\phi)\sin(\psi)
u(2+\rho u)
\partial_\psi \, ,\\
\chi_D(g)&=ug(\phi)\partial_u
-\rho \left[ g(\phi)+\rho u g''(\phi) \right]\partial_\rho
-\rho u g'(\phi)\partial_\phi
\ ,
\end{aligned}
\end{equation}
where the functions that determine the vectors are given by
\begin{equation}\label{eq:196}
\begin{aligned}
f_0(\phi)&=\frac{a_0}{\cos(\psi)}+a_1\cos(\phi)+a_2\sin(\phi)\ ,\\
Y_0(\phi)&=b_0+
\frac{b_1\cos(\phi)+b_2\sin(\phi)}{\cos(\psi)}\ , \\
h_0(\phi)&=\frac{c_0}{\cos(\psi)}+c_1\cos(\phi)+c_2\sin(\phi)\ ,\\
g_0(\phi)&=d_0\ .
\end{aligned}
\end{equation}
It is straightforward to check these vectors generate exact isometries of the metric (\ref{eq:224n}) provided the functions are fixed according to (\ref{eq:196}). Moreover, when taking the boundary limit $\psi \rightarrow 0$ for arbitrary functions, we recover the boundary vectors in (\ref{eq:159})
\begin{equation}\label{eq:225}
\lim_{\psi \rightarrow 0}
\chi_p=\xi_p\ ,
\qquad
p=T,R,S,D\ .
\end{equation}
While this suggests the bulk vectors (\ref{eq:151}) are the extension of the boundary vectors into the bulk, it is clear these vectors are highly not unique. For instance, if we redefine the vector as $\chi_D(g)\rightarrow \chi_D(g)+\sin(\psi)g'(\phi)\partial_u$, then the resulting vector also gives an isometry when $g(\phi)=d_0$ and preserves the boundary limit in (\ref{eq:225}). However, we favor the bulk vectors in (\ref{eq:151}) over these other possibilities given that, when evaluated on the bulk Poincar\'e horizon $\rho=0$, i.e.
\begin{equation}\label{eq:154}
\begin{aligned}
\chi_T(f)\big|_{\rho=0}&=
\cos(\psi)f(\phi)\partial_u\ , \\
\chi_R(Y)\big|_{\rho=0}&=
\cos^2(\psi)
Y'(\phi)
u\partial_u+
Y(\phi)\partial_\phi
+\frac{1}{2}\sin(2\psi)
Y'(\phi)
\partial_\psi\ ,\\
\chi_S(h)\big|_{\rho=0}&=
\cos(\psi)h(\phi)u^2\partial_u
-
\frac{2uh'(\phi)}
{\cos(\psi)}\partial_\phi
-2\cos(\psi)
\left[
h(\phi)
+2h''(\phi)
\right]\partial_\rho-
2uh''(\phi)\sin(\psi)
\partial_\psi\ ,\\
\chi_D(g)\big|_{\rho=0}&=g(\phi)u\partial_u\ ,
\end{aligned}
\end{equation}
the vectors $\left\lbrace \chi_T(f),\chi_R(Y)\right\rbrace \cup \left\lbrace \chi_D(g)\right\rbrace$ satisfy a closed algebra that agrees exactly with the one satisfied by the boundary vectors, i.e. replacing $\xi_p\rightarrow \chi_p$ in (\ref{eq:165}).\footnote{Same as in the boundary analysis, including the vector $\chi_S(h)$ results in an algebra that does not close.} In other words, the boundary algebra is extended into the bulk by considering the bulk vectors (\ref{eq:151}) since other obvious choices typically spoil the algebra in the bulk.

Using the bulk vectors $\chi_p$ there is a natural proposal for the holographic description of the boundary states
\begin{equation}\label{eq:233}
\ket{f}\equiv e^{-i\mathcal{T}(f)}\ket{0}\ ,
\qquad \qquad
\ket{Y}\equiv e^{-i\mathcal{R}(Y)}\ket{0}\ ,
\qquad \qquad
\ket{g}\equiv e^{-i\mathcal{D}(g)}\ket{0}\ .
\end{equation}
The equivalent action of the boundary operator $e^{-i\mathcal{T}(f)}$ is obtained by computing the finite geometric transformation associated to the bulk vector $\chi_T(f)$, implemented by the differential operator $e^{\chi_T(f)}$. An explicit expression for the action of $e^{\chi_T(f)}$ is obtained by computing the integral curves associated to $\chi_T(f)$. Our proposal is that the bulk description of the boundary states (\ref{eq:233}) is given by the metric obtained by acting with $e^{\chi_p}$ on the ${\rm AdS}_4$ metric (\ref{eq:224n}), so that we get
\begin{equation}\label{eq:234}
g_{\mu \nu}(\chi_T)\equiv e^{\chi_T}(g_{\mu \nu}^{\rm AdS})\ ,
\qquad \qquad
g_{\mu \nu}(\chi_R)\equiv e^{\chi_R}(g_{\mu \nu}^{\rm AdS})\ ,
\qquad \qquad
g_{\mu \nu}(\chi_D)\equiv e^{\chi_D}(g_{\mu \nu}^{\rm AdS})\ .
\end{equation}
The usual AdS/CFT dictionary then maps the boundary expectation values to the bulk Noether charges, see last row in table \ref{table:7}, where we summarize the holographic dictionary.

\begin{table}[t]
\setlength{\tabcolsep}{10 pt} 
\centering
\begin{tabular}{  Sc | Sc  }
\specialrule{.13em}{0em}{0em}
Boundary CFT$_d$  &
Semi-classical gravity dual
\\
\specialrule{.05em}{0em}{0em}
$\displaystyle \ket{0}$  &
$\displaystyle g_{\mu \nu}^{\rm AdS}$ 
\\
$\displaystyle \xi_p$  &
$\displaystyle \chi_p$ 
\\
$\displaystyle 
\ket{\xi_p}\equiv e^{-i\widehat{Q}[\xi_p]}\ket{0}$  &
$\displaystyle 
g_{\mu \nu}(\chi_p)\equiv
e^{\chi_p}(g_{\mu \nu}^{\rm AdS})	$ 
\\
$\displaystyle
\braket{\xi_p|
\widehat{Q}[\xi_q]
|\xi_p}$  &
$\displaystyle 
Q_{g_{\mu \nu}(\chi_p)}[\chi_q]$ 
\\
\specialrule{.13em}{0em}{0em}
\end{tabular}
\caption{
Summary of our holographic proposal relating the boundary states (\ref{eq:233}) to the bulk metrics (\ref{eq:234}), where $p=T,R,D$. For the boundary charge $\widehat{Q}$ on the first column we add a hat to remind ourselves it is an operator, written in terms of the stress tensor as in (\ref{eq:182}). The metric $g_{\mu \nu}(\chi_p)$ in the second column is given in (\ref{eq:234}), obtained by acting on the pure ${\rm AdS}_{d+1}$ metric with the (finite) transformation generated by $\chi_p$. $Q_{g_{\mu \nu}(\chi_p)}[\chi_q]$ corresponds to the Noether charge associated to the vector $\chi_q$ computed in the metric $g_{\mu \nu}(\chi_p)$.
}\label{table:7}
\end{table}

To test this proposal we must compare the bulk and boundary results obtained when computing the last two rows in table \ref{table:7} from either side of the duality. For the boundary CFT this corresponds to the results summarized in table \ref{table:3}. In the bulk we must compute two different quantities, the transformation of the metric itself, given by $g_{\mu \nu}(\chi_p)$ in (\ref{eq:234}) and the gravitational Noether charges.

\subsubsection{Bulk metric transformation}

To calculate the transformation of the metric under the action of the vectors $\chi_p$ (as in (\ref{eq:234})), we first need to compute their associated integral curves. While for general values of $\rho$ the vectors in (\ref{eq:151}) have complicated expressions, making it very difficult to compute their integral curves explicitly, they are much simpler when evaluated at the bulk Poincar\'e horizon $\rho=0$ (\ref{eq:154}). This is analogous to the situation on the boundary, where we can only easily study the states (\ref{eq:233}) on the surface $\rho=0$, since the quantum charges (\ref{eq:182}) away from $\rho=0$ have complicated expressions. The pure ${\rm AdS}_4$ metric evaluated at $\rho=0$ is given by
\begin{equation}\label{eq:227}
ds^2_{{\rm AdS}_4}\big|_{\rho=0}=
\frac{d\psi^2+\cos^2(\psi)d\phi^2}{\sin^2(\psi)}=
\frac{dx^2}{x^2(1+x^2)}+
\frac{d\phi^2}{x^2}\ ,
\end{equation}
where in the second equality we have redefined the bulk coordinate $\psi$ according to $x \equiv\tan(\psi)\ge 0$. Note that although we have not fixed the time coordinate $u$, the induced metric is independent of $u$. The integral curves of the bulk vectors at $\rho=0$ can be computed analytically, so that the action of $e^{\chi_p}$ on the coordinates $(u,\phi,x)$ is given by
\begin{equation}\label{eq:226}
\begin{aligned}
e^{\chi_T(f)}\big|_{\rho=0}&:
\quad
(u,\phi,x)
\quad \longrightarrow \quad
(u+\cos(\psi)f(\phi),\phi,x)\ ,\\
e^{\chi_R(Y)}\big|_{\rho=0}&:
\quad
(u,\phi,x)
\quad \longrightarrow \quad
\left(
\frac{\sqrt{1+x^2}}
{\sqrt{1+\alpha'(\phi)^2x^2}}
\alpha'(\phi)u
,\alpha(\phi),
\alpha'(\phi)x
\right)\ , \\
e^{\chi_D(g)}\big|_{\rho=0}&:
\quad
(u,\phi,x)
\quad \longrightarrow \quad
(e^{g(\phi)}u,\phi,x)\ ,
\end{aligned}
\end{equation}
where the function $\alpha(\phi)$ is implicitly defined through\footnote{See appendix \ref{zapp:5} for details regarding the integral curves associated to the bulk vector $\chi_R(Y)\big|_{\rho=0}$. As a concrete example, if we take $Y(\phi)=\cos^2(n\phi)$, the function $\alpha(\phi)$ is defined from $\tan(n\alpha(\phi))=\tan(n\phi)+n$.}
\begin{equation}
1=\int_\phi^{\alpha(\phi)}\frac{d\phi'}{Y(\phi')}\ .
\end{equation}
As a check, we take the boundary limit $x\rightarrow 0$ and find $e^{\chi_T(f)}$ and $e^{\chi_R(Y)}$ in (\ref{eq:226}) agree with the standard supertranslation and superrotation transformations in three dimensional Minkowski (see section 9.1.2 in \cite{Oblak:2016eij}).

Let us now analyze how the ${\rm AdS}_4$ metric at $\rho=0$ (\ref{eq:227}) behaves under these transformations. For the special case in which we take the functions $\left\lbrace f,Y,g \right\rbrace=\left\lbrace f_0,Y_0,g_0 \right\rbrace$ as in (\ref{eq:196}), the vectors $\chi_p$ are exact Killing vectors of ${\rm AdS}_4$, meaning the full metric is invariant. For arbitrary functions $\left\lbrace f,Y,g \right\rbrace$, the supertranslated and superdilated metrics at $\rho=0$ are also invariant 
\begin{equation}\label{eq:235}
\begin{aligned}
ds^2_T\big|_{\rho=0}
&\equiv 
g_{\mu \nu}(\chi_T)dx^\mu dx^\nu\big|_{\rho=0}=
\frac{dx^2}{x^2(1+x^2)}+\frac{d\phi^2}{x^2}=
ds^2_{{\rm AdS}_4}\big|_{\rho=0}\ ,\\
ds^2_D\big|_{\rho=0}
&\equiv 
g_{\mu \nu}(\chi_D)dx^\mu dx^\nu\big|_{\rho=0}=
\frac{dx^2}{x^2(1+x^2)}+\frac{d\phi^2}{x^2}=
ds^2_{{\rm AdS}_4}\big|_{\rho=0}\ ,
\end{aligned}
\end{equation}
since the ${\rm AdS}_4$ metric at $\rho=0$ is independent of $u$ (\ref{eq:227}). On the other hand, the superrotated bulk metric transforms in an interesting way
\begin{equation}\label{eq:237}
ds^2_R\big|_{\rho=0}
\equiv 
g_{\mu \nu}(\chi_R)dx^\mu dx^\nu\big|_{\rho=0}=
\frac{( \alpha'(\phi)dx+\alpha''(\phi) x d\phi)^2}{\alpha'(\phi)^2x^2(1+\alpha'(\phi)^2x^2)}
+\frac{d\phi^2}{x^2}
\neq 
ds^2_{{\rm AdS}_4}\big|_{\rho=0}\ .
\end{equation}

The transformation of the ${\rm AdS}_4$ under the action of $\chi_T(f)$ and $\chi_R(Y)$ agree with the boundary CFT prediction, where we have
\begin{equation}
\ket{f}=e^{-i\mathcal{T}(f)}\ket{0}=\ket{0}\ ,
\qquad \qquad
\ket{Y}=e^{-i\mathcal{R}(Y)}\ket{0}\neq \ket{0}\ .
\end{equation}
However, this is not the case for superdilations, as the bulk metric in (\ref{eq:235}) is invariant but the boundary state is not, i.e. $\ket{g}=e^{-i\mathcal{D}(g)}\ket{0}\neq \ket{0}$. As we will see below, the vector $\chi _D$ also exhibits another pathology related to the Noether charges, which diverges for certain cases. We discuss the disagreement between boundary and bulk results involving $\chi_D(g)$ at the end of this subsection.

We can then apply a second transformation on the metrics $g_{\mu \nu}(\chi_p)$ in (\ref{eq:235}) and (\ref{eq:237}), using (\ref{eq:226}). While applying another transformations to the metrics $g_{\mu \nu}(\chi_T)$ and $g_{\mu \nu}(\chi_D)$ at $\rho=0$ is not different from considering pure AdS, the case of $g_{\mu \nu}(\chi_R)$ is more interesting since the space-time at $\rho=0$ is distinct (\ref{eq:237}). For instance, note that $g_{\mu \nu}(\chi_R)$ is not invariant under the rigid rotation $\phi \rightarrow \phi+\phi_0$, as the metric (\ref{eq:237}) depends explicitly on $\phi$. This is in exact agreement with the boundary CFT result in table \ref{table:3}, where we have $\mathcal{R}(Y_0)\ket{Y}\neq 0$.

The bulk results are summarized in table \ref{table:2}, where all quantities are understood to be evaluated at $\rho=0$. On the columns we write the action of the Lie derivative $\mathcal{L}_{\chi_p}(\,\cdot\,)$, which is obtained by expanding the finite action in (\ref{eq:226}) to first order. Although the metric $g_{\mu \nu}(\chi_D)$ at $\rho=0$ in (\ref{eq:235}) is the same as pure AdS (so that the first and third columns in table \ref{table:2} are exactly the same), we have included it explicitly for comparison with the CFT results in table \ref{table:3}, where $\ket{0}\neq \ket{g}$. We add a box on the entries in table \ref{table:2} that do not match with the CFT results.

\begin{table}[t]
\setlength{\tabcolsep}{4.6pt} 
\centering
\begin{tabular}{ Sl || Sc | Sc | Sc || Sc | Sc | Sc  }
\specialrule{.13em}{0em}{0em}
 &
$\mathcal{L}_{\chi_T(f_0)}(\,\cdot\,)$ &
$\mathcal{L}_{\chi_R(Y_0)}(\,\cdot\,)$ &
$\mathcal{L}_{\chi_D(g_0)}(\,\cdot\,)$ &
$\mathcal{L}_{\chi_T(f)}(\,\cdot\,)$ &
$\mathcal{L}_{\chi_R(Y)}(\,\cdot\,)$ &
$\mathcal{L}_{\chi_D(g)}(\,\cdot\,)$ 
\\
\specialrule{.05em}{0em}{0em}
$\,\,\,\,\,g_{\mu\nu}^{\rm AdS}$  &
\color{ForestGreen}{$0$} &
\color{ForestGreen}{$0$} &
\color{ForestGreen}{$0$} &
\color{ForestGreen}{$0$} &
\color{Red}{$\neq0$} &
$\colorboxed{black}{\color{ForestGreen}{0}}$ 
\\
$g_{\mu\nu}(\chi_R)$ &
\color{ForestGreen}{$0$} &
\color{Red}{$\neq 0$} &
\color{ForestGreen}{$0$} &
\color{ForestGreen}{$0$} &
\color{Red}{$\neq 0$} &
$\colorboxed{black}{\color{ForestGreen}{0}}$
\\
$g_{\mu\nu}(\chi_D)$ &
\color{ForestGreen}{$0$} &
$\colorboxed{black}{\color{ForestGreen}{0}}$ &
\color{ForestGreen}{$0$} &
\color{ForestGreen}{$0$} &
\color{Red}{$\neq 0$} &
$\colorboxed{black}{\color{ForestGreen}{0}}$
\\
\specialrule{.13em}{0em}{0em}
\end{tabular}
\caption{Action of the vectors $\chi_p$ with $p=T,R,D$ on the bulk metric, where all the quantities in this table are evaluated at $\rho=0$. We add a box on the entries of this table that do not agree with the boundary CFT results, given in the first six columns of table \ref{table:3}. Note that if we restrict to the ordinary supertranslations and superrotations BMS transformations, the bulk and boundary results match exactly.}\label{table:2}
\end{table}

\subsubsection{Gravitational Noether charges}

We now compute the gravitational Noether charges associated to the bulk vectors $\chi_p$ (\ref{eq:151}) and use the identifications given in table \ref{table:7} to compare with the boundary CFT results in the last three columns in table \ref{table:3}. To do so, we must fix a particular gravitational theory, that for simplicity we take as Einstein gravity
\begin{equation}
I[g_{\mu \nu}]=\frac{1}{16\pi G}
\int_{\mathcal{M}} d^4x\sqrt{-g}\left(R+6\right)+
\frac{1}{8\pi G}
\int_{\partial \mathcal{M}}
d^{3}y\sqrt{-h}K\ ,
\end{equation}
where we have included the appropriate boundary term, written in terms of the extrinsic curvature~$K$. 

As a first step we need to compute the gravitational charges associated to the vectors $\chi_p$ on the pure ${\rm AdS}_4$ space-time (\ref{eq:224n}). An effective method for doing so that unambiguously fixes its value (without the need of any vacuum substraction) is obtained from the Brown-York quasi-local stress tensor \cite{Brown:1992br} regularized regulated using the counter-term method \cite{Balasubramanian:1999re,Emparan:1999pm}. The quasi-local stress tensor is defined as
\begin{equation}\label{eq:236}
T^{\alpha \beta}_{{\rm quasi-local}}
=
\frac{-2}{\sqrt{-h}}
\frac{\delta I_{\rm on-shell}}{\delta h_{\alpha \beta}}\ ,
\end{equation}
where $I_{\rm on-shell}$ is the regularized on-shell action and $h_{\alpha \beta}$ is the induced metric  on the surface $\psi=\psi_0$, that we ultimately remove by taking $\psi_0\rightarrow 0$. After computing this quantity for a bulk metric $g_{\mu \nu}$, the Noether charge $Q_{g_{\mu \nu}}[\chi_p]$ associated to the vector $\chi_p$ is computed by appropriately contracting with $\chi_p$ and integrating (see \cite{Balasubramanian:1999re} for details). For the case of the Poincar\'e patch of pure ${\rm AdS}_4$ the tensor (\ref{eq:236}) was computed in section 4 of \cite{Balasubramanian:1999re}, where it was shown to vanish. This means the charges on the pure AdS background are given by
\begin{equation}\label{eq:240}
T^{\alpha \beta}_{{\rm quasi-local}}
[g_{\mu \nu}^{\rm AdS}]=0
\qquad \quad\Longrightarrow \quad\qquad
Q_{g_{\mu \nu}^{\rm AdS}}[\chi_p]=0\ ,
\end{equation}
in agreement with the CFT result given in the first row and last three columns in table \ref{table:3}.

A less trivial calculation corresponds to computing the gravitational charges of the vectors $\chi_q$ but on the deformed metric $g_{\mu \nu}(\chi_p)$ (\ref{eq:234}) instead. Since working with the full metric $g_{\mu \nu}(\chi_p)$ is very complicated, we consider its leading order behavior in $\chi_p$, given by
\begin{equation}\label{eq:238}
g_{\mu \nu}(\chi_p)=
g_{\mu \nu}^{\rm AdS}
+\mathcal{L}_{\chi_p}(g_{\mu \nu}^{\rm AdS})
+\mathcal{O}(\chi_p)^2=
g_{\mu \nu}^{\rm AdS}
+\delta_{\chi_p}g_{\mu \nu}
+\mathcal{O}(\chi_p)^2\ .
\end{equation}
We can then calculate the Noether charge to leading order in $\chi_p$
\begin{equation}\label{eq:239}
Q_{g_{\mu \nu}^{\rm AdS}+\delta_{\chi_p}g_{\mu \nu}}[\chi_q]\ ,
\end{equation}
where the subscript in $Q$ indicates the background metric in which the charge is computed. To do so, it is not convenient to use the Brown-York tensor (\ref{eq:236}) but instead the more powerful covariant formalism (see appendix \ref{zapp:6}). As a first step we need to write the first order metric variation in (\ref{eq:238}). Using the full expression for the vectors $\chi_p$ in (\ref{eq:151}) and the AdS metric in (\ref{eq:224n}), we find
\begin{eqnarray}\label{eq:241}
\begin{aligned}
\delta_{\chi_T} ds^2&=-2\rho^2 \frac{\cot \psi}{\sin \psi}\beta'(\phi) du d\phi+\rho^2 \frac{2}{\sin \psi}\beta(\phi) du d\psi-\frac{2}{\sin \psi}\beta(\phi)  d\rho d\psi-2\rho \frac{\beta'(\phi)}{\sin \psi}d\phi d\psi+\dots\\
\delta_{\chi_R} ds^2&=-2 \rho^2 \cot^2\psi\, \gamma(\phi)  du^2-2u \rho^2\cot^2\psi\, \gamma'(\phi) du d\phi-2\rho \cot \psi \,[\gamma(\phi)+Y'(\phi)]du d\psi\\
&\quad \quad \quad - 2u \cot \psi \,[\gamma(\phi)+Y'(\phi)] d\rho d\psi+2\cot \psi \,Y''(\phi) d\phi d\psi-2 Y'(\phi) d\psi^2+\dots\\
\delta_{\chi_D} ds^2&=-\rho^2\frac{2}{\sin^2 \psi} g''(\phi) du^2-u \rho\frac{4}{\sin^2 \psi} g''(\phi)du d\rho- 2\rho (\cot^2\psi+\frac{1}{\sin^2 \psi})g'(\phi) du d\phi\\
& \quad  \quad \quad +2u g'(\phi) d\rho d\phi-2 u \rho \cot^2\psi g''(\phi)d\phi^2+\dots\\
\end{aligned}
\end{eqnarray}
where we have defined $\beta(\phi)=f(\phi)+f''(\phi)$ and $\gamma(\phi)=Y'(\phi)+Y'''(\phi)$. 

Using this we can compute the charges in (\ref{eq:239}) for the different combinations of $p,q=T,R,D$ using the covariant formalism. Although this formalism only gives the variation of the charge between two metrics, since the pure AdS metric has vanishing charge (\ref{eq:240}), we are actually computing (\ref{eq:239}) directly. An important point is that to compare with the boundary CFT results we need to evaluate the charges both at the boundary $\psi\rightarrow 0$ and the surface $\rho\rightarrow 0$. This second limit is required because in previous sections we have only studied the boundary states in (\ref{eq:233}) on the surface $\rho=0$.

The final results for the charges are given in table \ref{table:8}, where the first row corresponding to pure AdS comes from the Brown-York stress tensor in (\ref{eq:240}). We find that all the charges vanish except for some cases involving the superdilation vector $\chi_D(g)$, in which we get a divergence when taking the boundary limit $\psi \to 0$. Comparing with the boundary CFT results given in the last three columns in table \ref{table:3} we find the bulk and boundary computations match perfectly for quantities involving the ordinary BMS transformations $\chi_T(f)$ and $\chi_R(Y)$, but disagree in some instances when $\chi_D(g)$ is involved.

\begin{table}[t]
\setlength{\tabcolsep}{4.6pt} 
\centering
\begin{tabular}{ Sl || Sc | Sc | Sc  }
\specialrule{.13em}{0em}{0em}
 &
$Q_{(\,\cdot\,)}[\chi_T(f)]$ &
$Q_{(\,\cdot\,)}[\chi_R(Y)]$ &
$Q_{(\,\cdot\,)}[\chi_D(g)]$ 
\\
\specialrule{.05em}{0em}{0em}
$\qquad g_{\mu\nu}^{\rm AdS}$  &
\color{ForestGreen}{$0$} &
\color{ForestGreen}{$0$} &
\color{ForestGreen}{$0$} 
\\
$g_{\mu \nu}^{\rm AdS}
+\delta_{\chi_T}g_{\mu \nu}$ &
\color{ForestGreen}{$0$} &
\color{ForestGreen}{$0$} &
\color{ForestGreen}{$0$} 
\\
$g_{\mu \nu}^{\rm AdS}
+\delta_{\chi_R}g_{\mu \nu}$ &
\color{ForestGreen}{$0$} &
\color{ForestGreen}{$0$} & 
$ \colorboxed{black}{\color{red} \infty} $
\\
$g_{\mu \nu}^{\rm AdS}
+\delta_{\chi_D}g_{\mu \nu}$ &
\color{ForestGreen}{$0$} &
$ \colorboxed{black}{\color{red} \infty} $ & 
 \color{ForestGreen}{$0$}
\\
\specialrule{.13em}{0em}{0em}
\end{tabular}
\caption{Gravitational Noether charges of the vectors $\chi_p$ with $p=T,R,D$ in (\ref{eq:151}) computed in a pure AdS background and the perturbed metrics in (\ref{eq:241}). The divergences in some of the charges arises when taking the boundary limit $\psi\rightarrow 0$. Comparing with the boundary CFT results, given in the last three columns of table \ref{table:3}, we find perfect agreement when considering quantities that do not involve the superdilation vector $\chi_D(g)$. We add a box on the entries where there is disagreement.}
\label{table:8}
\end{table}

\subsubsection{Conclusion from holographic analysis}

The result of the bulk computations summarized in tables \ref{table:2} and \ref{table:8} match exactly with the boundary CFT results in table \ref{table:3} when considering quantities involving the standard BMS supertranslations $\chi_T(f)$ and superrotations $\chi_R(Y)$. This provides strong evidence in favor of our holographic proposal for the description of the boundary states $\ket{f}$ and $\ket{Y}$ in (\ref{eq:233}).

This is not the case for the superdilation vector $\chi_D(g)$ where we find several discrepancies, indicated in the entries of tables \ref{table:2} and~\ref{table:8} with a box. One might wonder whether there is a way of redefining the superdilation vector $\chi _D(g)$ such that it reproduces the boundary result and also satisfies the properties discusses at the beginning of this subsection. However, we have not found any consistent way of doing so. For this reason, we do not interpret the discrepancies as a failure of the holographic prescription, but rather as evidence that the boundary superdilation transformations generated by $\xi_D(g)$ is not well behaved. This is supported by the observation that the fall-off condition preserved by $\xi_D(g)$ at the boundary metric at $\rho=0$ are more singular than those satisfied by ordinary BMS transformations (see tables \ref{table:6} and \ref{table:5}). The divergence of the Noether charge is usually a sign that the fall-off condition preserved by the associated transformation is too permissible. Conversely, one might ask whether there is a way of understanding the impossibility of extending BMS in a way of including superdilations. It would be interesting to understand how the singular behavior of superdilations is captured by the boundary state $\ket{g}=e^{-i\mathcal{D}(g)}\ket{0}$.


\subsection{Arbitrary dimensions}

The above construction can be generalized to arbitrary dimensions, where the ${\rm AdS}_{d+1}$ metric is given by (\ref{eq:158}). The isometries can be once again obtained from (\ref{eq:149}), so the resulting vectors $\chi_p$ that generalize (\ref{eq:151}) are compactly written as
\begin{equation}\label{eq:230}
\begin{aligned}
\chi_T(f)&=
\cos(\psi)\left[
f\partial_u
-\frac{\rho^2}{d-2}
(D^2f)
\partial_\rho
\right]
-\frac{\rho}{\cos(\psi)}
(D^Af)\partial_A
-
\frac{\rho \sin(\psi)}{d-2}
(D^2f)\partial_\psi\ , \\
\chi_R(Y)&=
\cos^2(\psi)
\frac{(D\cdot Y)}{d-2}
u\partial_u+
\frac{\rho\cos^2(\psi)}{d-2}
\left[
(D\cdot Y)
-\frac{\rho u}{d-2}
D^2(D\cdot Y)
\right]\partial_\rho
+\\
&\hspace{15mm}+\left[
Y^A-
\frac{\rho u}{d-2}
D^A(D\cdot Y)
\right]\partial_A+
\frac{\sin(2\psi)}{2(d-2)}
\left[
(D\cdot Y)
-\frac{\rho u}{d-2}
D^2(D\cdot Y)
\right]\partial_\psi\ ,\\
\chi_S(h)&=
\cos(\psi)h\,u^2\partial_u
-
\cos(\psi)\left[
2(1+\rho u)h+
\frac{(2+\rho u)^2}{d-2}
(D^2 h)
\right]\partial_\rho
+\\
&\hspace{15mm}-
\frac{u(2+\rho u)}{\cos(\psi)}
(D^A h)\partial_A
-\sin(\psi)
\frac{u(2+\rho u)}{d-2}
(D^2 h)\partial_\psi
\ ,\\
\chi_D(g)&=
g \,u\partial_u
-\rho
\left[g+\frac{\rho u}{d-2}(D^2 g)\right]\partial_\rho
-\rho u (D^A g)\partial_A
\ ,
\end{aligned}
\end{equation}
where $D_A$ is the covariant derivative on the unit sphere, $S^{d-2}$. These vectors exactly solve the Killing equation in ${\rm AdS}_{d+1}$ when the functions are given by
\begin{equation}\label{eq:231}
\begin{aligned}
f_0(\vec{y}\,)&=\frac{a_0}{\cos(\psi)}+
\sum_{B=1}^{d-2} a_B
\left(
\frac{y^B}{|\vec{y}\,|^2+1}
\right)+
a_{d-1}
\left(
\frac{|\vec{y}\,|^2-1}
{|\vec{y}\,|^2+1}
\right)\ ,\\
Y^A_0(\vec{y}\,)&=
\frac{b_0y^A}{\cos(\psi)}
+\sum_{B=1}^{d-2}
p_B
\frac{
2y^By^A-\delta^{AB}(|\vec{y}\,|^2+1)
}{\cos(\psi)}
+
\sum_{B=1}^{d-2}
\Big\lbrace
w^A_{\,\,\,\,B}y^B+
\tilde{p}_B\left[
2y^By^A-\delta^{AB}(|\vec{y}\,|^2-1)
\right]
\Big\rbrace
\ ,\\
h_0(\vec{y}\,)&=\frac{c_0}{\cos(\psi)}+
\sum_{B=1}^{d-2} c_B
\left(
\frac{y^B}{|\vec{y}\,|^2+1}
\right)+
c_{d-1}
\left(
\frac{|\vec{y}\,|^2-1}
{|\vec{y}\,|^2+1}
\right)\ ,\\
g_0(\vec{y}\,)&=d_0\ ,
\end{aligned}
\end{equation}
where $w^A_{\,\,\,B}=-w^B_{\,\,\,A}$ is an antisymmetric matrix. As we take the limit $\psi \rightarrow 0$, both the vectors (\ref{eq:230}) and functions (\ref{eq:231}) reduce to the boundary quantities in (\ref{eq:178}) and (\ref{eq:169}) that generate conformal transformations on the boundary. Moreover, when evaluating the vectors at the bulk Poincar\'e horizon $\rho=0$ we find
\begin{equation}
\begin{aligned}
\chi_T(f)\big|_{\rho=0}&=
\cos(\psi)
f(\vec{y}\,)\partial_u\ , \\
\chi_R(Y)\big|_{\rho=0}&=
\cos^2(\psi)
\frac{(D\cdot Y)}{d-2}
u\partial_u
+
Y^A\partial_A+
\frac{\sin(2\psi)}{2(d-2)}
(D\cdot Y)\partial_\psi\ ,\\
\chi_S(h)\big|_{\rho=0}&=
\cos(\psi)h(\vec{y}\,)u^2\partial_u-
\frac{2u(D^A h)}{\cos(\psi)}
\partial_A
-
2\cos(\psi)\left[
h(\vec{y}\,)+
\frac{2}{d-2}
(D^2 h)
\right]\partial_\rho
-\frac{2u\sin(\psi)}{d-2}
(D^2 h)\partial_\psi
\ ,\\
\chi_D(g)\big|_{\rho=0}&=
g(\vec{y}\,)u\partial_u\ .
\end{aligned}
\end{equation}
Computing the algebra associated to these vectors (without the vector $\chi_S(h)$) we find it is exactly the same as the boundary algebra, i.e. it becomes (\ref{eq:189}) after replacing $\xi\rightarrow \chi$. Following the same criteria as in the three dimensional case, the boundary vectors $\xi_p$ are extended inside the bulk by $\chi_p$ in (\ref{eq:231}). 

The holographic description of the boundary states $\left\lbrace \ket{f},\ket{Y},\ket{g} \right\rbrace$ parallels the one described for the AdS$_4$ case, summarized in table \ref{table:7}. While the computation of the transformed bulk metric $g_{\mu \nu}(\chi_p)$ in (\ref{eq:234}) and the Noether charges is more involved, we expect analogous results as those given in tables \ref{table:2} and \ref{table:8}.

\section{Discussion: BMS and black holes}
\label{sec:5}

Before concluding, we would like to make some comments about the possible applications that our results could have to investigate black hole physics. In fact, our main motivation for studying the action induced by the asymptotic (conformal) Killing vectors on the Hilbert space of a CFT is to get further insight on quantum aspects of black holes. The study of BMS symmetries in relation to black holes goes back to the proposal made in \cite{Hawking:2016msc}, and it is natural to ask whether the particular realization of the infinite-dimensional symmetry we studied here could have something to do with it. The connection arises for (near-)extremal black holes, that has a near horizon limit that corresponds to ${\rm AdS}_2\times S^{d-2}$. 

For concreteness, let us focus on the case of asymptotically flat electrically charged black holes in four dimensional Einstein-Maxwell theory
$$I[g_{\mu \nu},A_\mu]=\frac{1}{16\pi G}
  \int d^4x\sqrt{-g}\,
  \mathcal{R}-
  \frac{1}{4}
  \int d^4x\sqrt{-g}\,
  F_{\mu \nu}F^{\mu \nu}
  \ ,$$
whose metric is given by the Reissner-Nordst\"om geometry
\begin{equation}\label{eq:132}
ds^2=-f(r)dt^2+
  \frac{dr^2}{f(r)}+r^2d\Omega_2^2\ ,
  \qquad \qquad
  f(r)=
  \frac{(r-r_+)(r-r_-)}{r^2}\ ,
\end{equation}
where $r_\pm$ are the inner and outer horizons, which can be written in terms of the mass and charge of the black hole. The extremal and near-extremal black holes correspond to $r_+=r_-$ and $r_+\simeq r_-$ respectively. If we study the near horizon limit $r\sim r_+$, in both cases we find it is given by ${\rm AdS}_2\times S^{d-2}$. However, since the bifurcation surface in the extremal and near-extremal case ends up in a different place of the ${\rm AdS}_2$ factor, we must analyze each case separately. A detailed analysis of the near horizon limit of this black hole can be found in \cite{Spradlin:1999bn}.

\begin{figure}
\centering
\includegraphics[scale=0.30]{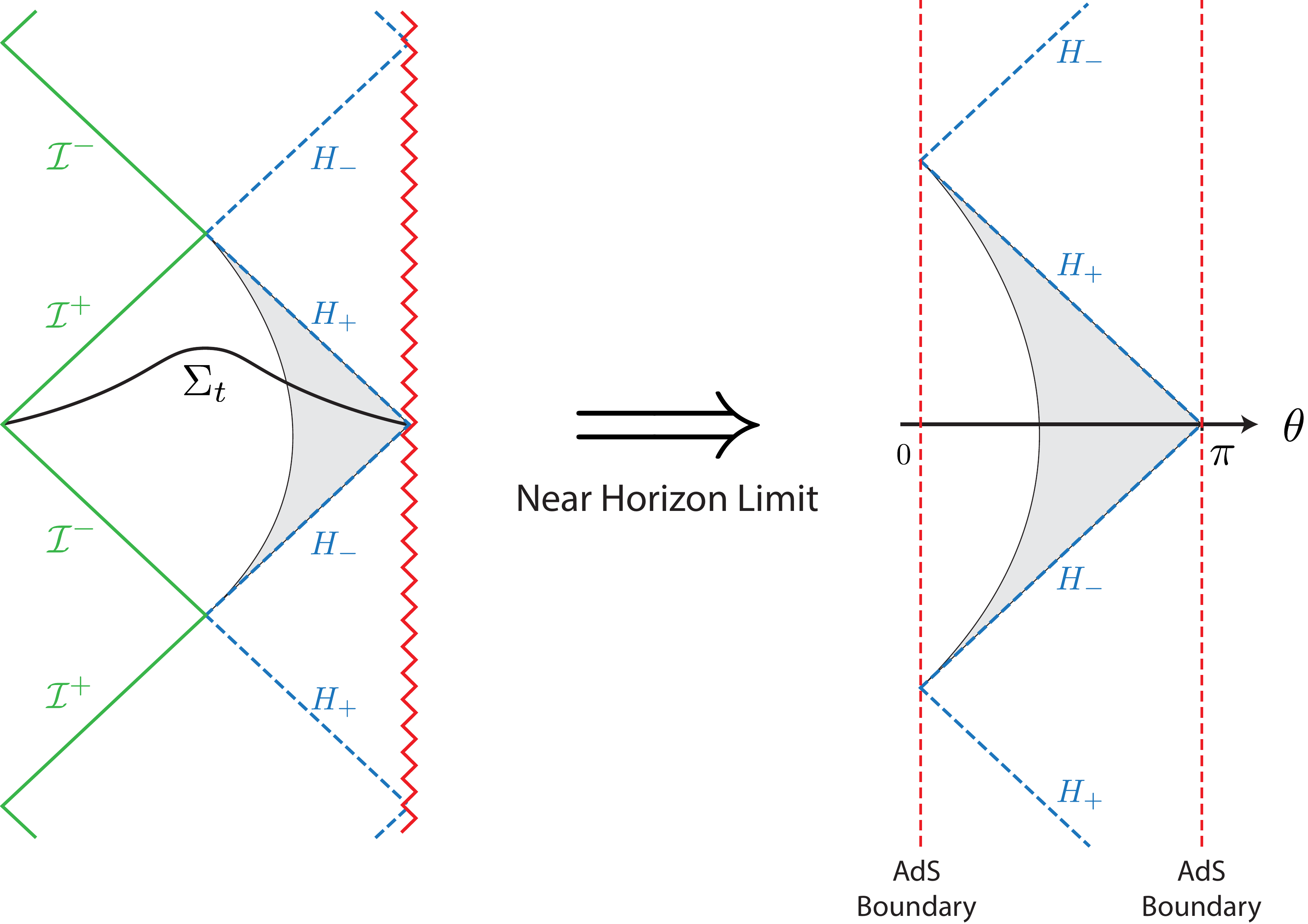}
\caption{On the left we see the maximally extended Penrose diagram of the extremal Reissner-Nordstrom black hole. Taking the near horizon limit we obtain the ${\rm AdS}_2\times S^{d-2}$ space-time on the right, see \cite{Spradlin:1999bn}. The gray shaded region corresponds to the section of the black hole that is well approximated by ${\rm AdS}_2\times S^{d-2}$.}\label{fig:2}
\end{figure}

Let us start by considering the extremal black hole $r_+=r_-$ that has a maximally extended Penrose diagram shown in the left diagram of figure \ref{fig:2}. The diamond region corresponds to the exterior of the black hole where the initial value problem can be defined. We sketch a Cauchy surface $\Sigma_t$ given by a constant $t$ surface on the black hole space-time (\ref{eq:132}) with $r_+=r_-$. There are four boundaries, the green lines on the left corresponding to the future and past null infinity $\mathcal{I}^\pm$, and the blue dashed lines to the future and past black hole horizons $H_\pm$. Taking the near horizon limit of the extremal solution (\ref{eq:132}) we get an ${\rm AdS}_2\times S^{d-2}$ metric, where the black hole horizons $H_\pm$ correspond to the surfaces $\theta_\pm=\pi$ in the global coordinates given in (\ref{eq:78}). As shown in the right diagram of figure \ref{fig:2}, the near horizon limit is equivalent to slicing the Penrose diagram of the black hole by inserting the two ${\rm AdS}_2$ boundaries, one on the singularity would be and the other along the surface where $\mathcal{I}^\pm$ and $H_\pm$ meet. 

Let us now consider an arbitrary CFT defined on the exterior region of the extremally charged black hole. For any Cauchy surface $\Sigma_t$ we have an associated Hilbert space $\mathcal{H}_t$. We can move through different foliations, i.e. different values of $t$, by acting on the $t$-translation operator as ${e^{-\Delta t\partial_t}:\, \Sigma_{t_1}\to \Sigma_{t_2}}$, and so $e^{-i\Delta tQ[\partial_t]}:\, \mathcal{H}_{t_1}\to \mathcal{H}_{t_2}$, where $Q[\partial_t]$ is the conserved charge (\ref{eq:182}) written in terms of the stress tensor operator of the CFT. Since $t$-translation is an isometry of the black hole metric~(\ref{eq:132}), the vacuum of the theory, which we denote $\ket{\Omega}$, is invariant under such a transformation, namely $ e^{-iQ[\partial_t]}\ket{\Omega}=e^{-iE_0}\ket{\Omega}\sim \ket{\Omega}$, where $E_0$ is the energy of the ground state. There are two Cauchy surfaces that are of particular interest to us. These are\footnote{The factorization of the space on the Cauchy surface has been recently discussed in \cite{Laddha:2020kvp, Haco:2018ske}.} $\Sigma_{\rm initial}\equiv \Sigma_{t\to -\infty}=\mathcal{I}^-\cup H_-$ and $\Sigma_{\rm final}\equiv \Sigma_{t\to +\infty}=\mathcal{I}^+\cup H_+$, cf. \cite{Hawking:2016msc}. On these Cauchy surfaces we expect the vacuum state $\ket{\Omega}$ to coincide with the vacuum of the CFT defined in Minkowski and ${\rm AdS}_2\times S^{d-2}$, so that we can act with the asymptotic charges $\mathcal{R}(Y)$ and $\mathcal{D}(g)$ to generate the zero energy eigenstates $\ket{Y}$ and $\ket{g}$ with respect to the Hamiltonians that generates the time evolution in the $u$ and $v$ coordinates. 

\begin{figure}
\centering
\includegraphics[scale=0.35]{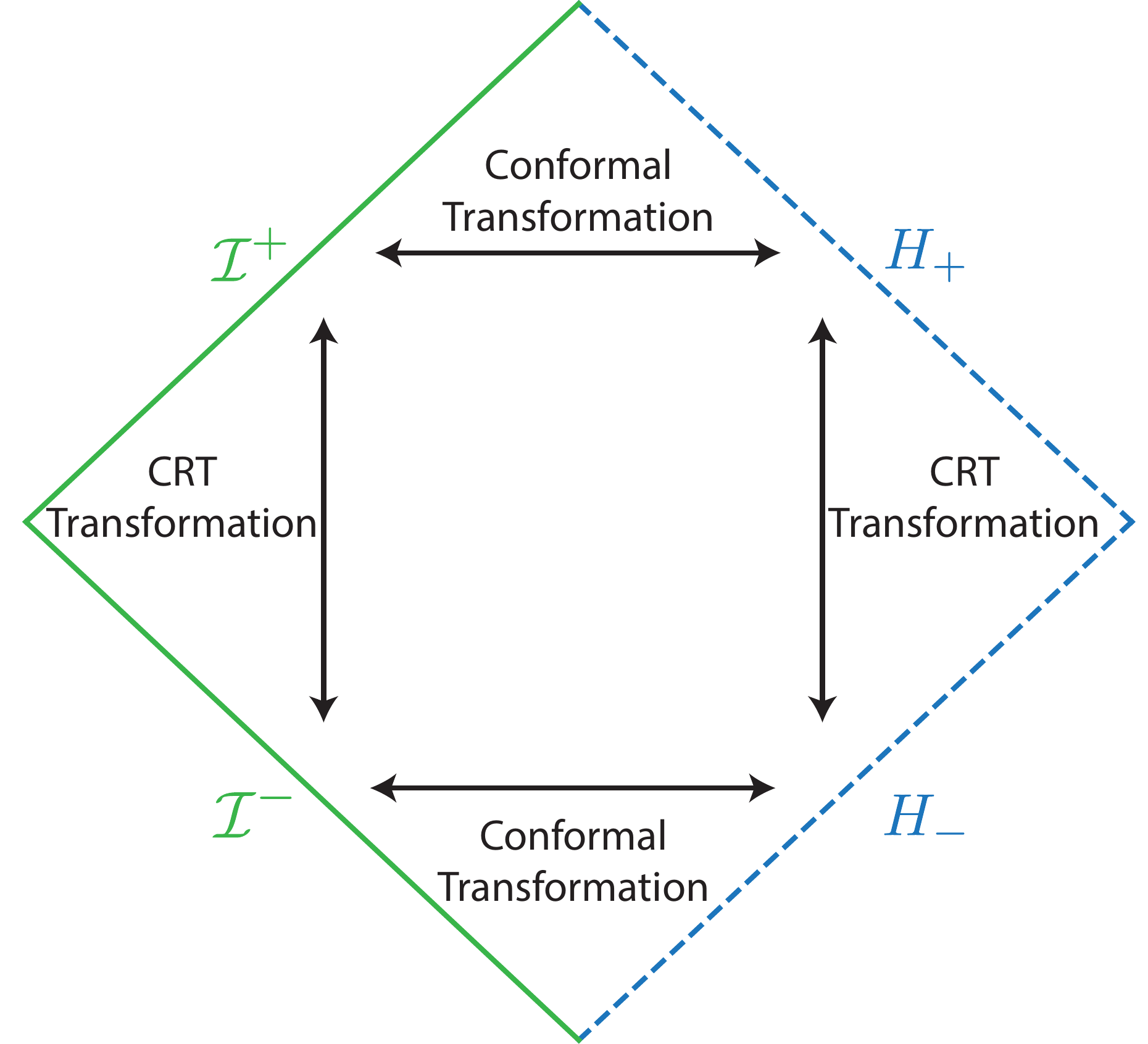}
\caption{Diagram showing the transformations that map between the four boundaries of the extremal black hole (\ref{eq:132}): $\mathcal{I}^\pm$ and $H_\pm$. The mapping of the charges under the conformal and CRT transformations where explored on the first two subsections of section \ref{sec:2}.}\label{fig:6}
\end{figure}

Interestingly, the states on the surfaces $\mathcal{I}^\pm$ and $H_\pm$ are not independent but related by conformal symmetry (mapping the Minkowski surface $\mathcal{I}^\pm$ to the Poincar\'e horizons $H_\pm$ in ${\rm AdS}_2\times S^{d-2}$) and CRT symmetry (mapping between the asymptotic regions $\mathcal{I}^\pm$ themselves). The transformations relating the different surfaces on the exterior of the black hole are summarized in figure \ref{fig:6}. Using these transformations we can relate the zero energy states $\ket{Y}$ and $\ket{g}$ on the different surfaces (see (\ref{eq:242})). Note that when applying the CRT transformation the functions $Y^A(\vec{y}\,)$ and $g(\vec{y}\,)$ are not invariant but transform in the way specified in (\ref{eq:221}).

One might still worry about the transformations in figure \ref{fig:6} not being well behaved as they only involve the near horizon and asymptotic regions, i.e. they are not true transformations of the whole black hole space-time. However, for this particular type of black hole, we can actually go even further and apply a conformal transformation on the \textit{full metric} (\ref{eq:132}) that maps the asymptotic regions $\mathcal{I}^\pm$ to the horizons $H_\pm$. This is obtained by considering \cite{conf}
\begin{equation}
\bar{r}=\frac{r}{r/r_+-1}\, ,
\end{equation}
which yields
\begin{equation}
ds^2=
\frac{1}{(\bar{r}/r_+-1)^2}
\left[
-f(\bar{r})dt^2+\frac{d\bar{r}^2}{f(\bar{r})}+
\bar{r}^2d\Omega_2^2
\right]\ ,
\end{equation}
where we have to remember that we are considering the extremal case $r_+=r_-$; see also \cite{TheNew}. Applying a Weyl transformation which removes the conformal factor, we recover the four dimensional black hole, with the difference that the horizon $r=r_+$ has been mapped to the asymptotic region $\bar{r} \rightarrow +\infty$ and viceversa. This transformation is very special to the extremal four dimensional Reissner-Nordstr\"om solution, and it does not generalize to higher-dimensional or non-extremal black holes, at least not in an obvious manner. Therefore, at least in the case $d=4$, it is natural to ask whether this conformal symmetry together with the symmetries studied in this paper can be used to construct the zero energy states on the full black hole geometry. 

\begin{figure}
\centering
\includegraphics[scale=0.30]{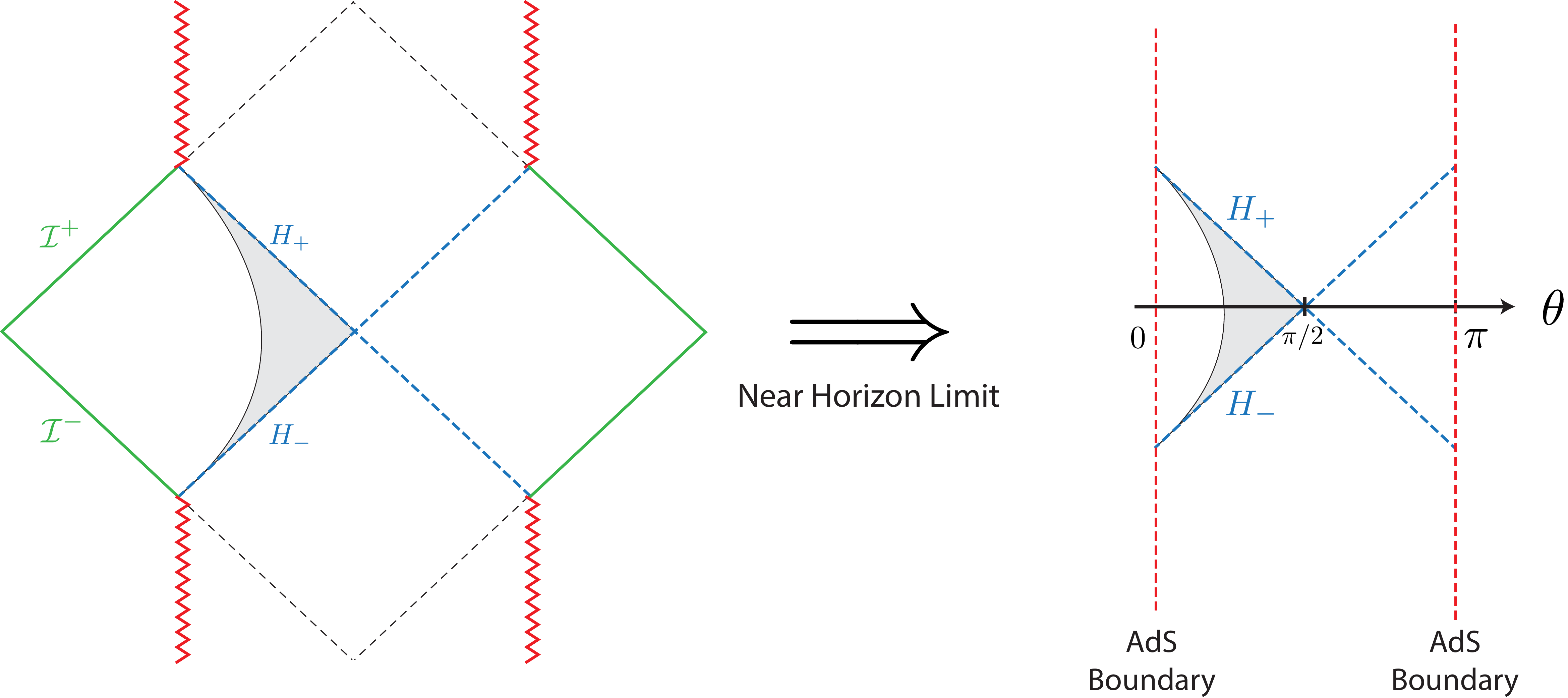}
\caption{On the left we see the maximally extended Penrose diagram of the near-extremal Reissner-Nordstrom black hole. Notice that in this case the future (past) horizon $H_+$ ($H_-$) is actually the past (future) horizon for the asymptotic region on the right. Taking the near horizon limit we obtain the ${\rm AdS}_2\times S^{d-2}$ space-time on the right, see \cite{Spradlin:1999bn}. The gray shaded region corresponds to the section of the black hole that is well approximated by ${\rm AdS}_2\times S^{d-2}$.}\label{fig:4}
\end{figure}

A similar discussion applies to near-extremal black holes with $r_+\simeq r_-$, that are more interesting as they have finite temperature $T_+$. The Penrose diagram of the black hole solution is given on the left diagram in figure \ref{fig:4}, where we see that in this case there are two asymptotically flat regions that are causally disconnected from each other. The exterior black hole space-time described by the metric (\ref{eq:132}) is shaded on the left side.

The near horizon limit $r\sim r_+$ in this case also yields an ${\rm AdS}_2\times S^{2}$ metric but only after we apply a change of coordinates (see (2.6) in \cite{Spradlin:1999bn}). As we see on the right diagram in figure \ref{fig:4}, the diffeomorphism changes the location of the horizons $H_\pm$, as they now intersect in the interior of ${\rm AdS}_2$, at $(\sigma,\theta)=(0,\pi/2)$ in the global coordinates (\ref{eq:78}). The exterior of the black hole covered by the ${\rm AdS}_2$ does not correspond to the Poincar\'e patch as in the extremal case, but to the smaller region $\theta_\pm<\pi/2$.

The position for the horizons in ${\rm AdS}_2\times S^2$ is obtained by taking $\theta_0=\pi/2$ when defining the coordinates $(u,\rho)$ and $(v,\varrho)$ in (\ref{eq:116}) and (\ref{eq:125}) respectively. The horizons $H_\pm$ appearing in the right diagram of figure \ref{fig:4} correspond to $\rho=0$ and $\varrho=0$ in each of these coordinates. Note that these coordinates cover a larger region than the one corresponding to the exterior of the black hole $\theta_\pm<\pi/2$.

Similarly to the extremal case, we can define the BMS transformations on $H_\pm$ and $\mathcal{I}^\pm$ and a family of zero energy states by acting with $\mathcal{R}(Y)$ and $\mathcal{D}(g)$ on the vacuum. The states defined in each of these null surfaces are related between themselves in a similar way as described for the extremal case in figure \ref{fig:6}.

\bigskip
\smallskip
\leftline{\bf Acknowledgements}
\smallskip
\noindent 
F.R. thanks David Simmons-Duffin for correspondence and Alan Garbarz for encouragement. F.R. is supported by the DOE grant DE-SC0011687, the Heising-Simons Foundation and NSF Grant No. PHY-1748958. L.D. is supported by the European Union's Horizon 2020 research and innovation programme under the Marie Sklodowska-Curie grant agreement No. 746297. G.G. thanks Hern\'an Gonz\'alez for interesting comments; he also thanks the support of CONICET through the grant PIP 1109-2017.

\appendix

\section{Light-ray operator algebra}
\label{zapp:1}

In this appendix we show the charges associated to the asymptotic (conformal) Killing vectors at $\rho=0$ in (\ref{eq:180}) satisfy the algebra in (\ref{eq:190}) for the CFT in ${\rm AdS}_2\times S^{d-2}$. We do this using some results of \cite{Cordova:2018ygx} involving the algebra of light-ray operators in the null plane together with a conformal transformation previously studied in \cite{Rosso:2019txh}.

Let us start by considering the $d$-dimensional Minkowski metric written in Cartesian coordinates $ds^2=-dt^2+dx_1^2+d\vec{x} \cdot d\vec{x}$. We can parametrize the null plane as
\begin{equation}\label{eq:25}
x^\mu(\lambda,\vec{x}_\perp)=
  (\lambda,\lambda,\vec{x}_\perp)\ ,
  \qquad \quad
  (\lambda,\vec{x}_\perp)
  \in \mathbb{R}\times \mathbb{R}^{d-2}\ ,
\end{equation}
where $(\lambda,\vec{x}_\perp)$ are parametrization coordinates, with $\lambda$ affine. The light-ray operators on the null plane analogous to those defined at Minkowski null infinity are given by
\begin{equation}\label{eq:30anterior}
\widehat{\mathcal{E}}(\vec{x}_\perp)\equiv
\int_{-\infty}^{+\infty}d\lambda\, 
  \widehat{T}_{\lambda \lambda}(\lambda,\vec{x}_\perp)\ ,\qquad \qquad
\widehat{\mathcal{K}}(\vec{x}_\perp)\equiv
  \int_{-\infty}^{+\infty}d\lambda\,
  \lambda\,
  \widehat{T}_{\lambda \lambda}(\lambda,\vec{x}_\perp)
  \ ,
\end{equation}
and
\begin{equation}\label{eq:30}
\widehat{\mathcal{N}}_A(\vec{x}_\perp)\equiv
\int_{-\infty}^{+\infty}d\lambda\,
\widehat{T}_{\lambda A}(\lambda,\vec{x}_\perp)=
\int_{-\infty}^{+\infty}d\lambda\,
\frac{dx^\mu }{d\lambda}
\frac{dx^\nu}{dx_\perp^A}
\widehat{T}_{\mu \nu}(\lambda,\vec{x}_\perp)\ ,
\end{equation}
where $A$ is the index in $\vec{x}_\perp\in \mathbb{R}^{d-2}$ and we have added the hat to differentiate from the light-ray operators defined previously in (\ref{eq:184}) and (\ref{eq:191}). In \cite{Cordova:2018ygx} it was shown that under some general assumptions these operators satisfy the following algebra\footnote{The last commutator is written in a different way from the one given in \cite{Cordova:2018ygx}, using the following expansion around $\vec{x}_2^{\,\perp}=\vec{x}_1^{\,\perp}-\vec{x}_{12}^{\,\perp}$
$$
\partial_B\delta(\vec{x}_{12}^{,\perp})
\widehat{\mathcal{N}}_A(\vec{x}_2^{\,\perp})=
\partial_B\delta(\vec{x}_{12}^{,\perp})\left[
\widehat{\mathcal{N}}_A(\vec{x}_1^{\,\perp})
-(x_{12}^\perp)^C
\partial_C	\widehat{\mathcal{N}}_A(\vec{x}_1^{\, \perp})+\dots
\right]=
\partial_B\delta(\vec{x}_{12}^{,\perp})
\widehat{\mathcal{N}}_A(\vec{x}_1^{\,\perp})
+
\delta(\vec{x}_{12}^{,\perp})
\partial_B\widehat{\mathcal{N}}_A(\vec{x}_1^{\, \perp})\ ,
$$
where in the second equality we have integrated by parts and used that quadratic and higher order contributions to the expansion vanish due to the action of the derivative of Dirac's delta. This equivalent way of writing the commutator has the advantage that both sides are explicitly anti-symmetric under the change $(A,\vec{x}_1^{\,\perp})\leftrightarrow (B,\vec{x}_2^{\, \perp})$.}
\begin{equation}\label{eq:1}
\begin{aligned}
\left[
\widehat{\mathcal{E}}(\vec{x}_1^{\, \perp}),
\widehat{\mathcal{E}}(\vec{x}_2^{\, \perp})
\right]&=0\ ,
\\
\left[
\widehat{\mathcal{K}}(\vec{x}_1^{\, \perp}),
\widehat{\mathcal{K}}(\vec{x}_2^{\, \perp})
\right]&=0\ , 
\\
\left[
\widehat{\mathcal{K}}(\vec{x}_1^{\, \perp}),
\widehat{\mathcal{E}}(\vec{x}_2^{\, \perp})
\right]&=
-i\delta(\vec{x}_{12}^{\, \perp})
\widehat{\mathcal{E}}(\vec{x}_2^{\, \perp})\ ,
\\ 
\left[
\widehat{\mathcal{N}}_A(\vec{x}_1^{\, \perp}),
\widehat{\mathcal{E}}(\vec{x}_2^{\, \perp})
\right]&=
-i\delta(\vec{x}_{12}^{\, \perp})
\partial_A\widehat{\mathcal{E}}
(\vec{x}_2^{\, \perp})+
i\partial_A\delta(\vec{x}_{12}^{\, \perp})
\widehat{\mathcal{E}}(\vec{x}_2^{\, \perp})\ ,
\\
\left[
\widehat{\mathcal{N}}_A(\vec{x}_1^{\, \perp}),
\widehat{\mathcal{K}}(\vec{x}_2^{\, \perp})
\right]&=
-i\delta(\vec{x}_{12}^{\, \perp})
\partial_A\widehat{\mathcal{K}}
(\vec{x}_2^{\, \perp})+
i\partial_A\delta(\vec{x}_{12}^{\, \perp})
\widehat{\mathcal{K}}(\vec{x}_{2}^{\, \perp})\ ,
\\
\left[
\widehat{\mathcal{N}}_A(\vec{x}_1^{\, \perp}),
\widehat{\mathcal{N}}_B(\vec{x}_{2}^{\, \perp})
\right]&=
-2i\delta(\vec{x}_{12}^{\, \perp})
\partial_{[A}\widehat{\mathcal{N}}_{B]}
+
i\partial_A
\delta(\vec{x}_{12}^{\, \perp})
\widehat{\mathcal{N}}_B(\vec{x}_2^{\, \perp})+
i\partial_B\delta(\vec{x}_{12}^{,\perp})
\widehat{\mathcal{N}}_A(\vec{x}_1^{\,\perp})
\ ,
\end{aligned}
\end{equation}
where $\delta(\vec{x}_{12}^\perp)\equiv\delta^{(d-2)}(\vec{x}_1^\perp-\vec{x}_2^\perp)$ and derivatives are with the respective arguments.

We now apply a conformal transformation relating the Minkowski null plane and the null surface $H_+$ (figure \ref{fig:1}) in ${\rm AdS}_2\times S^{d-2}$. This is done in two steps, let us say $(A)$ and $(B)$, which are given by 
\begin{equation}\label{eq:120}
\mathbb{R}\times \mathbb{R}^{d-1}
  \qquad 
  \xrightarrow[]{(A)} 
  \qquad
  \mathbb{R}\times S^{d-1}
  \qquad 
  \xrightarrow[]{(B)} 
  \qquad
  {\rm AdS}_2\times S^{d-2}\ .
\end{equation}
The advantage of this procedure is that the transformation $(A)$ was already worked out in section 3.1 of \cite{Rosso:2019txh},\footnote{The mapping described in section 3.1 of \cite{Rosso:2019txh} involves a special conformal transformation mapping the Minkowski null plane to the Minkowski null cone. The necessity for considering the Lorentzian cylinder comes from the fact that special conformal transformations are not globally well defined in Minkowski.} so that in terms of the global coordinates $(\theta_+,\theta_-,\vec{y}\,)$ in the Lorentzian cylinder metric (\ref{eq:192}), the null plane (\ref{eq:25}) is mapped to\footnote{These expressions are obtained from (3.17) and (3.15) of \cite{Rosso:2019txh} after making some redefinitions to match with the conventions used in this paper. The time coordinates are related according to $\sigma_{\rm here}=\sigma_{\rm there}/R+\theta_0$. The function $p(\vec{x}_\perp)$, defining the stereographic coordinates on $S^{d-2}$ in (3.8) of \cite{Rosso:2019txh}, is taken with $R=1/2$ in (\ref{eq:36}) to match with our definition in (\ref{eq:176}).}
\begin{equation}\label{eq:36}
(\theta_+,\theta_-,\vec{y}\,)=
\left(\theta_0,
2\,{\rm arctan}\left[
\frac{2p(\vec{x}_\perp)}
{\lambda+p(\vec{x}_\perp)}
\right]-\theta_0,
\vec{x}_\perp
\right)\ ,
\qquad \qquad
p(\vec{x}_\perp)\equiv
\left(
\frac{1+|\vec{x}_\perp|^2}{2}
\right)\ ,
\end{equation}
where the conformal factor $w_A(x^\mu)$ defined from $ds^2_{\rm Mink}=w^2_A(x^\mu)ds^2_{\mathbb{R}\times S^{d-1}}$ and evaluated on the null surface is
\begin{equation}
w_A(\lambda,\vec{x}_\perp)^2=
p(\vec{x}_\perp)^2
(1+\lambda^2)\ .
\end{equation}
Once we have the surface in the cylinder it is straightforward to apply the mapping $(B)$ by rewriting the metric in the cylinder as 
$$ds_{\mathbb{R}\times S^{d-1}}^2=\sin^2(\theta)
  \left[
  \frac{-d\sigma^2+d\theta^2}
  {\sin^2(\theta)}+
  d\Omega_{d-2}^2(\vec{y}\,)
  \right]\ .$$
Applying the Weyl rescaling $w^2_B(x^\mu)=\sin^2(\theta)$ we obtain the ${\rm AdS}_2\times S^{d-2}$ metric in global coordinates (\ref{eq:78}), so that the overall conformal factor resulting from both transformations in (\ref{eq:120}) is given by
\begin{equation}\label{eq:121}
w^2(\lambda,\vec{x}_\perp)=
w_A(\lambda,\vec{x}_\perp)^2
w_B(\lambda,\vec{x}_\perp)^2
=
p(\vec{x}_\perp)^2
(1+\lambda^2)
  \sin^2(\theta(\lambda))=
p(\vec{x}_\perp)^2\ .
\end{equation}
Since it is independent of $\lambda$ it implies as in (\ref{eq:185}) that $\lambda$ is also an affine parameter for the surface (\ref{eq:36}) in ${\rm AdS}_2\times S^{d-2}$, that travels between the two ${\rm AdS}_2$ boundaries along the future horizon $H_+$ at $\theta_+=\theta_0$ seen in figure \ref{fig:1}. The dependence of the conformal factor in the transverse coordinates $\vec{x}_\perp$ allows us to relate it to the determinant of the induced metric of the surface (\ref{eq:36}) on ${\rm AdS}_2\times S^{d-2}$, a unit sphere $S^{d-2}$ 
\begin{equation}\label{eq:84}
ds^2\big|_{x^\mu(\lambda,\vec{x}_\perp)}=
  d\Omega^2_{d-2}=
  \frac{4d\vec{x}_\perp \cdot d\vec{x}_\perp}
  {(1+|\vec{x}_\perp|)^2}
  \qquad \Longrightarrow \qquad
  \sqrt{h_{\rm ind}}=\frac{1}{w(\vec{x}_\perp)^{d-2}}\ .
\end{equation}

We can now map the light-ray operators in the null plane in (\ref{eq:30anterior}) and (\ref{eq:30}) using that the stress tensor transforms as (\ref{eq:195}), which gives
\begin{equation}\label{eq:28}
U
  \widehat{T}_{\lambda \lambda}
  (\lambda,\vec{x}_\perp)U^\dagger=
  \sqrt{h_{\rm ind}}
  \bar{T}_{\lambda \lambda}
  (\lambda,\vec{x}_\perp)\ ,
  \qquad \qquad
  U\widehat{T}_{\lambda A}(\lambda,\vec{x}_\perp)U^\dagger=
  \sqrt{h_{\rm ind}}
  \bar{T}_{\lambda A}(\lambda,\vec{x}_\perp)\ ,
\end{equation}
where we have used (\ref{eq:84}) and the projection in the components $(\lambda,A)$ are with respect to the surface (\ref{eq:36}). This gives a simple way of mapping the light-ray operators in (\ref{eq:30anterior}) and (\ref{eq:30}): we simply replace the hats by bars and multiply by the overall factor $\sqrt{h_{\rm ind}}$.

Applying this to the first three relations in the light-ray algebra (\ref{eq:1}) is straightforward, where the extra factor of $\sqrt{h_{\rm ind}}$ in the third relation allows us to replace the flat space Dirac delta $\delta(\vec{x}_{12}^{\, \perp})$ by the appropriate one associated to the sphere $S^{d-2}$, given by
\begin{equation}\label{eq:3}
\bar{\delta}(\vec{x}_{12}^{\,\perp})\equiv
  \frac{\delta(\vec{x}_{12}^{\,\perp})}
  {\sqrt{h_{\rm ind}}}=
  \frac{1}{\sqrt{h_{\rm ind}}}
  \delta^{(d-2)}(\vec{x}_1^\perp-\vec{x}_2^\perp)
  \ .
\end{equation}
The last three relations in (\ref{eq:1}) require additional care as derivative of the light-ray operators appear on the right-hand side. For instance, applying the adjoint action of $U$ to the first term on the fourth commutator in (\ref{eq:1}) gives
\begin{equation}
U\left(
-i\delta(\vec{x}_{12}^{\, \perp})
\partial_A\widehat{\mathcal{E}}(\vec{x}_2^{\, \perp})
\right)
U^\dagger=
-i\delta(\vec{x}_{12}^{\, \perp})
\partial_A
\left(
\sqrt{h_{\rm ind}}\bar{\mathcal{E}}(\vec{x}_2^{\, \perp})
\right)\ .
\end{equation}
Further expanding the right-hand side we get an additional term involving the derivative of $\sqrt{h_{\rm ind}}$; however, this is compensated by the second term on the fourth commutator in (\ref{eq:1}), where we must replace the flat space Dirac delta by the one associated to $S^{d-2}$ in (\ref{eq:3}). The end result is that the structure of the last three relations in (\ref{eq:1}) is preserved under the mapping, so that we obtain the following algebra for the light-ray operators in ${\rm AdS}_2\times S^{d-2}$
\begin{equation}\label{eq:58}
\begin{aligned}
\left[
\mathcal{\bar{E}}(\vec{x}_1^{\, \perp}),
\mathcal{\bar{E}}(\vec{x}_2^{\, \perp})
\right]&=0\ ,
\\
\left[
\mathcal{\bar{K}}(\vec{x}_1^{\, \perp}),
\mathcal{\bar{K}}(\vec{x}_2^{\, \perp})
\right]&=0\ , 
\\
\left[
\mathcal{\bar{K}}(\vec{x}_1^{\, \perp}),
\mathcal{\bar{E}}(\vec{x}_2^{\, \perp})
\right]&=
-i\bar{\delta}(\vec{x}_{12}^{\, \perp})
\mathcal{\bar{E}}(\vec{x}_2^{\, \perp})\ ,
\\ 
\left[
\mathcal{\bar{N}}_A(\vec{x}_1^{\, \perp}),
\mathcal{\bar{E}}(\vec{x}_2^{\, \perp})
\right]&=
-i\bar{\delta}(\vec{x}_{12}^{\, \perp})
D_A\mathcal{\bar{E}}
(\vec{x}_2^{\, \perp})+
i
\mathcal{\bar{E}}(\vec{x}_2^{\, \perp})
D_A
\bar{\delta}(\vec{x}_{12}^{\, \perp})\ ,
\\
\left[
\mathcal{\bar{N}}_A(\vec{x}_1^{\, \perp}),
\mathcal{\bar{K}}(\vec{x}_2^{\, \perp})
\right]&=
-i\bar{\delta}(\vec{x}_{12}^{\, \perp})
D_A\mathcal{\bar{K}}
(\vec{x}_2^{\, \perp})+
i
\mathcal{\bar{K}}
(\vec{x}_{2}^{\, \perp})
D_A\bar{\delta}
(\vec{x}_{12}^{\, \perp})\ ,
\\
\left[
\mathcal{\bar{N}}_A(\vec{x}_1^{\, \perp}),
\mathcal{\bar{N}}_B(\vec{x}_{2}^{\, \perp})
\right]&=
-2
i\bar{\delta}(\vec{x}_{12}^{\, \perp})
D_{[A}\mathcal{\bar{N}}_{B]}
+
i
\mathcal{\bar{N}}_A(\vec{x}_1^{\, \perp})
D_B
\bar{\delta}(\vec{x}_{12}^{\, \perp})+
i
\mathcal{\bar{N}}_B(\vec{x}_2^{\, \perp})D_A
\bar{\delta}(\vec{x}_{12}^{\, \perp})
\ .
\end{aligned}
\end{equation}
We have replaced the ordinary derivatives $\partial_A$ by covariant derivatives $D_A$ on the unit sphere $S^{d-2}$. While this is trivial for the derivatives acting on scalars, it is also allowed for the ones acting on $\bar{\mathcal{N}}_A$ since the anti-symmetric combination means the connection of $S^{d-2}$ does not contribute.

We can now construct the analogous charges to (\ref{eq:181}) for the ${\rm AdS}_2\times S^{d-2}$ case
\begin{equation}\label{eq:210}
\begin{aligned}
\bar{\mathcal{T}}(f)&=
\int_{S^{d-2}}d\Omega(\vec{x}_\perp)
f(\vec{x}_\perp)\bar{\mathcal{E}}(\vec{x}_\perp)\ ,\\
\bar{\mathcal{R}}(Y)&=
\int_{S^{d-2}}d\Omega(\vec{x}_\perp)
\left[
\frac{(D\cdot Y)}{d-2}
\bar{\mathcal{K}}(\vec{x}_\perp)+
Y^A\bar{\mathcal{N}}_A(\vec{x}_\perp)
\right]\ ,\\
\bar{\mathcal{D}}(g)&=
\int_{S^{d-2}}d\Omega(\vec{x}_\perp)
g(\vec{x}_\perp)
\bar{\mathcal{K}}(\vec{x}_\perp)\ ,
\end{aligned}
\end{equation}
and use (\ref{eq:58}) to compute its algebra, so that we finally find (\ref{eq:190}). As an example let us write the computation of the second commutator explicitly
\begin{equation}
\begin{aligned}
\left[
\mathcal{\bar{T}}(f),
\bar{\mathcal{R}}(Y)
\right]&=
i
\int_{S^{d-2}}
d\Omega(\vec{x}_\perp)
\frac{(D\cdot Y)}{d-2}
f(\vec{x}_\perp)
\mathcal{\bar{E}}(\vec{x}_\perp)
+
\int_{S^{d-2}}
d\Omega(\vec{x}_1^{\,\perp})
d\Omega(\vec{x}_2^{\,\perp})
f(\vec{x}_1^{\,\perp})
Y^A(\vec{x}_2^{\,\perp})
\left[
\bar{\mathcal{E}}(\vec{x}_1^{\,\perp}),
\bar{\mathcal{N}}_A(\vec{x}_2^{\,\perp})
\right]\\
&=i
\int_{S^{d-2}}
d\Omega(\vec{x}_\perp)
\frac{(D\cdot Y)}{d-2}
f(\vec{x}_\perp)
\mathcal{\bar{E}}(\vec{x}_\perp)
-i
\int_{S^{d-2}}
d\Omega(\vec{x}_\perp)
Y^A(\vec{x}_\perp)
D_Af(\vec{x}_\perp)=
i\bar{\mathcal{T}}(\widehat{f})\ ,
\end{aligned}
\end{equation} 
where we have been careful with the signs when integrating $D_A\bar{\delta}(\vec{x}_{12}^{\, \perp})$ and we have defined $\widehat{f}(\vec{x}_\perp)$ as in the second line in (\ref{eq:190}). The rest of the algebra in (\ref{eq:190}) follows from analogous computations. The most involved commutator is the one involving two operators $\bar{\mathcal{R}}$, where we must be careful with signs and the commutation of covariant derivatives. 

Finally, we can relate the parametrization coordinates $(\lambda,\vec{x}_\perp)$ to the space-time coordinates $(u,\rho,\vec{y}\,)$ used to described the ${\rm AdS}_2\times S^{d-2}$ metric in (\ref{eq:78}). We can do this using the description of the surface in the global coordinates in (\ref{eq:36}) together with the relations between the coordinates in (\ref{eq:116}), which gives
\begin{equation}
(\theta_+,\theta_-,\vec{y}\,)=
\left(\theta_0,
2\,{\rm arctan}\left[
\frac{2p(\vec{x}_\perp)}
{\lambda+p(\vec{x}_\perp)}
\right]-\theta_0,
\vec{x}_\perp
\right)
\quad \Longrightarrow \quad
(u,\rho,\vec{y}\,)=\left(1/2+
\frac{\lambda}{2p(\vec{x}_\perp)}
,0,\vec{x}_\perp
\right).
\end{equation}
Using this we can write the light-ray operators in (\ref{eq:210}) in terms of the space-time coordinates $(u,\rho,\vec{y}\,)$ as in (\ref{eq:184})-(\ref{eq:191}). Shifting $u\rightarrow u-1/2$ we see that $\lambda$ is proportional to $u$, meaning that only the light-ray operator shifts according to $\bar{\mathcal{E}}(\vec{x}_\perp)\rightarrow 2p(\vec{x}_\perp)\bar{\mathcal{E}}(\vec{x}_\perp)$, which can be absorbed in the supertranslations charge $\bar{\mathcal{T}}(f)$ in (\ref{eq:210}) by the appropriate definition of the function $f$. The parametrization coordinate $\vec{x}_\perp$ is the same as $\vec{y}$, so that after the charges (\ref{eq:210}) become (\ref{eq:181}) (without the factor $1/\rho^{d-2}$, that is not present for the ${\rm AdS}_2\times S^{d-2}$ case).

\section{Discrete symmetry of quantum theory}
\label{zapp:2}

In this appendix we study discrete symmetries of QFTs in Minkowski and ${\rm AdS}_2\times S^{d-2}$ that arise from the isometries of the Euclidean theory. Let us start by considering a theory defined in the Euclidean plane in Cartesian coordinates $(t_E,x_1,\vec{x}\,)$ which has a symmetry group given by ${\rm SO}(d)$. In particular we can consider a $\pi$ rotation in the plane $(t_E,x_1)$ given by
\begin{equation}\label{eq:222}
{\rm Euclidean\,\,rotation}:
  \qquad(t_E,x_1,\vec{x}\,)
  \quad \longrightarrow \quad
  (-t_E,-x_1,\vec{x}\,)\ .
\end{equation}
Upon analytic continuation $t_E=it$ this gives the discrete ${\rm CRT}$ symmetry given in (\ref{eq:220}). If we change from Cartesian coordinates in Minkowski to spherical, defined as
\begin{equation}
x_1=r
  \left(
  \frac{|\vec{y}\,|^2-1}
  {|\vec{y}\,|^2+1}
  \right)\ ,
  \qquad \qquad
\vec{x}=r
  \left(\frac{2\vec{y}}
  {|\vec{y}\,|^2+1}\right)\ ,
\end{equation}
the Minkowski metric becomes
$$ds^2=-dt^2+dx_1^2+|d\vec{x}\,|^2=
  -dt^2+dr^2+r^2d\Omega_{d-2}^2\ ,$$
where $d\Omega_{d-2}$ is written in stereographic coordinates as in (\ref{eq:176}). The Minkowski ${\rm CRT}$ transformation (\ref{eq:220}) in these coordinates becomes
\begin{equation}\label{eq:106}
{\rm CRT}:
\qquad
(t,r,\vec{y}\,)
\quad \longrightarrow \quad
\left(
-t,r,\frac{\vec{y}}{|\vec{y}\,|^2}
\right)\ .
\end{equation}
For even space-time dimension we can equivalently consider the ${\rm CPT}$ symmetry instead, obtained by applying a $\pi$ rotation on the remaining spatial Cartesian coordinates $\vec{x}$, so that ${\rm CRT}$ becomes ${\rm CPT}$, i.e. $(t,x_1,\vec{x}\,)\rightarrow -(t,x_1,\vec{x}\,)$. In terms of the transformation written in spherical coordinates in (\ref{eq:106}), this corresponds to adding an additional minus sign on the $\vec{y}$ inversion in (\ref{eq:106}). For the unit sphere $S^{d-2}$ the $\vec{y}$ inversion with the minus sign corresponds to the antipodal map, as can be seen by noting the unit vector $\vec{n}\in \mathbb{R}^{d-1}$ defining the sphere $S^{d-2}$ transforms as
$${\rm CPT}:
  \qquad
  \vec{n}=
  \left(
  \frac{|\vec{y}\,|^2-1}
  {|\vec{y}\,|^2+1},
  \frac{2\vec{y}}{|\vec{y}\,|^2+1}
  \right)
  \quad \longrightarrow \quad
  -\vec{n}\ .$$
It is only for even dimension that we have this CPT transformation, as for odd $d$ the reflection in the Cartesian coordinates $\vec{x}\rightarrow -\vec{x}$ we started from is not a part of the connected group of the Euclidean symmetry group.

Let us now consider a QFT in ${\rm AdS}_2\times S^{d-2}$ and show the ${\rm CRT}$ transformation in (\ref{eq:106}) is also a symmetry of the QFT. While for conformal theories this follows after using the space-times are related by a conformal transformation (see subsection \ref{subsec:1}), we want to give a more explicit proof working directly in ${\rm AdS}_2\times S^{d-2}$. We do this using the embedding space formalism of the conformal group, whose main idea is to embed the space-time of the CFT into a larger space where conformal transformations act linearly. Since the conformal group is isomorphic to ${\rm SO}(d,2)$ we define the embedding coordinates~$X\in\mathbb{R}^{d,2}$ 
$$X=(X^0,X^i,X^d,X^{d+1})\ ,$$
in the space
\begin{equation}\label{eq:87}
ds^2=
  -(dX^0)^2
  +\sum_{i=1}^{d-1}(dX^i)^2
  +\left[
  (dX^{d})^2-(dX^{d+1})^2
  \right]
  \ .
\end{equation}
Every group element $g\in {\rm SO}(d,2)$ has a representation in terms of a matrix $\Lambda_g$ which has a linear action in the embedding coordinates given by ordinary matrix multiplication $X'=\Lambda_g\cdot X$. The relation with the $d$-dimensional space-time is obtained by considering the projective null cone
\begin{equation}
\mathcal{PC}=
  \frac{\left\lbrace  
  X\in \mathbb{R}^{2,d}\,\,:
  \quad
  (X \cdot X)=0
  \right\rbrace}
  {X\sim c \,X\ ,\,\, c\in \mathbb{R}_{+}}\ ,
\end{equation}
where $(X\cdot X)$ is computed using the embedding metric (\ref{eq:87}). The denominator means there is a gauge redundancy in the scaling of $X$. This gauge freedom can be used to fix one of the components in the vector $X$ arbitrarily, which is usually taken as
\begin{equation}\label{eq:99}
X^+=X^d+X^{d+1}={\rm fixed}\ .
\end{equation}
This condition is equivalent to fixing the conformal frame where the $d$-dimensional theory is defined.

To describe a CFT in the metric ${\rm AdS}_2\times S^{d-2}$ we consider the following parametrization of the projective null cone
\begin{equation}\label{eq:103}
X=
  \left(
  \frac{\sin(\sigma)}{\sin(\theta)},
  \vec{n},
  \cot(\theta),
  \frac{\cos(\sigma)}{\sin(\theta)}
  \right)\ ,
\end{equation}
where $\vec{n}\in \mathbb{R}^{d-1}$ has unit norm $|\vec{n}|=1$. It is straightforward to check this vector is null in the embedding space, with gauge fixing
\begin{equation}\label{eq:102}
X^+=\frac{\cos(\theta)+\cos(\sigma)}{\sin(\theta)}\ .
\end{equation}
A convenient parametrization for the unit vector $\vec{n}$ is obtained by taking stereographic coordinates $\vec{y}\in \mathbb{R}^{d-2}$, so that (\ref{eq:103}) becomes
\begin{equation}\label{eq:88}
X(\sigma,\theta,\vec{y}\,)=
  \left(
  \frac{\sin(\sigma)}{\sin(\theta)},
  \frac{2y^A}{1+|\vec{y}\,|^2},
  \frac{|\vec{y}\,|^2-1}
  {|\vec{y}\,|^2+1},
  \cot(\theta),
  \frac{\cos(\sigma)}{\sin(\theta)}
  \right)\ .
\end{equation}
Computing the induced metric we find
$$ds^2_{\rm ind}=
  dX(\sigma,\theta,\vec{y}\,)
  \cdot dX(\sigma,\theta,\vec{y}\,)=
  \frac{-d\sigma^2+d\theta^2}
  {\sin^2(\theta)}+
  \frac{4d\vec{y}.d\vec{y}}
  {(1+|\vec{y}\,|)^2}\ ,$$
that is precisely the ${\rm AdS}_2\times S^{d-2}$ metric we are interested in. 

Let us now consider a $\pi$ rotation in the embedding space between the coordinates $(X^0,X^{d-1})$
\begin{equation}\label{eq:101}
(X^0,X^{d-1})
  \quad \longrightarrow \quad
  -(X^0,X^{d-1})\ .
\end{equation}
While this is not a transformation of the Lorentzian conformal group ${\rm SO}(d,2)$ it is in the Euclidean group ${\rm SO}(d-1,1)$, similarly to the Minkowski case in (\ref{eq:222}). Using this on (\ref{eq:88}) we get
\begin{equation}\label{eq:90}
\tilde{X}(\sigma,\theta,\vec{y}\,)=
  \left(
  \frac{\sin(-\sigma)}{\sin(\theta)},
  \frac{2y^A}{1+|\vec{y}\,|^2},
  -\frac{|\vec{y}\,|^2-1}
  {|\vec{y}\,|^2+1},
  \cot(\theta),
  \frac{\cos(\sigma)}{\sin(\theta)}
  \right)\ .
\end{equation}
Comparing with (\ref{eq:88}) we see the transformation induced in the $d$-dimensional coordinates $(\sigma,\theta,\vec{y}\,)$ is precisely given by the ${\rm CRT}$ transformation in (\ref{eq:128}). Moreover, since the transformation (\ref{eq:101}) does not change the gauge fixing condition in (\ref{eq:102}) the ${\rm CRT}$ transformation is an exact isometry of ${\rm AdS}_2\times S^{d-2}$, meaning it is actually a symmetry of the vacuum state of any QFT, not necessarily conformal.

\section{Commutator identity}
\label{zapp:4}

In this appendix we compute the following commutator
\begin{equation}\label{eq:51}
\left[
 W,e^{-iV}
 \right]=-
 \sum_{n=1}^{\infty}
 \frac{(-i)^n}{n!}
 \left[
 V^n,W
 \right]\ .
\end{equation}
To do so, we use induction to prove the following identity
\begin{equation}\label{eq:41}
\left[
  V^n,
  W
  \right]=
\sum_{m=1}^n
(-1)^{m+1}
\binom{n}{m}
  V^{n-m}
  \mathscr{L}^m_{V}
  \left(
  W
  \right)\ .
\end{equation}
For $n=1$ the identity is obviously true, so let us assume it holds for $n$ and show it does for $n+1$, where we have
$$\left[
  V^{n+1},W
  \right]=
  V^n
  \mathscr{L}_V(W)+
  \sum_{m=1}^{n}(-1)^{m+1}
  \binom{n}{m}
  V^{n-m}
  \mathscr{L}_V^m(W)
  V\ ,$$
and we have already used our hypothesis (\ref{eq:41}). Rearrange the terms to find
$$
\begin{aligned}
\left[
  V^{n+1},W
  \right]&=
  V^n
  \mathscr{L}_V(W)+
  \sum_{m=1}^{n}(-1)^{m+1}
  \binom{n}{m}
  V^{n-m}
  \left(
  V
  \mathscr{L}_V^m(W)
  -\mathscr{L}_V^{m+1}(W)
  \right)\\
  &=(n+1)V^n\mathscr{L}_V(W)+
  \sum_{k=2}^{n}
  (-1)^{k+1}
  \binom{n}{k}
  V^{(n+1)-k}
  \mathscr{L}_V^k(W)+
  \sum_{k=2}^{n+1}
  (-1)^{k+1}
  \binom{n}{k-1}
  V^{(n+1)-k}
  \mathscr{L}_V^k(W)\ .
\end{aligned}
$$
where in the first line we have redefined the summation indices $k=m$ and $k=m+1$ for each term. Rearranging one more time, we find
$$
\begin{aligned}
\left[
  V^{n+1},W
  \right]=(n+1)V^n\mathscr{L}_V(W)+
  \sum_{k=2}^{n}
  (-1)^{k+1}
  \left\lbrace
  \binom{n}{k}+
  \binom{n}{k-1}
  \right\rbrace
  V^{(n+1)-k}
  \mathscr{L}_V^k(W)+
  (-1)^{n}
  \mathscr{L}_V^{n+1}(W)\ ,
\end{aligned}
$$
The term between curly brackets is the binomial $\binom{n+1}{k}$, while the two additional terms give the $k=1$ and $k=n+1$ contributions so that we recover (\ref{eq:41}) with $n+1$. Using this in (\ref{eq:51}) we get the following identity
\begin{equation}\label{eq:66}
\left[
 W,e^{-iV}
 \right]=
 \sum_{n=1}^{\infty}
 \sum_{m=1}^n
 \binom{n}{m}
 \frac{(-i)^n
 (-1)^{m}}{n!}
  V^{n-m}
  \mathscr{L}^m_{V}
  \left(
  W
  \right)\ ,
\end{equation}
that we apply in our analysis in (\ref{eq:53}).

\section{Bulk superrotation integral curves}
\label{zapp:5}

The integral curves associated to bulk vector $\chi_R(Y)$ at $\rho=0$ (\ref{eq:154}) are determined by the following system of equations
\begin{equation}
\begin{aligned}
u'(s)&=
\frac{Y'(\phi(s))}{1+x(s)^2}
u(s)\ ,
\\
\phi'(s)&=
Y(\phi(s))\ ,
\\
x'(s)&=Y'(\phi(s))x(s)
 \ ,
\end{aligned}
\end{equation}
where $s$ is the parameter along the curve. We must solve for the three functions $(u(s),\phi(s),x(s))$ with initial conditions such that $s=1$ gives the identity. The second differential equation can be written in terms of a simple integral of $Y(\phi)$
\begin{equation}
s=\int_{\phi}^{\phi(s)}
\frac{d\phi'}{Y(\phi')}
\qquad \Longrightarrow \qquad
\phi(s)\equiv \alpha(\phi,s)\ .
\end{equation}
For any well behaved function $Y(\phi)$ the integral can be solved and inverted so that the function $\alpha(\phi,s)$ is written explicitly. The remaining equations can be easily solved in terms of $\alpha(\phi,s)$ as
\begin{equation}\label{eq:156}
\begin{aligned}
x(s)&=
(\partial_\phi \alpha)
x\\
u(s)&=
\frac{\sqrt{1+x^2}}
{\sqrt{1+(\partial_\phi \alpha)^2 x^2}}
(\partial_\phi \alpha)
u
\end{aligned}
\ ,
\qquad \qquad
(\partial_\phi \alpha)=
\frac{Y(\phi(s))}{Y(\phi)}\ ,
\end{equation}
where we have already imposed the appropriate initial conditions at $s=0$. The superrotation transformation at $\rho=0$ is implemented by evaluating at $s=1$, so that we get the final result in (\ref{eq:226}), where we have defined $\alpha(\phi)=\alpha(s=1,\phi)$.

\section{Gravitational surface charges}
\label{zapp:6}

To an asymptotic Killing vector $\chi$ one can  associate a surface charge through the covariant phase space formalism \cite{Barnich:2001jy}. More precisely, one calculates via this approach the field-variation of a charge, which is a one-form in the configuration space:
\begin{equation} \label{deltaQ}
\delta Q[\chi]=\oint\mathbf{k}[\chi],
\end{equation}
where $\mathbf{k}[\chi]$ depends on the metric $g$ and its phase-space variation $\delta g\equiv h$. It is a one-form with respect to the field configuration space but a $(d-2)$-form w.r.t. the spacetime. Its explicit expression depends on the theory under consideration; for Einstein gravity (with or without cosmological constant) it is given by
\begin{equation}
\mathbf{k}[\chi]=\frac{\sqrt{-g}}{8\pi G}(d^{d-2}x)_{\mu \nu}\left(\chi^\mu \nabla_\rho h^{\nu \rho}-\chi^\mu \nabla^\nu h+\chi_\rho\nabla^\nu h^{\mu \rho}+\frac{1}{2}h\nabla^ \nu\chi^\mu -h^{\rho \nu}\nabla_\rho\chi^\mu \right),
\end{equation}
where $h=g^{\mu \nu}h_{\mu \nu}$ and $(d^{d-2}x)_{\mu\nu}=\frac{1}{2(d-2)!}\epsilon_{\mu  \nu \rho_1\ldots \rho_{d-2}}dx^{\rho_1}\wedge\ldots\wedge dx^{\rho_{d-2}}$. In the integrable case, i.e. when expression \eqref{deltaQ} is $\delta$-exact, $Q[\chi]$ represents the generator of the associated infinitesimal transformation $\chi$.
\bibliography{sample}

\providecommand{\href}[2]{#2}\begingroup\raggedright\begin{thebibliography}{10}

\bibitem{Sachs:1962wk}
R.~K. Sachs, {\it {Gravitational waves in general relativity. 8. Waves in
  asymptotically flat space-times}},  {\em Proc. Roy. Soc. Lond.} {\bf A270}
  (1962) 103--126.

\bibitem{Sachs:1962zza}
R.~Sachs, {\it {Asymptotic symmetries in gravitational theory}},  {\em Phys.
  Rev.} {\bf 128} (1962) 2851--2864.

\bibitem{Bondi:1962px}
H.~Bondi, M.~G.~J. van~der Burg, and A.~W.~K. Metzner, {\it {Gravitational
  waves in general relativity. 7. Waves from axisymmetric isolated systems}},
  {\em Proc. Roy. Soc. Lond.} {\bf A269} (1962) 21--52.

\bibitem{Barnich:2010eb}
G.~Barnich and C.~Troessaert, {\it {Aspects of the BMS/CFT correspondence}},
  {\em JHEP} {\bf 05} (2010) 062, [\href{http://arxiv.org/abs/1001.1541}{{\tt
  arXiv:1001.1541}}].

\bibitem{Barnich:2011mi}
G.~Barnich and C.~Troessaert, {\it {BMS charge algebra}},  {\em JHEP} {\bf 12}
  (2011) 105, [\href{http://arxiv.org/abs/1106.0213}{{\tt arXiv:1106.0213}}].

\bibitem{Barnich:2016lyg}
G.~Barnich and C.~Troessaert, {\it {Finite BMS transformations}},  {\em JHEP}
  {\bf 03} (2016) 167, [\href{http://arxiv.org/abs/1601.04090}{{\tt
  arXiv:1601.04090}}].

\bibitem{Compere:2018ylh}
G.~Comp\`ere, A.~Fiorucci, and R.~Ruzziconi, {\it {Superboost transitions,
  refraction memory and super-Lorentz charge algebra}},  {\em JHEP} {\bf 11}
  (2018) 200, [\href{http://arxiv.org/abs/1810.00377}{{\tt arXiv:1810.00377}}].

\bibitem{Brown:1986nw}
J.~D. Brown and M.~Henneaux, {\it {Central Charges in the Canonical Realization
  of Asymptotic Symmetries: An Example from Three-Dimensional Gravity}},  {\em
  Commun. Math. Phys.} {\bf 104} (1986) 207--226.

\bibitem{Guica:2008mu}
M.~Guica, T.~Hartman, W.~Song, and A.~Strominger, {\it {The Kerr/CFT
  Correspondence}},  {\em Phys. Rev. D} {\bf 80} (2009) 124008,
  [\href{http://arxiv.org/abs/0809.4266}{{\tt arXiv:0809.4266}}].

\bibitem{Strominger:2013jfa}
A.~Strominger, {\it {On BMS Invariance of Gravitational Scattering}},  {\em
  JHEP} {\bf 07} (2014) 152, [\href{http://arxiv.org/abs/1312.2229}{{\tt
  arXiv:1312.2229}}].

\bibitem{He:2014laa}
T.~He, V.~Lysov, P.~Mitra, and A.~Strominger, {\it {BMS supertranslations and
  Weinberg's soft graviton theorem}},  {\em JHEP} {\bf 05} (2015) 151,
  [\href{http://arxiv.org/abs/1401.7026}{{\tt arXiv:1401.7026}}].

\bibitem{Strominger:2014pwa}
A.~Strominger and A.~Zhiboedov, {\it {Gravitational Memory, BMS
  Supertranslations and Soft Theorems}},  {\em JHEP} {\bf 01} (2016) 086,
  [\href{http://arxiv.org/abs/1411.5745}{{\tt arXiv:1411.5745}}].

\bibitem{Strominger:2017zoo}
A.~Strominger, {\em {Lectures on the Infrared Structure of Gravity and Gauge
  Theory}}.
\newblock {Princeton University Press}, 2018.

\bibitem{Donnay:2015abr}
L.~Donnay, G.~Giribet, H.~A. Gonzalez, and M.~Pino, {\it {Supertranslations and
  Superrotations at the Black Hole Horizon}},  {\em Phys. Rev. Lett.} {\bf 116}
  (2016), no.~9 091101, [\href{http://arxiv.org/abs/1511.08687}{{\tt
  arXiv:1511.08687}}].

\bibitem{Donnay:2016ejv}
L.~Donnay, G.~Giribet, H.~A. González, and M.~Pino, {\it {Extended Symmetries
  at the Black Hole Horizon}},  {\em JHEP} {\bf 09} (2016) 100,
  [\href{http://arxiv.org/abs/1607.05703}{{\tt arXiv:1607.05703}}].

\bibitem{Hawking:2016msc}
S.~W. Hawking, M.~J. Perry, and A.~Strominger, {\it {Soft Hair on Black
  Holes}},  {\em Phys. Rev. Lett.} {\bf 116} (2016), no.~23 231301,
  [\href{http://arxiv.org/abs/1601.00921}{{\tt arXiv:1601.00921}}].

\bibitem{Hawking:2016sgy}
S.~W. Hawking, M.~J. Perry, and A.~Strominger, {\it {Superrotation Charge and
  Supertranslation Hair on Black Holes}},  {\em JHEP} {\bf 05} (2017) 161,
  [\href{http://arxiv.org/abs/1611.09175}{{\tt arXiv:1611.09175}}].

\bibitem{Haco:2017ekf}
S.~J. Haco, S.~W. Hawking, M.~J. Perry, and J.~L. Bourjaily, {\it {The
  Conformal BMS Group}},  {\em JHEP} {\bf 11} (2017) 012,
  [\href{http://arxiv.org/abs/1701.08110}{{\tt arXiv:1701.08110}}].

\bibitem{Adami:2020ugu}
H.~Adami, M.~Sheikh-Jabbari, V.~Taghiloo, H.~Yavartanoo, and C.~Zwikel, {\it
  {Symmetries at null boundaries: two and three dimensional gravity cases}},
  {\em JHEP} {\bf 10} (2020) 107, [\href{http://arxiv.org/abs/2007.12759}{{\tt
  arXiv:2007.12759}}].

\bibitem{Nguyen:2020hot}
K.~Nguyen and J.~Salzer, {\it {The Effective Action of Superrotation Modes}},
  \href{http://arxiv.org/abs/2008.03321}{{\tt arXiv:2008.03321}}.

\bibitem{Witten:2018lha}
E.~Witten, {\it {APS Medal for Exceptional Achievement in Research: Invited
  article on entanglement properties of quantum field theory}},  {\em Rev. Mod.
  Phys.} {\bf 90} (2018), no.~4 045003,
  [\href{http://arxiv.org/abs/1803.04993}{{\tt arXiv:1803.04993}}].

\bibitem{Cordova:2018ygx}
C.~Córdova and S.-H. Shao, {\it {Light-ray Operators and the BMS Algebra}},
  {\em Phys. Rev.} {\bf D98} (2018), no.~12 125015,
  [\href{http://arxiv.org/abs/1810.05706}{{\tt arXiv:1810.05706}}].

\bibitem{Kologlu:2019bco}
M.~Kologlu, P.~Kravchuk, D.~Simmons-Duffin, and A.~Zhiboedov, {\it {Shocks,
  Superconvergence, and a Stringy Equivalence Principle}},
  \href{http://arxiv.org/abs/1904.05905}{{\tt arXiv:1904.05905}}.

\bibitem{Faulkner:2016mzt}
T.~Faulkner, R.~G. Leigh, O.~Parrikar, and H.~Wang, {\it {Modular Hamiltonians
  for Deformed Half-Spaces and the Averaged Null Energy Condition}},  {\em
  JHEP} {\bf 09} (2016) 038, [\href{http://arxiv.org/abs/1605.08072}{{\tt
  arXiv:1605.08072}}].

\bibitem{Hartman:2016lgu}
T.~Hartman, S.~Kundu, and A.~Tajdini, {\it {Averaged Null Energy Condition from
  Causality}},  {\em JHEP} {\bf 07} (2017) 066,
  [\href{http://arxiv.org/abs/1610.05308}{{\tt arXiv:1610.05308}}].

\bibitem{Rosso:2020cub}
F.~Rosso, {\it {Achronal averaged null energy condition for extremal horizons
  and (A)dS}},  {\em JHEP} {\bf 07} (2020) 023,
  [\href{http://arxiv.org/abs/2005.06476}{{\tt arXiv:2005.06476}}].

\bibitem{Newman:1962cia}
E.~T. Newman and T.~W.~J. Unti, {\it {Behavior of Asymptotically Flat Empty
  Spaces}},  {\em J. Math. Phys.} {\bf 3} (1962), no.~5 891.

\bibitem{Barnich:2011ty}
G.~Barnich and P.-H. Lambert, {\it {A Note on the Newman-Unti group and the BMS
  charge algebra in terms of Newman-Penrose coefficients}},  {\em J. Phys.
  Conf. Ser.} {\bf 410} (2013) 012142,
  [\href{http://arxiv.org/abs/1102.0589}{{\tt arXiv:1102.0589}}].

\bibitem{Spradlin:1999bn}
M.~Spradlin and A.~Strominger, {\it {Vacuum states for AdS(2) black holes}},
  {\em JHEP} {\bf 11} (1999) 021,
  [\href{http://arxiv.org/abs/hep-th/9904143}{{\tt hep-th/9904143}}].

\bibitem{Akhmedov:2013vka}
E.~Akhmedov, {\it {Lecture notes on interacting quantum fields in de Sitter
  space}},  {\em Int. J. Mod. Phys. D} {\bf 23} (2014) 1430001,
  [\href{http://arxiv.org/abs/1309.2557}{{\tt arXiv:1309.2557}}].

\bibitem{Herzog:2013ed}
C.~P. Herzog and K.-W. Huang, {\it {Stress Tensors from Trace Anomalies in
  Conformal Field Theories}},  {\em Phys. Rev.} {\bf D87} (2013) 081901,
  [\href{http://arxiv.org/abs/1301.5002}{{\tt arXiv:1301.5002}}].

\bibitem{Rosso:2019txh}
F.~Rosso, {\it {Global aspects of conformal symmetry and the ANEC in dS and
  AdS}},  {\em JHEP} {\bf 03} (2020) 186,
  [\href{http://arxiv.org/abs/1912.08897}{{\tt arXiv:1912.08897}}].

\bibitem{Iizuka:2019ezn}
N.~Iizuka, A.~Ishibashi, and K.~Maeda, {\it {Conformally invariant averaged
  null energy condition from AdS/CFT}},
  \href{http://arxiv.org/abs/1911.02654}{{\tt arXiv:1911.02654}}.

\bibitem{Oblak:2016eij}
B.~Oblak, {\em {BMS Particles in Three Dimensions}}.
\newblock PhD thesis, Brussels U., 2016.
\newblock \href{http://arxiv.org/abs/1610.08526}{{\tt arXiv:1610.08526}}.

\bibitem{Campoleoni:2016vsh}
A.~Campoleoni, H.~A. Gonzalez, B.~Oblak, and M.~Riegler, {\it {BMS Modules in
  Three Dimensions}},  {\em Int. J. Mod. Phys. A} {\bf 31} (2016), no.~12
  1650068, [\href{http://arxiv.org/abs/1603.03812}{{\tt arXiv:1603.03812}}].

\bibitem{Brown:1992br}
J.~D. Brown and J.~W. York, Jr., {\it {Quasilocal energy and conserved charges
  derived from the gravitational action}},  {\em Phys. Rev.} {\bf D47} (1993)
  1407--1419, [\href{http://arxiv.org/abs/gr-qc/9209012}{{\tt gr-qc/9209012}}].

\bibitem{Balasubramanian:1999re}
V.~Balasubramanian and P.~Kraus, {\it {A Stress tensor for Anti-de Sitter
  gravity}},  {\em Commun. Math. Phys.} {\bf 208} (1999) 413--428,
  [\href{http://arxiv.org/abs/hep-th/9902121}{{\tt hep-th/9902121}}].

\bibitem{Emparan:1999pm}
R.~Emparan, C.~V. Johnson, and R.~C. Myers, {\it {Surface terms as counterterms
  in the AdS / CFT correspondence}},  {\em Phys. Rev.} {\bf D60} (1999) 104001,
  [\href{http://arxiv.org/abs/hep-th/9903238}{{\tt hep-th/9903238}}].

\bibitem{Laddha:2020kvp}
A.~Laddha, S.~G. Prabhu, S.~Raju, and P.~Shrivastava, {\it {The Holographic
  Nature of Null Infinity}},  \href{http://arxiv.org/abs/2002.02448}{{\tt
  arXiv:2002.02448}}.

\bibitem{Haco:2018ske}
S.~Haco, S.~W. Hawking, M.~J. Perry, and A.~Strominger, {\it {Black Hole
  Entropy and Soft Hair}},  {\em JHEP} {\bf 12} (2018) 098,
  [\href{http://arxiv.org/abs/1810.01847}{{\tt arXiv:1810.01847}}].

\bibitem{conf}
W.~Couch and R.~Torrence, {\it {Conformal invariance under spatial inversion of
  extreme Reissner-Nordström black holes}},  {\em General Relativity and
  Gravitation} {\bf 16} (1984).

\bibitem{TheNew}
K.~Fernandes, D.~Ghosh, and A.~Virmani, {\it {Horizon Hair from Inversion
  Symmetry}},  \href{http://arxiv.org/abs/2008.04365}{{\tt arXiv:2008.04365}}.

\bibitem{Barnich:2001jy}
G.~Barnich and F.~Brandt, {\it {Covariant theory of asymptotic symmetries,
  conservation laws and central charges}},  {\em Nucl. Phys. B} {\bf 633}
  (2002) 3--82, [\href{http://arxiv.org/abs/hep-th/0111246}{{\tt
  hep-th/0111246}}].

\end{thebibliography}\endgroup
\bibliographystyle{JHEP}

\end{document}